\documentclass[%
 reprint,
superscriptaddress,
 amsmath,amssymb,
 aps,
rmp,
]{revtex4-2}

\usepackage[utf8]{inputenc}
\usepackage[english]{babel}

\usepackage{graphicx}
\usepackage{dcolumn}
\usepackage{bm}
\usepackage{multirow, array, tabularx, booktabs, makecell}

\usepackage{physics}
\usepackage{amsthm}
\usepackage{quantikz}

\usepackage[dvipsnames]{xcolor}
\usepackage[normalem]{ulem}
\usepackage{enumitem}
\usepackage{rotating}
\usepackage{textcomp} 

\usepackage[colorlinks=true,linkcolor=blue,citecolor=blue,filecolor=blue,urlcolor=blue]{hyperref}

\begin{document}

\title{Quantum Deep Learning: A Comprehensive Review}

\author{Yanjun Ji}%
\email{y.ji@fz-juelich.de}
\affiliation{%
\mbox{Institute for Quantum Computing Analytics (PGI-12), Forschungszentrum Jülich, 52425 Jülich, Germany}
}%

\author{Zhao-Yun Chen}
\email{chenzhaoyun@iai.ustc.edu.cn}
\affiliation{%
\mbox{Institute of Artificial Intelligence, Hefei Comprehensive National Science Center, Hefei, Anhui, 230088, P. R. China}
}%

\author{Marco Roth}
\email{marco.roth@ipa.fraunhofer.de}
\affiliation{%
\mbox{Fraunhofer Institute for Manufacturing Engineering and Automation (IPA), Nobelstrasse 12, Stuttgart 70569, Germany}
}%

\author{David A. Kreplin}
\email{david.kreplin@hs-heilbronn.de}
\affiliation{%
\mbox{Heilbronn University of Applied Sciences, Max-Planck-Str. 39, 74081 Heilbronn, Germany}
}%

\author{Christian Schiffer}
\affiliation{\mbox{Institute of Neuroscience and Medicine (INM-1), Research Centre Jülich, Jülich, Germany}}
\affiliation{\mbox{Helmholtz AI, Research Centre Jülich, Jülich, Germany}}

\author{Martin King}
\affiliation{\mbox{Munich Center for Mathematical Philosophy, Ludwig Maximilian University of Munich, Munich, Germany}}
\affiliation{\mbox{Lichtenberg Group for History and Philosophy of Physics, University of Bonn, Bonn, Germany}
}

\author{Oliver Anton}
\email{oliver.anton@physik.hu-berlin.de}
\affiliation{Institut für Physik and Center for the Science of Materials Berlin (CSMB) Adlershof, \mbox{Humboldt-Universität zu Berlin, Newtonstrß 15, 12489 Berlin, Germany}}
\affiliation{\mbox{Ferdinand-Braun-Institut (FBH), Gustav-Kirchoff-Straße 4, 12489 Berlin, Germany}}

\author{M. Sahnawaz Alam}
\affiliation{%
\mbox{Institute for Automation of Complex Power Systems, RWTH Aachen University, Aachen, Germany}
}%

\author{Markus Krutzik}
\affiliation{Institut für Physik and Center for the Science of Materials Berlin (CSMB) Adlershof, \mbox{Humboldt-Universität zu Berlin, Newtonstrß 15, 12489 Berlin, Germany}}
\affiliation{\mbox{Ferdinand-Braun-Institut (FBH), Gustav-Kirchoff-Straße 4, 12489 Berlin, Germany}}

\author{Dennis Willsch}
\email{d.willsch@fz-juelich.de}
\affiliation{%
\mbox{Jülich Supercomputing Centre, Forschungszentrum Jülich, 52425 Jülich, Germany}
}%
\affiliation{
\mbox{Faculty of Medical Engineering and Technomathematics, University of Applied Sciences Aachen, 52428 Jülich, Germany}
}

\author{Ludwig Mathey}
\email{ludwig.mathey@uni-hamburg.de}
\affiliation{\mbox{Zentrum für Optische Quantentechnologien, Universität Hamburg, 22761 Hamburg, Germany}}
\affiliation{\mbox{Institut für Quantenphysik, Universität Hamburg, 22761 Hamburg, Germany}}
\affiliation{\mbox{The Hamburg Centre for Ultrafast Imaging, 22761 Hamburg, Germany}}

\author{Frank K. Wilhelm}
\email{f.wilhelm-mauch@fz-juelich.de}
\affiliation{%
\mbox{Institute for Quantum Computing Analytics (PGI-12), Forschungszentrum Jülich, 52425 Jülich, Germany}
}%
\affiliation{%
\mbox{Theoretical Physics, Saarland University, 66123 Saarbrücken, Germany}
}%

\author{Guo-Ping Guo}%
 \email{gpguo@ustc.edu.cn}
 \affiliation{%
\mbox{Institute of Artificial Intelligence, Hefei Comprehensive National Science Center, Hefei, Anhui, 230088, P. R. China}
}%
\affiliation{%
\mbox{Laboratory of Quantum Information, University of Science and Technology of China, Hefei, Anhui, 230026, P. R. China}
}%
\affiliation{
CAS Center for Excellence and Synergetic Innovation Center in Quantum Information and Quantum Physics, \mbox{University of Science and Technology of China, Hefei, Anhui 230026, P. R. China}
}
\affiliation{
\mbox{Hefei National Laboratory, University of Science and Technology of China, Hefei 230088, P. R. China}
}
\affiliation{
\mbox{Origin Quantum Computing Technology (Hefei) Co., Ltd., Hefei, Anhui, 230026, P. R. China}
}

\date{\today}

\begin{abstract}

Quantum deep learning (QDL) explores the use of both quantum and quantum-inspired resources to determine when deep learning's core capabilities, such as expressivity, generalization, and scalability, can be enhanced based on specific resource constraints.
Distinct from broader quantum machine learning, QDL emphasizes compositional depth at the pipeline level and the integration of quantum or quantum-inspired components within end-to-end workflows.
This review provides an operational definition of QDL and introduces a taxonomy comprising four primary paradigms: hybrid quantum-classical models, quantum deep neural networks, quantum algorithms for deep learning primitives, and quantum-inspired classical algorithms.
Theoretical principles are connected to advanced architectures, software toolchains, and experimental demonstrations across superconducting, trapped-ion, photonic, semiconductor spin, and neutral-atom systems, as well as quantum annealers. Claims of quantum advantage are critically assessed by distinguishing provable complexity-theoretic separations from empirical observations. The analysis characterizes trade-offs between model expressivity, trainability, and classical simulability, while systematically detailing the bottlenecks imposed by optimization landscapes, input-output access models, and hardware constraints.
Applications are surveyed in domains encompassing image classification, natural language processing, scientific discovery, quantum data processing, and quantum optimal control, underscoring fair benchmarking against optimized classical counterparts and a comprehensive assessment of resource requirements. This review serves as a tutorial entry point for graduate students while guiding readers to specialized literature. It concludes with a verification-aware roadmap to transition QDL from near-term demonstrations to scalable and fault-tolerant implementations.

\end{abstract}

\maketitle

\tableofcontents

\section{Introduction}

The intersection of artificial intelligence (AI) and quantum computing (QC) has grown into a rapidly expanding interdisciplinary domain spanning quantum learning theory, algorithm design, and hardware-driven resource constraints \cite{biamonte2017quantum,arunachalam2017guest,acampora2025quantum,alexeev2025artificial,schuld2021machine,chang2025primerquantummachinelearning}.
While deep learning (DL) provides the prevailing AI paradigm for representation learning and large-scale function approximation \cite{lecun2015deep,Goodfellow-et-al-2016}, QC has progressed from foundational theory \cite{nielsen2010quantum} to tangible experimental architectures \cite{arute2019quantum, wu2021strong}.
At the same time, continued performance gains increasingly demand disproportionate increases in computational resources, energy, and data \cite{Kaplan2020,hoffmann2022training,patterson2021carbon}, reflecting both the economics of large-scale training and slowing improvements in classical hardware scaling \cite{theis2017end}.
This dynamic motivates the central question that frames this review: can quantum and quantum-inspired resources change the cost-accuracy scaling of learning primitives once data access, readout, and noise are specified?
Against this backdrop of growing convergence, the present review uses the term quantum deep learning (QDL) in an operational sense. We distinguish it from AI for QC, an adjacent field where classical machine learning (ML) is used primarily for quantum-system design, control, compilation, or characterization \cite{alexeev2025artificial}.
Instead, our emphasis lies on learning primitives that explicitly exploit coherent quantum resources under realistic measurement and classical-control constraints.

\subsection{The convergence into QDL}\label{subsec:the_conver_qdl}

Classical DL originally drew on foundational neural concepts such as perceptrons \cite{rosenblatt1958perceptron}, multilayer networks \cite{ivakhnenko1965}, associative memory \cite{hopfield1982neural}, and energy-based probabilistic models \cite{ackley1985learning}. Subsequent progress relied on developments in gradient-based optimization and representational structure, including reverse-mode automatic differentiation \cite{linnainmaa1970representation}, error backpropagation for multilayer networks \cite{rumelhart1986learning}, and convolutional architectures ranging from the neocognitron \cite{fukushima1980neocognitron} to convolutional networks trained via backpropagation \cite{lecun1989backpropagation}. The 2024 Nobel Prize in Physics recognized foundational contributions to artificial neural networks, awarding Hopfield and Hinton for discoveries enabling ML with neural networks \cite{nobel2024physics}.

Modern DL’s resurgence was enabled by large labeled datasets, graphics processing unit (GPU)-accelerated training, and architectural and optimization advances, marked by AlexNet's accuracy gains on the ILSVRC-2012 ImageNet benchmark \cite{krizhevsky2012imagenet,Deng2009} over hand-engineered pipelines, followed by architectural innovations including VGGNet \cite{simonyan2014very}, GoogLeNet \cite{szegedy2015going}, and ResNet \cite{he2016deep}.
These breakthroughs rapidly expanded beyond computer vision, delivering major advances in natural language processing (NLP) via attention-based sequence modeling \cite{vaswani2017attention}, protein structure prediction \cite{jumper2021highly,tunyasuvunakool2021highly}, and materials discovery \cite{merchant2023scaling}.

In the domain of QC, modern theory took shape in the 1980s with Benioff's Hamiltonian formulation of reversible Turing machines \cite{benioff1980computer}, Manin's identification of complexity in simulating quantum dynamics \cite{manin1980vychislimoe}, and Feynman's argument that simulating generic quantum mechanical systems was plausibly hard for classical computation in general, motivating devices operating on quantum principles as an alternative \cite{feynman1982simulating,feynman1985quantum}.
Deutsch further formalized the universal quantum computer, establishing a theoretical framework for QC  \cite{deutsch1985quantum}.
Subsequent algorithmic advances included Shor’s polynomial-time quantum factoring and discrete logarithms \cite{shor1994algorithms,shor1997polynomial}, Grover’s quadratic speedup for unstructured search \cite{grover1996,grover1997search}, and Lloyd's efficient simulation of local quantum systems \cite{lloyd1996universal}.
Quantum error correction (QEC) \cite{shor1996fault,steane1996error} and threshold theorems \cite{knill1996thresholdaccuracyquantumcomputation,preskill1997faulttolerantquantumcomputation,knill1998resilient,aharonov1997fault} clarified the path toward physical realization, establishing that scalable fault-tolerant computation is possible under suitable locality assumptions if physical error rates lie below an architecture- and noise-model-dependent threshold.

As a theoretical prelude to QDL, early quantum neural proposals drew heuristic correspondences between neural primitives and quantum state manipulation \cite{kak1995quantum,chrisley1995quantum,menneer1995quantum,purushothaman1997QNNs}. Related associative-memory constructions leveraged amplitude-amplification-style primitives \cite{grover1997search,ventura2000quantum,trugenberger2001probabilistic}.
This early literature revealed constraints absent in classical  neural networks, particularly the limitations imposed by linear unitary evolution and the no-cloning theorem \cite{Wootters1982NoCloning,dieks1982communication}. Consequently, effective nonlinearity was typically realized via measurement-conditioned classical postprocessing and feedback \cite{schuld2021machine}, although coherent alternatives leveraging nonlinear media or many-body interactions remain a significant theoretical frontier.

Proposals oriented toward fault tolerance connected learning primitives to quantum subroutines, notably quantum linear-algebraic (QLA) routines \cite{harrow2009quantum} and quantum principal component analysis primitives \cite{Lloyd2014}, typically under strong input-access assumptions.
In the mid-2010s, \textcite{wiebe2014quantum} framed ``quantum deep learning" as accelerating deep Boltzmann machines under strong data-access assumptions.
Related studies explored classical classification as a task suitable for quantum implementation via kernel estimation or linear-algebraic primitives \cite{rebentrost2014quantum}.
The advent of noisy intermediate-scale quantum (NISQ) hardware \cite{preskill2018quantum} fundamentally recalibrated these expectations, as severe limitations in coherent circuit depth, gate fidelities, and sampling overhead strictly bounded practical algorithmic viability.
Attention shifted to variational hybrid loops \cite{cerezo2021variational,peruzzo2014variational,mcclean2016theory} in which a parameterized quantum circuit (PQC) prepares a state, measurements estimate an objective, and a classical optimizer updates parameters in a feedback loop \cite{tilly2022variational,sim2019expressibility}, with hardware-efficient ans\"atze reducing compiled depth under native constraints \cite{kandala2017hardware}.
Subsequent architectural diversification introduced concepts such as hierarchy, locality, and compositionality \cite{farhi2018classification,cong2019quantum}, data re-uploading \cite{perez2020data}, transfer learning \cite{mari2020transfer}, and representative generative and structured models \cite{lloyd2018quantum,dallaire2018quantum,bausch2020recurrent,verdon2019quantum}, alongside early experimental end-to-end training demonstrations \cite{beer2020training,hu2019quantum}.

However, trainability constraints emerged as circuit depth and complexity increased. These manifested primarily as barren plateau phenomena \cite{mcclean2018barren}, which motivate locality-aware costs and structured designs \cite{cerezo2021cost,pesah2021absence} (see Sec.~\ref{subsubsec:resou_codesign}).
In parallel, variational classifiers were theoretically reframed as kernel machines, with kernels defined by quantum feature maps \cite{schuld2021supervised}, e.g., the ZZ feature map \cite{havlivcek2019supervised}, which encodes data via entangling ZZ interactions between qubits.
Complexity-theoretic arguments identify regimes where the induced feature states are conjectured to be costly to simulate classically, while information-theoretic analyses delimit when any end-to-end advantage can survive finite data, finite shots, and noise \cite{huang2021information,huang2021power}.
Furthermore, dequantization and quantum-inspired classical algorithms emphasize that apparent separations can disappear when classical baselines are granted comparable access models \cite{tang2019quantum,tang2021quantum}.
This synthesis motivates the current focus on task-specific quantum utility under explicit resource contracts \cite{chang2025primerquantummachinelearning,preskill2025beyond}.
A central unresolved issue in the field is to characterize and control the three-way trade-off among expressivity, trainability, and classical simulatability. Architectures that are sufficiently expressive to represent useful hypothesis classes often become difficult to optimize due to vanishing-gradient phenomena. Conversely, architectures engineered for stable optimization are typically shallow or structured enough to admit strong classical simulation and competitive classical baseline performance.

Recent breakthroughs in quantum hardware are signaling a gradual transition from NISQ heuristics toward the fault-tolerant regime assumed by early QLA proposals. Experiments have successfully demonstrated encoded logical qubits and repeated error-correction cycles across multiple architectures, notably in trapped ions \cite{paetznick2024demonstration}, neutral atoms \cite{bluvstein2024logical}, and superconducting platforms \cite{google2025quantum}. Concurrently, the maturation of differentiable software toolchains \cite{bergholm2018pennylane, broughton2020tensorflow, kreplin2025squlearn, dou2022qpanda, bian2023vqnet} has transformed QDL from theoretical speculation into an operational reality, enabling end-to-end optimization across quantum-classical boundaries at small and intermediate scales. As physical deployments scale, navigating the associated trade-offs increasingly hinges on rigorous benchmarking against strong classical baselines. This requires leveraging high-performance classical simulators \cite{deraedt2019juqcs} and adhering to best-practice protocols for fair, resource-matched evaluation \cite{bowles2024better}.
However, rapid interdisciplinary progress has also produced a deeply fragmented literature, with results scattered across physics, computer science, and ML venues using incompatible terminology and evaluation methodology.
These coupled tensions motivate the central theme of this review: identifying regimes in which end-to-end quantum utility survives realistic access, noise, and benchmarking constraints, and articulating the corresponding resource contracts under which such claims are meaningful. To unify this fragmented landscape, we establish a rigorous terminology and structural taxonomy.

\subsection{Definition and taxonomy}

To formalize our scope, we adopt the following operational definition.

\emph{Definition~1 (QDL).---}QDL refers to compositional learning architectures and workflows exhibiting hierarchical depth at the level of the overall learning pipeline, in which at least one quantum or quantum-inspired component is functional and architecturally central. Concretely, the learning objective or update rule depends on this component's outputs, and its inclusion nontrivially determines the hypothesis class and/or estimator family under an explicitly stated access and resource model.

This definition distinguishes QDL from broader QML by requiring compositional depth at the pipeline level and by making access models and resource assumptions part of the problem specification.
Consequently, QDL includes both hardware-executed and quantum-inspired classical realizations of quantum components. However, it explicitly excludes AI-for-QC studies as well as standalone shallow quantum feature maps, kernels, or single-circuit predictors, provided they are used solely as end-to-end models rather than embedded components of a hierarchically deep pipeline. Examples of QDL under this framework include: (i) a PQC head on a frozen or trainable classical backbone; (ii) a quantum sampling subroutine used only at deployment within a deep generative pipeline; and (iii) dequantization baselines embedded in a matched-access evaluation.
We adopt this convention to align ``deep" with hierarchical composition and to keep access models and end-to-end costing explicit.

\begin{figure*}[tb]\centering
\includegraphics[width=.9\textwidth]{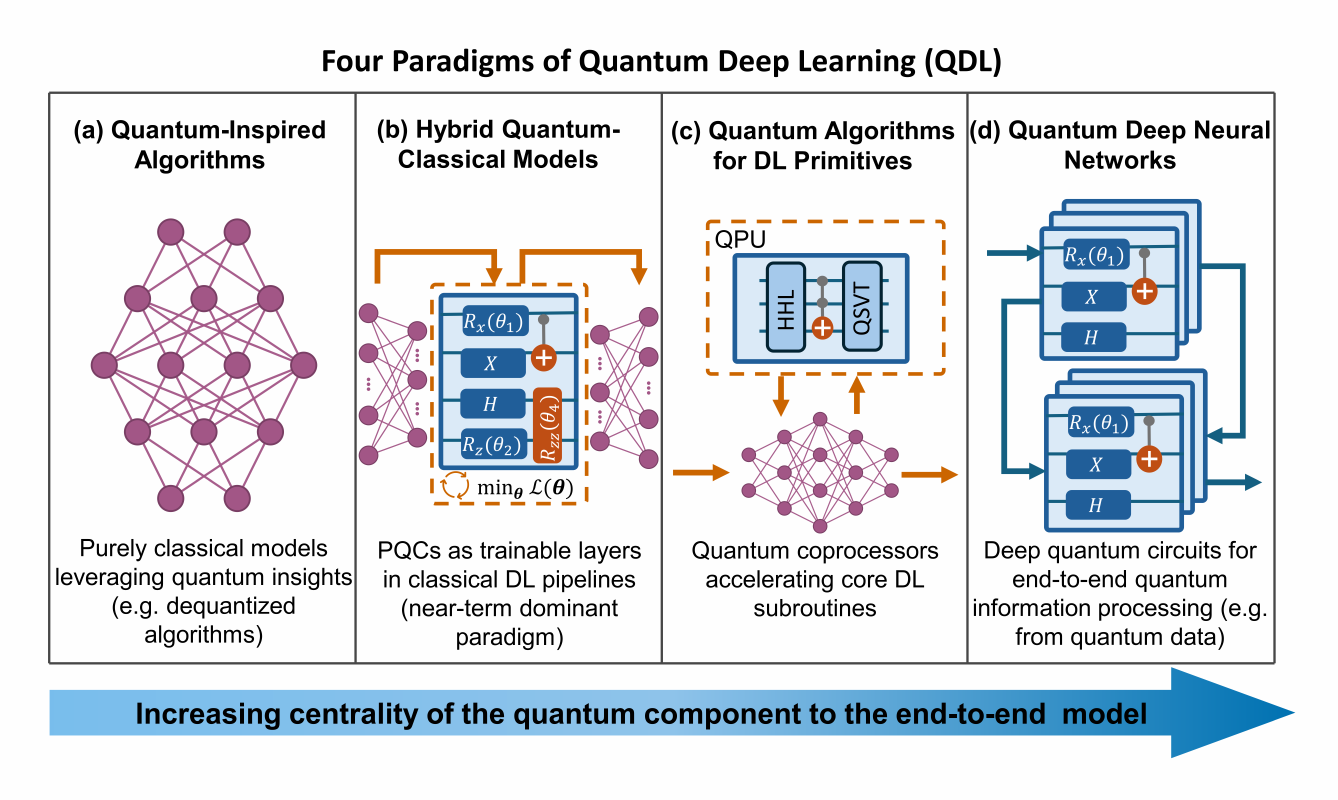}
\caption{Four paradigms of quantum deep learning (QDL), ordered left to right by increasing centrality of the quantum component to the end-to-end model. (a) Quantum-inspired algorithms: purely classical models in which the structure or training algorithms are motivated by quantum information theory. Quantum content is expressed through the model class, not through hardware-executed quantum states. (b) Hybrid quantum-classical models: a parameterized quantum circuit (PQC) with trainable gate parameters $\boldsymbol{\theta}$ is embedded as a differentiable module within a classical pipeline. The cycle denotes the classical outer optimization loop that updates $\boldsymbol{\theta}$ from measurement outcomes by minimizing a loss function $\mathcal{L}(\boldsymbol{\theta})$. (c) Quantum algorithms for deep learning (DL) primitives: the quantum processing unit (QPU) acts as a specialized coprocessor to accelerate core subroutines, e.g., linear-system solvers such as HHL and the quantum singular value transformation (QSVT),  within an overarching classical model under explicitly stated access and resource assumptions. (d) Quantum deep neural networks: deep, hierarchical quantum circuits designed for end-to-end quantum information processing, e.g., feature extraction directly from quantum data. $R_x(\theta_i)$ and $R_z(\theta_i)$: single-qubit rotation gates; $X$: Pauli-$X$ gate; $H$: Hadamard gate.}
\vspace{-12pt}
\label{fig:qdl_arc}
\end{figure*}

Complementary perspectives on learning quantum systems and AI for QC are reviewed in \cite{gebhart2023learning} and \cite{alexeev2025artificial}, respectively. For foundational overviews we recommend \textcite{nielsen2010quantum} and \textcite{Goodfellow-et-al-2016}.
The broader QML landscape is well documented, spanning canonical surveys and perspectives \cite{rodriguez2025survey,schuld2015introduction,biamonte2017quantum,dunjko2018machine,schuld2021machine,lamichhane2025quantum,cerezo2022challenges}, reviews \cite{wang2024comprehensive, zeguendry2023quantum, chen2024design, klusch2024quantum, mishra2021quantum, peral2024systematic, du2025quantum,siddi2025quantum}, pedagogical introductions \cite{alchieri2021introduction,khan2020machine}, and domain-focused analyses and roadmaps \cite{ullah2024quantum, kharsa2023advances, tian2023recent, jia2019quantum, acampora2025quantum,orka2025quantum}.
In contrast, our aim is to synthesize the QDL-specific lineage, architectural archetypes, implementation stack, and comparative-evaluation discipline.

Figure~\ref{fig:qdl_arc} summarizes four recurring paradigms we use throughout the review, conceptually ordered along the indicated axis by increasing centrality of the quantum component in the end-to-end model.
At the classical end of the spectrum (Fig.~\ref{fig:qdl_arc}(a)) lie quantum-inspired algorithms that import quantum-motivated structure or arise from dequantization \cite{tang2019quantum,tang2021quantum}. These algorithms use tools inspired by quantum information theory to build powerful classical models that can serve as strong, structure-exploiting baselines. In this regime, ``quantum" content is expressed through the model class, linear-algebraic primitives, or sampling and optimization structure, not through hardware-executed quantum states.
Figure~\ref{fig:qdl_arc}(b) shifts from inspiration to explicit quantum components embedded in otherwise classical pipelines. The schematic emphasizes the placement of a PQC as a trainable module between classical feature processing stages and the hybrid feedback loop required for end-to-end optimization. This paradigm is dominant in the NISQ era because it amortizes quantum resources into repeated circuit evaluations while delegating representation learning and scaling to classical deep networks. The third paradigm (Fig.~\ref{fig:qdl_arc}(c)) isolates a distinct, and often conflated, use case: quantum processors as coprocessors for specific DL primitives. The key scientific constraint, highlighted directly in the panel, is that any claimed speedup or advantage is inseparable from an explicit access and resource model. The same high-level primitive, such as Harrow-Hassidim-Lloyd (HHL) algorithm \cite{harrow2009quantum} and quantum singular value transformation (QSVT) \cite{gilyen2019quantum}, can map to qualitatively different end-to-end costs depending on data access, state preparation, and measurement constraints.
Finally, Fig.~\ref{fig:qdl_arc}(d) represents the quantum-native extreme, in which depth and hierarchy reside primarily in the quantum circuit itself: layered unitary blocks process quantum data inputs and yield quantum outputs and/or measurements. This setting naturally interfaces with learning tasks defined on quantum states or quantum processes, where the quantum circuit is not merely a subroutine but the principal representational backbone.

\subsection{Contributions and organization}

This review provides a synthesis of QDL, offering both a tutorial-style, textbook-quality entry point and a comprehensive guide to the specialized literature on the topic. It serves a dual audience: providing an accessible roadmap for graduate students and researchers entering the field, while delivering an authoritative, structured evaluation for experts in QC or AI.
Specifically, the core contributions of this review are to (i) weave the interdisciplinary literature into a coherent narrative that clarifies how QDL emerged and matured; (ii) introduce a precise operational definition of QDL and a unifying taxonomy that categorizes its four primary architectural paradigms; (iii) connect abstract learning principles to physical implementation by adopting a full-stack perspective spanning models, software toolchains, and hardware platforms; (iv) critically assess claims of quantum advantage by distinguishing between provable theoretical separations and heuristic empirical evidence, culminating in a rigorous protocol for fair, task-specific benchmarking; and (v) outline a forward-looking roadmap across near-term and fault-tolerant regimes.
These elements address the central questions driving the field: in which domains can quantum models provide a genuine, defensible learning advantage, and what are the fundamental resource constraints to realizing that potential?

The remainder of the review is organized as follows.
Section~\ref{sec:found_backg} establishes the foundational concepts of QC and classical DL, formally defining the framework of computational symbiosis that evaluates hybrid systems under explicit resource contracts. Building on this foundation, Sec.~\ref{sec:qdl_archi} constructs the theoretical blueprint of QDL architectures. It systematically surveys data encoding strategies, hybrid quantum-classical models, quantum deep neural networks, and quantum algorithms for classical deep learning primitives. It also examines quantum-inspired classical algorithms as essential baselines, details the prerequisite optimization techniques, and critically evaluates the theoretical limits of trainability and quantum advantage.
Section~\ref{sec:imple_and_chall} grounds these abstract models in physical reality. We introduce a five-layer QDL implementation stack and survey the leading quantum hardware platforms that form the bedrock of experimental execution. Following a review of the software frameworks that bridge algorithms and hardware, this section synthesizes milestone experimental demonstrations and critically analyzes the formidable scaling barriers imposed by hardware noise, connectivity constraints, and measurement bottlenecks.

In Sec.~\ref{sec:appli_and_compa}, we survey the application domains of QDL, distinguishing between classical data inputs (e.g., image classification and NLP) and coherent quantum data processing. We conclude this section by introducing a rigorous, four-pillar comparative protocol for evaluating empirical QDL performance against matched classical baselines.
Finally, Sec.~\ref{sec:concl_and_futur} synthesizes our key findings and presents a strategic research roadmap. We highlight the three coupled tensions, including the expressivity-trainability trade-off, the trainability-simulability constraint, and the quantum-classical interface bottleneck, that fundamentally govern the field's trajectory. The review concludes by outlining actionable milestones to navigate these challenges, charting the transition of QDL from near-term heuristics to early fault-tolerant systems and, ultimately, application-scale fault-tolerant quantum intelligence.

\section{Foundations and Preliminaries}\label{sec:found_backg}

This section reviews the quantum mechanical postulates and computational models that govern quantum information processing, alongside the mathematical frameworks that define classical DL. Building upon these independent foundations, we introduce the concept of computational symbiosis (Sec.~\ref{subsec:comp_symbiosis}), a rigorous preliminary framework that formalizes how quantum and classical resources interact, exchange data, and jointly optimize objective functions under realistic physical constraints.

\subsection{Quantum computing basics}

\subsubsection{Quantum states and resources for learning}

The basic unit of quantum information is a quantum bit, or qubit, represented in Dirac notation by a normalized vector in a two-dimensional complex Hilbert space, expanded in the computational basis $|0\rangle$ and $|1\rangle$ as \cite{dirac1930principles,nielsen2010quantum}:
\begin{equation}
|\psi\rangle = \alpha|0\rangle + \beta|1\rangle,
\end{equation}
where $\alpha, \beta \in \mathbb{C}$ satisfy $|\alpha|^2 + |\beta|^2 = 1$.
Pure single-qubit states admit a geometric parametrization on the Bloch sphere up to an overall physically irrelevant global phase \cite{bloch1946nuclear,nielsen2010quantum}:
\begin{equation}
|\psi\rangle = \cos(\vartheta/2)|0\rangle + e^{i\varphi}\sin(\vartheta/2)|1\rangle,
\end{equation}
with polar angle $\vartheta \in [0,\pi]$ and azimuthal angle $\varphi \in [0,2\pi)$.

For $n$ qubits, the Hilbert space dimension scales as $2^n$ and the tensor-product structure admits nonseparable (entangled) states in which correlations cannot be reproduced by product states \cite{amico2008entanglement,eisert2021entangling,horodecki2009quantum,jozsa2003role}.
Entanglement is analyzed as a circuit-template property alongside ansatz expressibility (see Sec.~\ref{subsec:comp_symbiosis}) \cite{sim2019expressibility}, but it can also degrade trainability in regimes where typical states approach volume-law entanglement, yielding entanglement-induced barren plateaus \cite{ortiz2021entanglement,patti2021entanglement}.
Decoherence, arising from coupling to uncontrolled environmental degrees of freedom, drives pure states toward mixed states and suppresses phase coherences. Key timescales include the energy-relaxation time $T_1$ and the dephasing time $T_2$, with $T_2^{-1}=(2T_1)^{-1}+T_{\phi}^{-1}$ in common Markovian models, where $T_{\phi}$ is the pure-dephasing time.

\subsubsection{Models of computation}\label{subsubsec:models_of_comput}

Quantum computation is commonly formulated either in a discrete-time gate model or in continuous-time Hamiltonian models, including adiabatic quantum computation (AQC) \cite{albash2018adiabatic} and quantum annealing, typically formulated as an open-system heuristic for optimization at finite temperature \cite{kadowaki1998quantum,rajak2022quantum}.
In closed-system AQC, the system starts in the ground state of an initial Hamiltonian $H_{\mathrm{init}}$ and evolves toward a problem Hamiltonian $H_{\mathrm{prob}}$. Adiabatic theorems ensure approximate tracking to the ground state of $H_{\mathrm{prob}}$ \cite{jansen2007bounds} under sufficiently slow schedules set by the minimum spectral gap and the schedule's smoothness.
In the gate-based model, dynamics are expressed as a sequence of unitaries acting on a qubit register \cite{deutsch1985quantum,barenco1995elementary,divincenzo1995two}.
Continuous-time evolution can be compiled into gate sequences via product-formula methods \cite{childs2021theory,suzuki1976generalized,suzuki1985decomposition,suzuki1976relationship,Faehrmann2022randomizingmulti}, while qubitization and quantum signal processing can achieve improved or optimal asymptotic query scaling for Hamiltonian simulation \cite{Low2019hamiltonian}, providing a bridge between Hamiltonian time evolution and gate-based implementations. Moreover, a rigorous polynomial equivalence between AQC and the gate-based model can be proven~\cite{aharonov2004adiabatic}.

Given this computational equivalence, and because modern QDL architectures are overwhelmingly constructed using PQCs, the remainder of this review focuses primarily on the discrete-time gate model. In this paradigm, arbitrary unitary evolutions are decomposed into sequences of elementary operations. Specifically, a universal gate set comprising arbitrary single-qubit unitaries and a fixed two-qubit entangling gate can approximate any $n$-qubit unitary to arbitrary precision \cite{barenco1995elementary,divincenzo1995two,nielsen2010quantum}.
Within these gate sets, a fundamental complexity-theoretic distinction exists between Clifford and non-Clifford operations. The $n$-qubit Clifford group, generated, for example, by Hadamard, phase, and CNOT gates, maps Pauli operators to Pauli operators under conjugation. This property underlies the stabilizer formalism, meaning that a stabilizer state is the simultaneous $+1$ eigenstate of an Abelian group of commuting Pauli operators. Consequently, by the Gottesman-Knill theorem, Clifford circuits combined with Pauli measurements can be efficiently simulated classically \cite{gottesman1998heisenberg,aaronson2004improved}. By contrast, non-Clifford resources, such as the $T$ gate, are strictly required for universal computation. In fault-tolerant architectures, these non-Clifford gates dominate computational overhead due to the necessity of magic state distillation for error-correcting codes where such gates are non-transversal \cite{bravyi2005universal}.

\subsubsection{Parameterized quantum circuits as trainable models}\label{subsubsec:pqcs_as_train}

Near-term QDL architectures predominantly employ continuously parameterized gates to enable gradient-based optimization. Standard single-qubit rotations $R_\alpha(\theta) = e^{-i\theta \sigma_\alpha/2}$ for $\alpha \in \{x, y, z\}$ combined with two-qubit entangling operations such as the controlled-NOT (CNOT)
\begin{equation}
\text{CNOT} = |0\rangle\langle0|\otimes I + |1\rangle\langle1|\otimes X,
\end{equation}
or parameterized $R_{ZZ}(\theta)$ constitute the elementary vocabulary from which PQC ans\"atze are assembled.
A PQC serves as the foundational hypothesis class for variational and hybrid QDL, specifying a parameter-dependent unitary evolution characterized by tunable parameters $\boldsymbol{\theta}$ and $L$ sequential ansatz layers \cite{cerezo2021variational,mcclean2016theory,benedetti2019parameterized}. The effective hypothesis class is determined by the induced input-output map once the data-encoding (state preparation), measurement observables, and any classical postprocessing are specified.
The specific structural decomposition of this unitary imposes an inductive bias that governs both the expressibility and trainability of the model \cite{sim2019expressibility,abbas2021power}, a critical distinction that is formalized in Sec.~\ref{subsec:comp_symbiosis}. Broadly, PQC architectures fall into two broad families: structure-driven constructions that encode specific problem symmetries or geometric priors \cite{larocca2022group,cong2019quantum}, and hardware-efficient ans\"atze designed to minimize compilation overhead by aligning strictly with native gate sets and physical qubit connectivity \cite{kandala2017hardware}.
A standard hardware-efficient ansatz alternates single-qubit parameterized rotations with topology-respecting entangling layers:
\begin{equation}
U_{\mathrm{HE}}(\boldsymbol{\theta})=\prod_{l=1}^{L}\left[\left(\bigotimes_{i=1}^{n} U_i(\theta_{l,i})\right) U_{\mathrm{ent}}^{(l)}\right],
\label{eq:hardw_effic_ansat}
\end{equation}
where $n$ is the number of qubits, $U_i(\theta_{l,i})$ denotes a parameterized single-qubit gate on qubit $i$ in layer $l$, and $U_{\mathrm{ent}}^{(l)}$ is a fixed or parameterized entangling layer composed of native two-qubit interactions respecting the hardware topology.
To balance expressivity with hardware constraints, the design of these circuits can also be partially automated using quantum architecture search or structure-learning methods, which dynamically optimize circuit connectivity and depth alongside the continuous parameter \cite{du2022quantum,ostaszewski2021structure,rapp2024reinforcement}.

\begin{figure*}[tb]
\includegraphics[width=.75\linewidth]{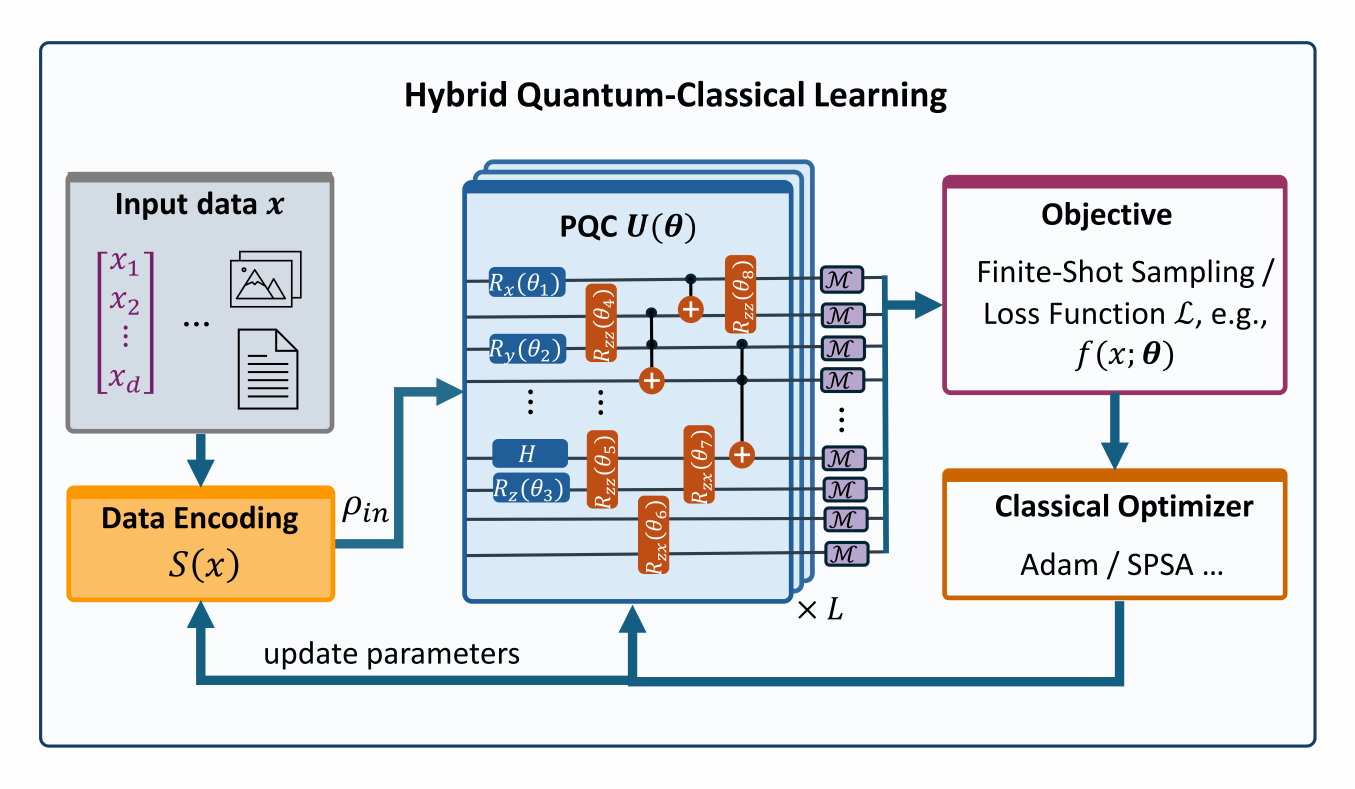}
\caption{The hybrid quantum-classical learning loop. Classical data $x$ are encoded into an input state $\rho_{\mathrm{in}}(x)$ via the data-encoding unitary $S(x)$. A parameterized quantum circuit (PQC) $U(\boldsymbol{\theta})$ is executed, measurements of an observable yield finite-shot statistics used to estimate a loss function (e.g., expectation value $f(x;\boldsymbol{\theta})$), and a classical optimizer updates $\boldsymbol{\theta}$ by minimizing an objective built from these estimates. The loop repeats until convergence to an optimal parameter set $\boldsymbol{\theta}^*$.}
\label{fig:vqa_loop}
\end{figure*}

Once the architecture is defined, the PQC is embedded within a hybrid quantum-classical optimization loop. In a standard supervised-learning framing, the computational pipeline composes a data-dependent state preparation map $\rho_{\mathrm{in}}(x)$, where $x$ denotes classical input data, the parameterized unitary evolution $U(\boldsymbol{\theta})$, and a final measurement step. Crucially, the direct output of the quantum device is a discrete measurement sample (bitstrings) drawn from the final state's probability distribution. These finite-shot statistics are subsequently processed on a classical host to estimate empirical quantities of interest, such as sample probabilities or expectation values. A common scalar feature used for model evaluation is the expectation value of a designated Hermitian observable $O$:
\begin{equation}
f(x;\boldsymbol{\theta})=\mathrm{Tr}\!\left[\,O\,\left(U(\boldsymbol{\theta})\,\rho_{\mathrm{in}}(x)\,U^\dagger(\boldsymbol{\theta})\right)\right],
\label{eq:pqc_model_output}
\end{equation}

Figure~\ref{fig:vqa_loop} makes this hybrid loop explicit.
A central practical consequence of this explicit feedback path is that wall-clock training time is heavily constrained by the execute-measure-communicate cycle. Optimization bottlenecks are typically dominated by the massive shot budgets required to suppress estimator variance, compounded by the quantum-classical latency inherent to the loop \cite{lubinski2022advancing,Menickelly2023latency}.
These stringent operational constraints strongly motivate the use of shallow circuits, noise-aware compilation and execution strategies \cite{ji2025algorithm,montanez2025optimizing} that maximize the effective accuracy per circuit evaluation \cite{endo2021hybrid}.

\subsubsection{Quantum measurement and classical interface}\label{subsubsec:quant_measu}

Measurement is the quantum-classical interface, mapping a quantum state to classical outcomes and, except for quantum-nondemolition settings, disturbs the post-measurement state \cite{nielsen2010quantum}.
For an observable $O = \sum_i \lambda_i M_i$ and a quantum state $\rho$, projective measurement yields outcome $\lambda_i$ with probability $p(\lambda_i) = \mathrm{Tr}(M_i\rho)$ and collapses the state to $M_i \rho M_i / p(\lambda_i)$, where $M_i$ are orthogonal projectors \cite{von1955mathematical,nielsen2010quantum}.
More general measurements are described by positive operator-valued measures (POVMs) \cite{holevo1973statistical, peres2002quantum, nielsen2010quantum} specified by effects $\{E_m\}$ with $E_m \geq 0$, and $\sum_m E_m = I$, yielding
\begin{equation}
p(m) = \mathrm{Tr}(E_m \rho).
\end{equation}

In hybrid algorithms, measured observables are typically decomposed into sums of Pauli strings:
\begin{equation}
    O = \sum_k c_k P_k,
\end{equation}
where $P_k=\bigotimes_{j=1}^n \sigma_{k_j}^{(j)}$ with $\sigma_{k_j}^{(j)} \in \{I, X, Y, Z\}$ and $c_k\in \mathbb{R}$. Here, $I$ is the identity and $X$, $Y$, $Z$ are the Pauli matrices.
Because measurement outcomes are stochastic and noncommuting terms require different measurement bases, objectives such as $\langle O \rangle=\mathrm{Tr}(O\rho)$ are estimated from repeated state preparations and measurements, by grouping commuting terms.
For independent single-copy measurements and bounded-outcome estimators, achieving additive precision $\epsilon$ at failure probability at most $\delta$ requires a number of shots that scales as $\mathcal{O}(\log(1/\delta)/\epsilon^2)$ for a single bounded observable. For sums of many Pauli terms, the shot cost additionally depends on coefficient weights and the chosen grouping strategy, and can become the dominant runtime bottleneck \cite{gonthier2022measurements,chen2024optimal}.

Advanced schemes target this sampling bottleneck. Randomized measurement schemes (classical shadows) framework \cite{aaronson2018shadow,huang2020predicting} constructs a compact classical record (``shadow") from randomized single-copy measurements.
To estimate $C$ expectation values $\{\mathrm{tr}(O_i\rho)\}_{i=1}^C$ to additive error $\epsilon$ with failure probability at most $\delta$, it suffices to take
\begin{equation}
N \;=\; \mathcal{O}\!\left(\frac{\log(C/\delta)}{\epsilon^{2}}\;\max_{i\le C}\,\|O_i\|_{\mathrm{shadow}}^{2}\right),
\end{equation}
where $\|\cdot\|_{\mathrm{shadow}}$ depends on the chosen measurement ensemble \cite{huang2020predicting}.
For random single-qubit Pauli measurements, $\|O_i\|_{\mathrm{shadow}}^{2}$ scales at most exponentially with locality $k$. In particular, for an observable acting nontrivially on at most $k$ qubits $\|O_i\|_{\mathrm{shadow}}^{2}\le 4^{k}\|O_i\|_{\infty}^{2}$ \cite{huang2020predicting}, implying
\begin{equation}
N \;=\; \mathcal{O}\!\left(\frac{4^{k}\log(C/\delta)}{\epsilon^{2}}\right)
\end{equation}
at fixed $k$ and bounded operator norm.
Classical shadows also serve as a canonical ``measure-first" protocol \cite{gyurik2025limitations} in quantum-data applications. In this framework, quantum states are first converted into classical representations via a fixed measurement strategy before being processed by a downstream classical learning algorithm.

\subsubsection{Near-term and fault-tolerant quantum computing}\label{subsubsec:nisq_fault}

As introduced in Sec.~\ref{subsec:the_conver_qdl}, the current landscape is dominated by the NISQ era \cite{preskill2018quantum}. This regime is characterized by architectures containing $10^2$--$10^3$ physical qubits in leading platforms with imperfect operations restricting reliable execution to relatively shallow circuits  \cite{chen2023complexity,leymann2020bitter,kim2023evidence}.
Within QDL, these hardware constraints are particularly severe because most trainable models are instantiated as PQCs \cite{cerezo2021variational} and their performance degrades markedly under hardware noise \cite{buonaiuto2024effects,skolik2023robustness}.
Furthermore, compilation overheads required to map algorithmic ans\"atze onto restricted physical topologies further increase circuit depth, compounding accumulated errors \cite{buonaiuto2024effects,zhang2025experimental}. Consequently, the maximum feasible model depth is determined by a strict bottleneck between algorithmic connectivity demands and physical hardware fidelities.

To combat these effects without full fault tolerance,  quantum error mitigation (QEM) techniques, such as zero-noise extrapolation and probabilistic error cancellation, attempt to estimate ideal expectation values from noisy circuit executions by trading sampling complexity for improved accuracy \cite{cai2023quantum}. However, this sampling overhead generally scales exponentially with circuit depth and noise rates. For QDL, these steep mitigation costs severely constrain the utility of QEM for deep or highly entangled networks, heavily incentivizing architectures that are shallow and natively hardware-aligned.

Overcoming these fundamental depth limits requires a transition to fault-tolerant quantum computing (FTQC). In FTQC, logical information is redundantly encoded across multiple physical qubits and protected via repeated syndrome extraction and correction \cite{shor1995scheme,steane1996error,terhal2015quantum,gottesman2009introduction}.
Below a critical physical error threshold, topological schemes such as the surface code can suppress logical error rates exponentially with code distance \cite{fowler2012surface}. However, this protection incurs massive resource overheads, typically demanding $10^2$--$10^4$ physical qubits per logical qubit depending on the target logical error and ancilla requirements. Furthermore, non-Clifford resources, most notably $T$-gate distillation factories, often dominate the total spatial footprint and runtime of the architecture \cite{fowler2012surface,gidney2025yoked}, while the classical control stack demands high-throughput, low-latency decoders to keep pace with rapid syndrome cycles \cite{bausch2024learning}.
To bridge the gap between NISQ and FTQC, researchers are also exploring ``almost fault-tolerant'' or early-fault-tolerant regimes, which selectively reduce distillation overheads by leaving certain non-Clifford resources less strictly protected \cite{kang2025almost}.

Contextualizing this trajectory, \textcite{eisert2025mind} outline four transitional gaps the field needs to cross to reach practical utility: (i) from error mitigation to active error detection and correction, (ii) from rudimentary error correction to scalable fault tolerance, (iii) from early heuristics to mature, verifiable algorithms, and (iv) from exploratory simulators to credible advantage in quantum simulation.
For the specific domain of QDL, gaps (i) and (ii) strictly bound the feasible circuit depth, and therefore the representational capacity, of any quantum model under a fixed resource contract. Meanwhile, gaps (iii) and (iv) underscore the imperative for resource-efficient validation and rigorous benchmarking against matched classical baselines. Ultimately, achieving end-to-end practical performance in QDL requires deep codesign across the entire stack, optimizing algorithms, compilation strategies, and QEC architectures in tandem \cite{babbush2025grand}.

\subsection{Classical DL basics}

\subsubsection{Model classes and energy-based formalisms}\label{subsubsec:model_class_energy_based}

DL studies high-capacity parametric hypothesis classes, most commonly multilayer neural networks, trained by variants of empirical risk minimization on differentiable objectives that enable efficient gradient-based optimization \cite{lecun2015deep,Goodfellow-et-al-2016}. Formally, a learning problem specifies a parametric family $f_\theta:\mathcal{X}\to\mathcal{Y}$, an objective, and a data-generating distribution or empirical sample over inputs and targets.
Universal approximation theorems \cite{cybenko1989approximation,hornik1989multilayer,lu2017expressive} establish approximation capacity at sufficient width or depth under standard regularity assumptions, but do not address optimization, data requirements, or generalization. Practical performance is governed by trainability, sample efficiency, and architectural constraints that encode inductive bias.

Energy-based models (EBMs) are particularly relevant to QDL because they share structural features with quantum many-body systems and provide a natural interface for quantum sampling primitives \cite{amin2018quantum}.
These models specify a scalar energy function $E_\theta(z)$ over configurations $z$, such as observations, labels, and latent variables. In probabilistic EBMs, this energy induces an unnormalized density with explicit normalization by a partition function \cite{hopfield1982neural,ackley1985learning,lecun2006tutorial}. For binary configurations $z\in\{0,1\}^n$ this induces the Gibbs distribution
\begin{equation}
    p_\theta(z)=\frac{1}{Z_\theta}\exp(-\beta E_\theta(z)),
\end{equation}
where $Z_\theta=\sum_{z}\exp(-\beta E_\theta(z))$ is the  partition function and $\beta>0$ is an inverse temperature parameter.

Hopfield and Boltzmann-type models provide a useful historical and conceptual bridge between classical DL and statistical-mechanics formalisms. The original Hopfield network can be cast as deterministic dynamics that perform energy minimization, yielding a prototypical associative-memory mechanism \cite{hopfield1982neural}. Boltzmann machines generalize this picture by introducing stochasticity and making expectation estimation under the induced Gibbs distribution a central computational primitive, which is precisely where likelihood training becomes expensive \cite{ackley1985learning,hinton2006fast}.
Restricted Boltzmann machines (RBMs) and related EBMs leverage tractable conditional structure to enable approximate learning via gradient surrogates \cite{hinton2006fast,lecun2006tutorial,hinton2002training}.
Modern Hopfield formulations have connected content-addressable retrieval to attention-style readout rules, providing a concrete link between associative-memory energy landscapes and contemporary sequence modeling \cite{ramsauer2020hopfield,vaswani2017attention}.

\subsubsection{Training dynamics and theoretical regimes}

Deep networks are trained by stochastic optimization of nonconvex objectives using gradients computed by reverse-mode differentiation through composed operations \cite{Goodfellow-et-al-2016}.
In large-scale settings, gradients are estimated on minibatches, and adaptive methods such as Adam \cite{Kingma2014} modify updates using running moment estimates.
Modern practice frequently operates in an overparameterized regime, with parameter count comparable to or exceeding samples in supervised settings. In such settings, many low-loss (and often interpolating) solutions can coexist, and the particular solution reached can depend strongly on implicit regularization arising from optimization dynamics, initialization, and algorithmic choices \cite{bahri2020statistical,advani2020high}.

Width scaling provides analytically tractable reference regimes.
In the infinite-width limit, under appropriate initialization and scaling, training can enter the ``lazy" or neural tangent kernel (NTK) regime, where the network is well approximated by its linearization around initialization.
For a model $f(x;\theta)\in\mathbb{R}^m$, where $m$ is the output dimension, let $J_\theta f(x)\in\mathbb{R}^{m\times p}$ denote the Jacobian with respect to parameters $\theta$, with $p$ the number of parameters. The NTK is
\begin{equation}
\Theta(x,x';\theta)=J_\theta f(x)\,J_\theta f(x')^\top.
\end{equation}
In the strict NTK limit, $\Theta$ converges to a deterministic kernel that remains effectively constant during training\cite{jacot2018neural}, so gradient descent is well approximated by kernel regression dynamics. \textcite{chizat2019lazy} clarify when this approximation is appropriate and emphasize that substantial representation change (feature learning) requires departing from the strictly kernelized regime.
At finite widths and standard learning rates, networks can enter a ``rich" regime in which internal representations evolve substantially and the effective kernel becomes time-dependent and data-adaptive \cite{bordelon2025feature,mei2018mean}.

\subsubsection{Symmetry and inductive bias}\label{subsubsec:symmet_induc_bias}

In high-capacity settings, empirical risk minimization alone does not uniquely specify a solution: many functions can interpolate a finite training set.
Consequently, achieving true \emph{generalization}, i.e., the ability of a model to maintain predictably low expected error on novel, unseen data drawn from the underlying distribution, rather than merely memorizing the training examples, cannot rely on fitting the data alone.
Because the learning algorithm inherently interpolates behavior between sparsely observed data points, successful generalization depends strictly on \emph{inductive bias}: the structural constraints or algorithmic preferences that favor particular solutions within the hypothesis space.
Architectures encode inductive bias by restricting hypothesis classes through symmetry, locality, and connectivity constraints.
Formally, when data live on domains with group actions, inductive biases can be implemented by designing networks to be invariant or equivariant under those actions \cite{bronstein2021geometric}.
Convolutional neural networks, for example, encode a bias toward local, translation-equivariant features via weight sharing and local receptive fields \cite{lecun1989backpropagation}. Group-equivariant generalizations extend this principle to broader transformation groups by enforcing equivariance at the level of the hypothesis class \cite{cohen2016group}.

For structured non-Euclidean data, graph neural networks implement permutation-equivariant message passing by aggregating neighbor information through permutation-invariant pooling operations \cite{kipf2016semi,gilmer2017neural,battaglia2018relational}.
For sequence and language modeling, the dominant architectural prior shifts from local receptive fields to global context aggregation.
Transformer architectures implement this via self-attention, which computes data-dependent pairwise interactions across tokens in a sequence \cite{vaswani2017attention}.
This structural and group-theoretic framing provides the essential lens for evaluating hybrid quantum-classical models. Establishing a genuine quantum advantage requires explicitly articulating how the quantum module alters the underlying inductive bias. Consequently, validating these theoretical separations demands rigorous empirical benchmarking against classical baselines equipped with equivalent geometric symmetries.

\subsubsection{Generative modeling and reinforcement learning}

Beyond supervised prediction, deep networks are widely used for generative modeling and sequential decision-making, two settings in which probabilistic inference and sampling play a central role and are often discussed as potential targets for quantum acceleration. Generative modeling requires sampling from complex distributions and, in many formulations, estimating expectations or score functions. As noted in Sec.~\ref{subsubsec:model_class_energy_based}, likelihood-based training of EBMs requires expectations under the model distribution and therefore relies on approximate inference.
A broad class of modern generative methods avoids explicit energy normalization. Denoising diffusion probabilistic models and related score-based methods reparameterize learning in terms of non-equilibrium stochastic dynamics, learning to reverse a prescribed forward noising process by estimating the score function $\nabla_x\log p(x)$  or an equivalent denoising objective \cite{ho2020denoising,song2021scorebased}.

Reinforcement learning (RL) uses deep networks as function approximators to solve sequential decision problems, commonly formalized as a Markov decision process \cite{Sutton2018,mnih2015dqn,Puterman1994MDP}. In this framework, an agent interacts with its environment by observing a state $s \in \mathcal{S}$ and selecting an action (control) $c \in \mathcal{C}$. The objective of RL is to learn a parameterized stochastic policy $\pi_\theta(c|s)$, where $\theta$ denotes the trainable parameters (e.g., network weights), that maximizes the expected cumulative reward over time, typically when the environment dynamics are unknown and learning relies on sampled trajectories rather than differentiating through the environment. Because the environment's transition dynamics cannot be analytically differentiated, policy-gradient methods \cite{Williams1992Simple,Sutton1999Policy} instead estimate policy gradients from sampled trajectories using log-derivative identities. This sampling-centric structure motivates connections to quantum RL and quantum optimal control (QOC) \cite{Dong2008qrl,dunjko2016quantum}, where measurement outcomes provide intrinsic stochastic samples that can be used to estimate gradients or returns under explicit access models.

\subsubsection{Scaling laws and emergent regimes}

Classical DL provides a moving baseline because performance often improves approximately with resources over measured ranges. In large-scale language modeling and related settings, empirical scaling laws describe approximate power-law relationships between loss or downstream performance and resources such as model size, dataset size, and training compute over broad ranges \cite{Kaplan2020,hoffmann2022training}.
Analyses aim to explain these scalings by distinguishing noise-limited from resolution-limited regimes and relating them to target-function structure and data geometry \cite{bahri2024explaining}. In contemporary practice, self-supervised pretraining at scale underpins transfer across tasks, notably in foundation-model settings \cite{Chen2020,Bommasani2022}.

Scaling can also induce qualitative regime transitions.
When capacity increases at fixed data, the model may cross an interpolation threshold where training error first becomes approximately zero. Double descent refers to the resulting nonmonotonic test error curve as a function of capacity or effective degrees of freedom: error decreases in the underparameterized regime (effective degrees of freedom below the number of training examples), peaks near the interpolation threshold (where training error first becomes approximately zero), and decreases again in the overparameterized regime (effective degrees of freedom above the number of training examples) as optimization selects interpolating solutions with improved generalization \cite{belkin2019reconciling,nakkiran2021deep}.
Grokking highlights a distinct axis: at fixed architecture and data, continued optimization, often with regularization, can yield delayed generalization, where a model fits the training set early yet only abruptly transitions to a generalizing solution late in training \cite{power2022grokking}. Phase-transition analogies provide a compact language for organizing regime diagrams \cite{ziyin2023zeroth}, but the mathematical structure of these transitions differs from equilibrium statistical mechanics and does not imply universality in the sense of critical exponents or renormalization-group scaling.

\subsection{Computational symbiosis}\label{subsec:comp_symbiosis}

Hybrid QDL architectures compose quantum and classical modules that are codesigned and jointly optimized under a shared objective \cite{cong2019quantum, long2025hybrid, grant2018hierarchical, beer2020training, skolik2021layerwise}.
To rigorously evaluate these systems, we operate hybrid learning as minimizing an expected loss over a task distribution $\mathcal{D}$ with a fixed evaluation protocol and primary metric, subject to a resource contract $\mathcal{R}$ and an access-and-readout model $\mathcal{A}$. We denote this triple of specifications collectively as $(\mathcal{D},\mathcal{R},\mathcal{A})$ and use it to parametrize learning systems under comparison.
This framework allows us to formally define the target operational regime of QDL.

\emph{Definition~2 (Computational symbiosis).---}A learning system exhibits computational symbiosis if (i) it composes at least one quantum module and one classical module coupled through a shared objective; (ii) functional roles are partitioned via explicit interface variables; and (iii) the system is defined and evaluated relative to a specified task distribution $\mathcal{D}$, including the evaluation protocol and primary metric, together with a jointly specified pair ($\mathcal{R}$, $\mathcal{A}$) consisting of a resource contract $\mathcal{R}$ that fixes budget ceilings on end-to-end resources and an access-and-readout model $\mathcal{A}$ that fixes allowed data-access and measurement permissions.

Computational symbiosis is a systems-level comparative property. It is not an intrinsic property of the quantum circuit alone but of the full hybrid system evaluated under the stated $(\mathcal{D},\mathcal{R},\mathcal{A})$.
The access-and-readout model $\mathcal{A}$ specifies allowed data-access and query operations together with measurement permissions, whereas the resource contract $\mathcal{R}$ specifies ceilings on end-to-end variables, including qubit count, circuit depth, total circuit executions, wall-clock time, hyperparameter tuning budget, classical memory, and error mitigation overhead, subject to a target statistical precision.
In deployed settings, abstract ceilings map to operational budgets such as compiled circuit depth $D_{\mathrm{eff}}$, time per circuit execution $t_{\mathrm{shot}}$, iteration latency $T_{\mathrm{iter}}$, and a stability window $\tau_{\mathrm{stable}}$ over which device behavior is approximately stationary (see Sec.~\ref{subsubsec:hardware_constraints}).

Symbiosis is evidenced on $\mathcal{D}$ when, under fixed $(\mathcal{R},\mathcal{A})$, the hybrid yields a statistically significant improvement in a pre-specified primary metric, or matches the metric while reducing at least one ceiling in $\mathcal{R}$, compared against the strongest admissible classical baseline. Crucially, this requires that the quantum module implement a clearly specified encoding-ansatz-readout protocol in which the signal remains learnable within $D_{\mathrm{eff}}$ and $\tau_{\mathrm{stable}}$, allowing any metric gain or budget reduction to be confidently credited to the quantum component.

Achieving computational symbiosis in practice requires navigating the interplay among model expressivity, trainability, and classical simulability under finite measurement cost and bounded resources.
It is crucial to distinguish \emph{ansatz expressibility}, a circuit-ensemble notion quantifying how broadly a PQC explores state space \cite{sim2019expressibility}, from \emph{model expressivity} or \emph{expressive power}, which is the task-level representational capacity of the induced hypothesis class \cite{abbas2021power, wu2021expressivity}. Highly expressive unstructured circuits are prone to gradient concentration, i.e., the barren-plateau phenomenon (Sec.~\ref{subsubsec:resou_codesign}), creating a fundamental tension between representational power and optimization feasibility.
This tension directly impacts trainability, which denotes the physical feasibility of extracting an optimization signal to a target precision within the operational budgets of $\mathcal{R}$~\cite{mcclean2018barren,cerezo2021cost,holmes2022connecting}. Increased compiled depth further suppresses the signal-to-noise ratio by exposing the system to gate and decoherence errors, raising the shot budget required to estimate gradients to a fixed precision \cite{arrasmith2021effect,chinzei2025trade}.
Shifting representational capacity into the classical component can improve robustness to finite-shot noise, but it reallocates cost to classical compute and memory, both of which are counted against $\mathcal{R}$.

Addressing trainability, however, often exacerbates the \emph{trainability-simulability tension}. Provable trainability is often achieved by imposing structural restrictions on the quantum circuit that can also enable efficient classical evaluation of the same objective.
In several analyses, the objective becomes classically efficient to evaluate after a one-time quantum data-acquisition stage \cite{cerezo2025does}. Consequently, trainability certificates alone do not establish classical intractability unless compared against classical algorithms granted identical $(\mathcal{R},\mathcal{A})$ assumptions.
Recent analyses emphasize that many existing trainability guarantees apply only to restricted circuit families or regimes that also admit efficient classical simulation~\cite{cerezo2025does}. While this implication is not known to be universal, it underscores the need to strictly specify the architecture class when claiming ``trainable yet beyond-classical" behavior.

Computational symbiosis is a comparative systems property demonstrable at current scales, whereas establishing formal \emph{quantum advantage} is a significantly stronger claim that requires scalable asymptotic separation relative to an explicit classical comparator \cite{lanes2025framework}.

\emph{Definition~3 (QDL advantage).---}A QDL advantage claim requires specifying: (i) a task and data-generating model defining $\mathcal{D}$; (ii) an access-and-readout model $\mathcal{A}$; (iii) the classical comparator class restricted to the same access-and-readout assumptions and evaluated under the same resource contract $\mathcal{R}$; (iv) the error metric, target accuracy $\epsilon$, and success probability $1-\delta$; (v) an end-to-end resource contract $\mathcal{R}$; and (vi) a validation or verification procedure credible at the target scale under $\mathcal{A}$ together with an explicit separation statement identifying which ceilings in $\mathcal{R}$ are improved at fixed $(\epsilon,\delta)$ relative to the comparator.

Throughout this review, we reserve the term ``QDL advantage" for statements intended to satisfy Definition 3. Otherwise, we describe empirically observed improvements as computational symbiosis or task- and budget-conditional performance gains.

\section{QDL Architectures}\label{sec:qdl_archi}

This section surveys the core architectures and algorithmic building blocks that define QDL, concluding with a critical evaluation of the theoretical advantages and inherent limitations.

\subsection{Data encoding and preparation}\label{subsec:data_encoding}

Classical data, such as real vectors or matrices, require encoding into quantum states for processing on quantum hardware \cite{schuld2021effect,lloyd2020quantum}. The encoding scheme determines the model's inductive bias, the geometry of the induced feature map \cite{larocca2022group,nguyen2024theory}, and the dominant state-preparation and readout costs. Encodings exhibit dataset- and implementation-dependent accuracy-runtime tradeoffs. Benchmark analyses indicate strong task-, noise-, and implementation-dependence, with no encoding consistently dominating across studied regimes \cite{rath2024quantum,bowles2024better}.

\subsubsection{Basis and amplitude encodings}

Basis encoding discretizes the input data into classical bit strings and then transfers these bits one by one into computational basis states in the quantum system \cite{schuld2018supervised}. Given a classical data point with an $N$-bit binary representation $x=(x_1, x_2, \dots, x_N)$ with $x_j\in\lbrace 0,1\rbrace$, the basis-encoded state is
\begin{equation}
    \ket{x} = \ket{x_1,x_2,\dots,x_N}\,.
\end{equation}
Initialization of $\ket{x}$ from $\ket{0}^{\otimes N}$ only requires at most $N$ Pauli-$X$ operations. Furthermore, because these encoded states are classical bit strings forming an orthogonal computational basis, they permit copying via basis-preserving unitaries, such as a CNOT fan-out operation ($\text{CNOT}|x_j\rangle|0\rangle = |x_j\rangle|x_j\rangle$), without violating the no-cloning theorem, which only strictly prohibits the universal copying of unknown, arbitrary superpositions.
Despite these initialization advantages, basis encoding suffers from severe spatial bottlenecks when applied to data that is not originally binary. Because each real-valued feature is discretized into multiple bits at a fixed numerical precision, the encoding requires linear qubit overhead. For example, if a continuous variable requires a $\tau$-bit binary representation to achieve the target precision, an $n$-qubit register can encode at most $\lfloor n/\tau \rfloor$ scalar features, making this approach prohibitively expensive for high-dimensional classical deep learning tasks.

Amplitude encoding is a common conceptual primitive in QLA-based approaches to QML~\cite{wiebe2012fitting, lloyd2013qasup, Lloyd2014, rebentrost2014quantum}. In this scheme, a $d$-dimensional data vector $\boldsymbol{x}=(x_0, x_1, \dots, x_{d-1})$ is normalized and encoded as amplitudes of an $n=\lceil\log_2 d\rceil$ qubit state
\begin{equation}
    \ket{\psi_{\mathrm{amp}}(\boldsymbol{x})} \;=\; \sum_{j=0}^{2^n-1} \frac{x_j}{\|\boldsymbol{x}\|_2}\ket{j}.
\end{equation}
Although qubit count scales logarithmically with data dimension, this advantage is offset by state-preparation cost in the worst case: preparing a generic $n$-qubit amplitude-encoded state can require resources exponential in $n$, and even achieving a circuit depth of $\Theta(n)$, where the number of sequential gate layers scales linearly with the number of qubits, preparation may demand an exponential number of ancillas unless additional structure is exploited \cite{zhang2022quantum}. Moreover, extracting global quantities can require exponentially many shots in $n$ under standard access models~\cite{aaronson2015read, zhang2024circuit}.

\subsubsection{Data-access models and QRAM}\label{subsubsec:data_acces_qram}

Speedup claims in amplitude-encoding and block-encoding pipelines are conditional on the input-access model, which determines how classical data is loaded into quantum states or unitaries, and the output model, which determines what classical information is extracted. When inputs are provided by explicit classical descriptions, constructing the state-preparation and access primitives can itself dominate the cost.
\textcite{zhang2024circuit} prove near-optimal Clifford$+T$ circuit complexity bounds for constructing typical quantum access primitives from explicit classical descriptions and show that for unstructured matrices this cost can scale near linearly with matrix dimension in worst-case settings. Consequently, end-to-end resources can be set by data access unless stronger access assumptions or exploitable structure are present.

Many proposals therefore assume quantum random access memory (QRAM) \cite{giovannetti2008quantum,aaronson2015read} to enable coherent superposition queries to a classical dataset. However, QRAM feasibility and overhead are highly architecture-dependent. 
For standard circuit-model QRAM, architectural analyses show that a superposition query can activate $\mathcal{O}(N_\mathrm{mem})$ routers, where $N_\mathrm{mem}$ denotes the number of stored addresses. Because this circuit cost scales at least proportionally to the number of stored bits, it threatens to erase putative asymptotic advantages at large $N_\mathrm{mem}$~\cite{jaques2023qram}. Furthermore, causality-based bounds constrain how the QRAM size $N_\mathrm{mem}$ can scale for fixed clock-cycle time and spatial layout~\cite{wang2024fundamental}. Surprisingly, hardware noise is not necessarily the decisive obstruction in these limits; for instance, \textcite{hann2021resilience} proved that the tree-structured bucket brigade QRAM can retain favorable (poly)logarithmic-in-$N_\mathrm{mem}$ query-infidelity scaling even under arbitrary local error channels, distinguishing it from naive fanout designs.

Instead, the primary obstruction for large-scale QRAM is often the fault-tolerant gate compilation cost. Because QRAM routers typically require heavily controlled operations (e.g., Toffoli gates), they accumulate massive non-Clifford depth. As established in Sec.~\ref{subsubsec:models_of_comput}, non-Clifford gates require highly resource-intensive magic state distillation, which frequently dictates the dominant spatial footprint and temporal bottleneck of the entire FTQC architecture. Consequently, reducing non-Clifford depth is an imperative optimization target for the practical viability of data-loading oracles.
Nevertheless, recent polynomial-encoding circuit constructions have been developed specifically to reduce the non-Clifford depth of QRAM lookups~\cite{mukhopadhyay2025quantum}.
Parallel efforts seek to mitigate these access bottlenecks by bypassing exact QRAM entirely, utilizing approximate amplitude encoding to trade strict fidelity for dramatically shallower circuits \cite{nakaji2022approximate}, alongside alternative state-preparation protocols, such as repetitive amplitude encoding and structure-aware function-loading schemes \cite{li2025repetitive,Rosenkranz2025quantumstate,GonzalezConde2024efficientquantum,balewski2024quantum}.

\subsubsection{Dynamic encoding and the Fourier perspective}\label{subsubsec:dynamic_encoding_fourier}

Dynamic or Hamiltonian encoding embeds data via the unitary time evolution of Hermitian operators, fundamentally defining the model's inductive bias through the algebraic structure of the map \cite{schuld2021effect,gil2024understanding}. Multivariate inputs $x \in \mathbb{R}^d$ are encoded via factorized unitaries
\begin{equation}
    S(x) = \prod_{j=1}^d e^{-i h_j x_j t}\,,
\end{equation}
where $h_j$ are local generators and $t\in \mathbb{R}$ is a fixed or trainable scale parameter. Common choices include single-qubit Pauli operators $h_j\in\{X, Y, Z\}$, sometimes termed angle encoding \cite{schuld2021machine}.
To induce correlations in the feature map, the encoding can include multi-qubit interaction terms, as in Ising-type feature maps~\cite{havlivcek2019supervised}.

This encoding scheme admits a useful analytical framework. Unlike classical bias, which arises from architecture and regularization \cite{Goodfellow-et-al-2016}, QDL bias is strongly shaped by the spectral properties of $h_j$ and the circuit topology.
This framework admits a Fourier series representation of the model, where the accessible frequency spectrum is determined by the generators' eigenvalue gaps and the number and placement of data-dependent layers, and the chosen measurement \cite{schuld2021effect, jadeberg2024frequencies, shin2023exponential}.
In kernelized formulations the same bias is captured by the kernel eigenspectrum \cite{kubler2021inductive,thanasilp2024exponential}.
Repeating the encoding layers interleaved with trainable unitaries, a strategy known as data reuploading, expands the accessible trigonometric feature family and, in theory, can yield universal approximation guarantees for continuous functions on compact domains given sufficiently many reuploading layers and parameters \cite{perez2020data}, but in practice, the resulting extended circuits frequently encounter severe trainability bottlenecks. Concentration phenomena can render naive fidelity kernels uninformative, necessitating metrics that quantify when quantum decision geometries diverge meaningfully from classical baselines \cite{huang2021power,thanasilp2024exponential}. To mitigate this, one may impose structural constraints: utilizing equivariant constructions that enforce group structure to align the Hilbert space geometry with task symmetries \cite{larocca2022group,meyer2023exploiting,nguyen2024theory}.

Beyond trainability, increasing the complexity of the quantum encoding directly impacts model generalization. Specifically, encoding-dependent generalization bounds tend to worsen as the accessible frequency spectrum expands, unless explicit structural constraints or regularization techniques are applied \cite{caro2021encoding}. Fortunately, the training dynamics themselves can provide a form of implicit regularization: under standard gradient-based optimization, these models frequently exhibit a \emph{spectral bias}, preferentially learning low-frequency components before fitting high-frequency noise \cite{barthe2024gradients}. Alternatively, if high expressivity is required without incurring the generalization and trainability penalties of deep quantum circuits, practitioners can enhance the model's complexity entirely off-chip by composing a shallow quantum encoding with classical nonlinear preprocessing of the input \cite{gil2020input}.

\subsubsection{Emerging encoding paradigms}

Beyond fixed maps, neural quantum embedding schemes parameterize the encoding circuit and train it to improve separability in the induced quantum feature space before downstream classification layers \cite{hur2024neural,liu2025neural}. Alternatively, reservoir computing exploits the natural high-dimensional dynamics of quantum many-body systems for temporal feature injection \cite{fujii2017harnessing,mujal2021opportunities,hayashi2025effective}, while continuous-variable and photonic platforms encode data directly in field quadratures, yielding high-dimensional feature spaces without qubit discretization \cite{killoran2019continuous, anai2024continuous, maier2025continuous}.

\subsection{Hybrid quantum-classical models}\label{subsec:hybrid_models}

Building on the foundational hybrid loops introduced in Sec.~\ref{subsubsec:pqcs_as_train}, hybrid quantum-classical models integrate parameterized quantum subroutines into classical computational graphs, delegating data processing, optimization, and hardware orchestration to classical components \cite{mcclean2016theory,mitarai2018quantum,peruzzo2014variational,cerezo2021variational,benedetti2019parameterized}.
Architecturally, it is useful to partition these hybrid systems based on their training dynamics. ``Boundary" families utilize the quantum device strictly as a fixed feature extractor, restricting all gradient updates to the classical readout layers. Prominent examples of this approach include fixed-feature quantum kernel methods \cite{havlivcek2019supervised,schuld2018supervised,schuld2020measuring}, quantum reservoir computing \cite{mujal2021opportunities,fujii2017harnessing}, and fixed-feature quanvolution \cite{henderson2020quanvolutional}.
In contrast, core variational models establish an end-to-end feedback loop that actively updates the quantum circuit's internal parameters during training.
However, this boundary naturally dissolves when quantum modules are embedded within hierarchically deep pipelines. For instance, when the parameters of a quantum feature map or quantum kernel are jointly optimized in-loop alongside classical weights, these trainable variants cross the threshold, merging into the core regime of fully variational hybrid QDL \cite{rodriguez2025neural,alvarez2025benchmarking,gentinetta2023quantum}.

\subsubsection{Architectural paradigms for hybrid models}\label{subsubsec:archi_parad_for_hyb}

\begin{figure*}[t]
\centering
\includegraphics[width=\linewidth]{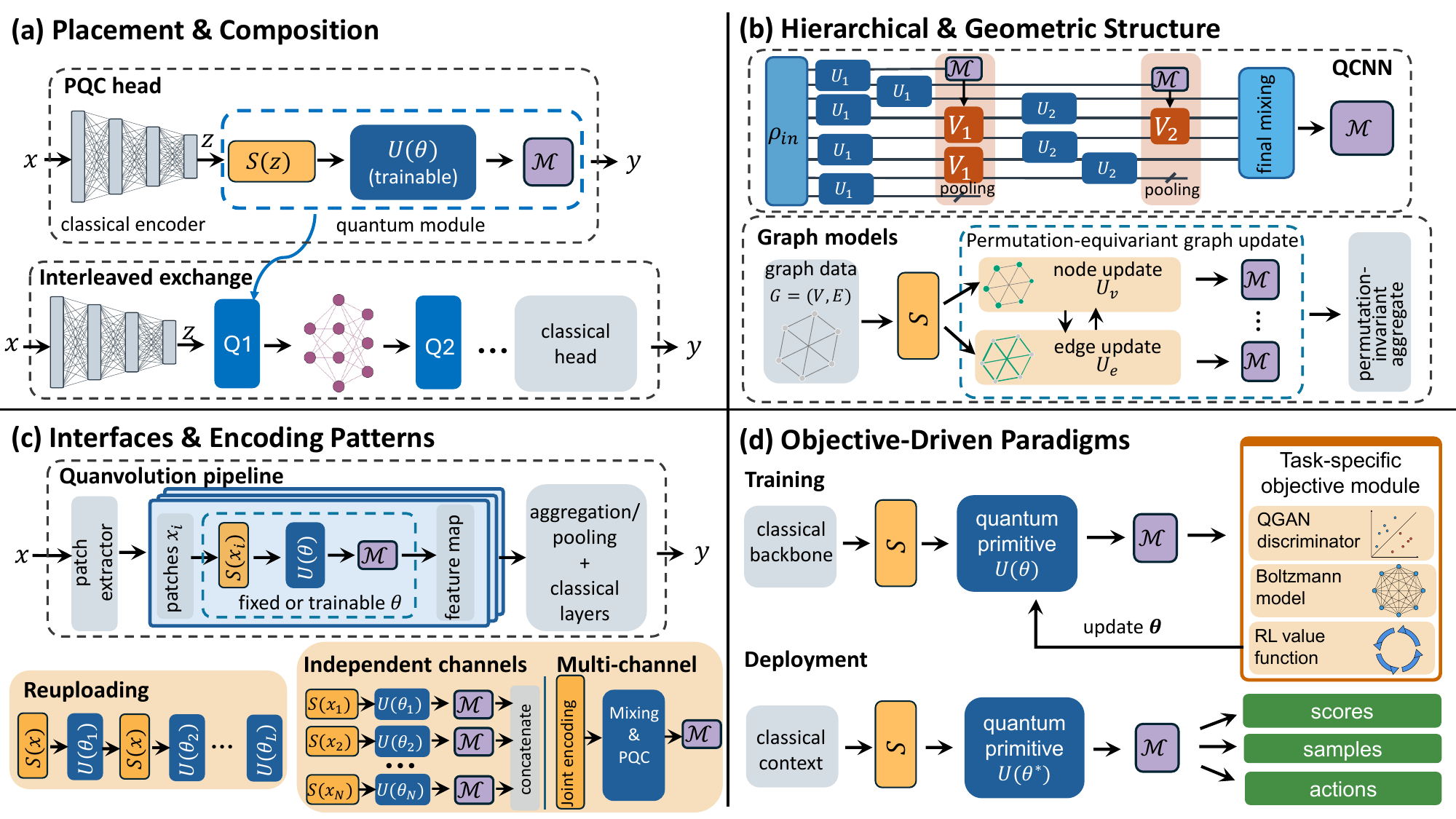}
\caption{Taxonomy of hybrid quantum-classical architectures across four paradigms.
(a) Placement and composition. A classical encoder compresses the input $x$ to a latent code $z$, which is embedded via $S(z)$; this encoding of $z$ rather than the raw data $x$ directly alters the induced hypothesis class. An interleaved-exchange variant (lower) routes $x$ through alternating quantum modules $Q_1, Q_2, \ldots$ and classical layers where each $Q_i$ subsumes the full encode-process-measure pipeline $S(\cdot) \to U(\boldsymbol{\theta}_i) \to \mathcal{M}$ of the upper subpanel. (b) Hierarchical and geometric structure. Quantum convolutional neural network (QCNN, upper): alternating trainable-unitary convolutional and measurement-based pooling layers act on $\rho_{\mathrm{in}}$, followed by a fully connected final mixing layer; Graph models (lower): node and edge unitaries $U_v$, $U_e$ implement permutation-equivariant message passing on $G{=}(V,E)$. (c) Interfaces and encoding patterns. Quanvolution (upper): local PQCs process image patches $x_i$ and output a spatial feature map for classical aggregation. Data reuploading (lower left): interleaving $S(x)$ with layers $U(\boldsymbol{\theta}_l)$ implements a Fourier-feature expansion. Independent and multi-channel (lower right): independent channels concatenate per-channel measurement outcomes classically, whereas in multi-channel, joint encoding applies coherent premeasurement mixing, enabling coherent cross-channel mixing in the quantum feature map prior to measurement. (d) Objective-driven paradigms. Training loop (upper): a classical backbone, quantum primitive $U(\boldsymbol{\theta})$, and task objective jointly update $\boldsymbol{\theta}$; instantiations include quantum generative adversarial networks (QGAN) discriminators, Boltzmann models, and reinforcement learning (RL) value functions. Deployment loop (lower): $U(\boldsymbol{\theta}^{*})$ operates at trained and fixed parameters under a sampling bottleneck on quantum devices, producing scores, samples, or actions.}
\label{fig:qdl_parads}
\end{figure*}

Hybrid quantum-classical architectures can be organized most cleanly by \emph{where} the quantum module appears in the computational graph, \emph{which} resources dominate end-to-end cost, e.g., circuit-evaluation budgets versus classical preprocessing, and \emph{what} inductive bias is enforced by the circuit topology, embedding map, and readout.
Figure~\ref{fig:qdl_parads} provides a schematic map of these design choices across four recurring paradigms at the level of computation graphs and estimator interfaces, while Table~\ref{tab:hybrid_architectures} lists representative instantiations and key references.

\begin{table*}[tb]\small
\caption{Representative taxonomy of hybrid quantum-classical models. Architectures are classified by the computational role of the quantum circuit and the primary structural constraint that shapes the hypothesis class and associated estimator family. Figure~\ref{fig:qdl_parads} depicts canonical \emph{core motifs} for each paradigm; the table lists representative instantiations and closely related variants. Abbreviations: PQC, parameterized quantum circuit; QCNN, quantum convolutional neural network; QNN, quantum neural network; RL, reinforcement learning; QGAN, quantum generative adversarial network.}
\label{tab:hybrid_architectures}
\centering

\newcommand{\colAa}{0.18\textwidth} 
\newcommand{\colBb}{0.24\textwidth} 
\newcommand{\colCc}{0.295\textwidth} 
\newcommand{\colDd}{0.22\textwidth} 

\newcommand{\cell}[2]{\parbox[t]{#1}{\raggedright #2}}

\begingroup
\setlength{\tabcolsep}{5pt}
\renewcommand{\arraystretch}{1.10}
\begin{ruledtabular}
\begin{tabular}{llll}
\cell{\colAa}{Paradigm} &
\cell{\colBb}{Quantum-module role} &
\cell{\colCc}{Inductive bias} &
\cell{\colDd}{Key refs.} \\
\hline

\noalign{\vskip 4pt}
\multicolumn{4}{l}{\textit{Placement and composition (graph location)}}\\
\noalign{\vskip 3pt}

\cell{\colAa}{PQC head} &
\cell{\colBb}{terminal readout / classifier} &
\cell{\colCc}{encoding-induced hypothesis restriction; restricted readout} &
\cell{\colDd}{\cite{farhi2018classification,schuld2020circuit}} \\

\cell{\colAa}{Transfer hybrid} &
\cell{\colBb}{quantum head on frozen classical features} &
\cell{\colCc}{representation bottleneck; low-dimensional embedding} &
\cell{\colDd}{\cite{mari2020transfer,chittem2025hybrid}} \\

\cell{\colAa}{Interleaved exchange} &
\cell{\colBb}{multiple quantum blocks separated by classical processing} &
\cell{\colCc}{non-sequential composition; measurement-mediated feature exchange across modules} &
\cell{\colDd}{\cite{liang2021hybrid,long2025hybrid,guo2025quantum,kabir2025lhqnn}} \\
\hline

\noalign{\vskip 4pt}
\multicolumn{4}{l}{\textit{Hierarchical and geometric structure (symmetry + locality priors)}}\\
\noalign{\vskip 3pt}

\cell{\colAa}{QCNN} &
\cell{\colBb}{hierarchical pooling} &
\cell{\colCc}{multiscale locality via pooling; parameter tying} &
\cell{\colDd}{\cite{cong2019quantum,herrmann2022realizing}} \\

\cell{\colAa}{Equivariant QNN} &
\cell{\colBb}{equivariant channels / symmetry-preserving layers} &
\cell{\colCc}{commutation with group action; parameter tying / twirling constructions} &
\cell{\colDd}{\cite{nguyen2024theory,schatzki2024theoretical}} \\

\cell{\colAa}{Graph models} &
\cell{\colBb}{graph-structured PQCs / graph learning blocks} &
\cell{\colCc}{permutation-equivariant graph embeddings; graph-structured interactions} &
\cell{\colDd}{\cite{verdon2019quantum,skolik2023equivariant}} \\
\hline

\noalign{\vskip 4pt}
\multicolumn{4}{l}{\textit{Interfaces and encoding patterns (encode-measure motifs)}}\\
\noalign{\vskip 3pt}

\cell{\colAa}{Data reuploading} &
\cell{\colBb}{repeated embedding; trainable layers} &
\cell{\colCc}{Fourier-feature periodic function class; universal approximation} &
\cell{\colDd}{\cite{perez2020data,schuld2021effect}} \\

\cell{\colAa}{Quanvolution} &
\cell{\colBb}{patchwise feature maps} &
\cell{\colCc}{local receptive fields; fixed random features or trainable filters} &
\cell{\colDd}{\cite{henderson2020quanvolutional,dou2023efficient}} \\

\cell{\colAa}{Channel-aware kernels} &
\cell{\colBb}{multi-channel circuit kernels} &
\cell{\colCc}{cross-channel mixing in circuit design} &
\cell{\colDd}{\cite{smaldone2023quantum}} \\
\hline

\noalign{\vskip 4pt}
\multicolumn{4}{l}{\textit{Objective-driven paradigms (estimator and readout regime)}}\\
\noalign{\vskip 3pt}

\cell{\colAa}{Transformer substitution} &
\cell{\colBb}{attention-like mixing block} &
\cell{\colCc}{structured mixing in classical transformer (no asymptotic claim)} &
\cell{\colDd}{\cite{chen2025quantummixed,chen2025quantum}} \\

\cell{\colAa}{Generative / unsupervised} &
\cell{\colBb}{generator/encoder modules} &
\cell{\colCc}{sampling-based or energy-based objectives} &
\cell{\colDd}{\cite{lloyd2018quantum,zoufal2021variational,benedetti2017quantum,shen2025variational}} \\

\cell{\colAa}{RL hybrids} &
\cell{\colBb}{policy/value module} &
\cell{\colCc}{stochastic policy/value via circuit + readout in RL loop} &
\cell{\colDd}{\cite{skolik2022quantum,jerbi2021quantum}} \\

\end{tabular}
\end{ruledtabular}
\endgroup
\end{table*}

\textit{(i) placement and composition.}---The most direct approach treats the PQC as a shallow and trainable quantum readout or classifier integrated as a compositional stage, often a terminal head, within a classical pipeline \cite{farhi2018classification,schuld2020circuit,schuld2014quest}. Figure~\ref{fig:qdl_parads}(a) fixes the corresponding computation graphs, emphasizing where measurement-derived features reenter the classical model in interleaved variants.
In transfer learning, a pretrained classical network serves as a frozen or lightly fine-tuned feature extractor, while a quantum head is trained on a low-dimensional vector \cite{mari2020transfer}. Domain-driven variants further emphasize latency and deployment constraints \cite{chittem2025hybrid}.
Recent proposals explore non-sequential layered hybrid architectures with parallel feature exchange, rather than a single pass from classical features to a quantum head~\cite{guo2025quantum,kabir2025lhqnn}.

\textit{(ii) Hierarchical and geometric structure.}---Hierarchical architectures encode symmetries and locality into circuit topology and parameter tying, imposing inductive biases that can improve trainability and sample efficiency when the enforced symmetry and locality match the task \cite{schatzki2024theoretical,grant2018hierarchical,meyer2023exploiting,bian2024symmetry}. Figure~\ref{fig:qdl_parads}(b) groups these approaches by how structure is enforced. Quantum convolutional neural networks (QCNNs) \cite{cong2019quantum} are structured PQC architectures that mirror the locality and hierarchy of classical CNNs for quantum data. Architecturally, a QCNN alternates between quantum convolutional layers, which apply parameterized, local entangling unitaries across neighboring qubits, and quantum pooling layers, which systematically reduce the system's dimensionality by measuring or discarding a subset of the qubits. By instantiating this hierarchical pooling for quantum inputs, QCNNs naturally align with the physical locality of quantum many-body systems and have been rigorously shown to avoid the barren-plateau phenomenon \cite{pesah2021absence}, while in some locally structured settings their induced decision rules can also admit efficient classical simulation (Sec.~\ref{subsubsec:dequantization}).
Representative extensions include higher-dimensional inputs \cite{sander2025quantum}, application-driven simulation studies \cite{chen2022quantumqcnn}, and alternative-platform realizations \cite{monbroussou2025photonic}.
Variants that modify the effective receptive field or adaptivity \cite{chen2022quantumdilated}, and noise-aware architecture optimization further explore robustness-resource trade-offs relevant for deployment \cite{maccormack2022branching,lee2025optimizing}.
Quantum blocks in classical CNN pipelines formalize convolutional hybrids in which local PQC modules act as trainable feature extractors coupled to classical nonlinearities and pooling \cite{dongfen2025quantum,liang2021hybrid,long2025hybrid}. Comparative studies emphasize strong sensitivity to qubit and layer choices and therefore the need for controlled, matched benchmarking \cite{zaman2024comparative}.

Mirroring the classical group-equivariant networks of Sec.~\ref{subsubsec:symmet_induc_bias}, equivariant QNNs construct channels or layers that commute with a target group action, enabling symmetry-preserving hypothesis classes with provable guarantees \cite{nguyen2024theory,schatzki2024theoretical}. Resource-efficient equivariant QCNN constructions can reduce measurement overhead via circuit parallelization \cite{chinzei2024resource}. The quantum analog of classical GNNs are quantum graph neural networks (QGNNs), with graph-interaction ans\"atze \cite{verdon2019quantum,lu2025quantum} and explicitly permutation-equivariant circuits via $S_n$-equivariant constructions or parameter tying \cite{skolik2023equivariant,bradshaw2025learning,liu2025rotation}.
Recent work also explores gate-level geometric constraints such as ``horizontal" gates on homogeneous spaces \cite{wiersema2025geometric}. Symmetry-informed encodings and architectures reduce hypothesis complexity and improve generalization and trainability~\cite{larocca2022group, bian2024symmetry, meyer2023exploiting}.

\textit{(iii) Classical-quantum data interfaces and encoding patterns.}---In hybrid visual or spatial models, the classical-quantum interface itself often serves as the primary inductive bias. As shown in Fig.~\ref{fig:qdl_parads}(c), quanvolutional architectures achieve this by encoding small, localized input patches into a quantum register, processing them via a local circuit, and measuring the output to produce a spatial feature map for subsequent classical layers \cite{henderson2020quanvolutional}. In the original proposal these patch circuits and quanvolution layers are fixed, meaning only the downstream classical postprocessing is trained. However, subsequent variants treat the patch kernel as a fully trainable PQC, effectively learning specialized local quantum filters \cite{dou2023efficient}. Recent work shows that making quanvolution layers trainable can substantially improve performance while raising nontrivial gradient-access challenges in deeper layers \cite{kashif2025deep}.

Figure~\ref{fig:qdl_parads}(c) also contrasts temporal data re-uploading with spatial multi-channel mixing.
From an architectural perspective, the classical data-access frequency defines the effective depth of the interface. Rather than encoding the classical patch only once, data reuploading repeatedly injects the input data interleaved with trainable unitaries \cite{perez2020data} (see Sec.~\ref{subsubsec:dynamic_encoding_fourier}.
Beyond depth, the interface also handles the structural complexity of the input, which is particularly challenging for multi-channel data. Naive encoding strategies face a strict physical bottleneck: processing each channel in an independent quantum circuit fails to capture critical inter-channel correlations prior to measurement, whereas classically concatenating all channels into a single, massive input vector causes the required qubit count to scale prohibitively. To resolve this, modern quantum convolutional kernels employ coherent pre-measurement mixing. As illustrated in Fig.~\ref{fig:qdl_parads}(c), this approach injects distinct channels into separate partitions of the qubit register and applies hardware-adaptable entangling operations to coherently mix the channel information before measurement \cite{smaldone2023quantum}. This ensures cross-channel correlations are captured at the quantum level without requiring exponential qubit overhead, yielding significant empirical improvements over single-channel QCNN baselines.

\textit{(iv) Objective-driven paradigms.}---While fundamentally relying on the architectural primitives described above, objective-driven hybrids are organized most cleanly by their readout regime and deployment loop. The choice of task, such as expectation estimation versus sampling, fixes the estimator family and the dominant circuit-evaluation cost. Transformer hybrids, for example, substitute a PQC for primitives in classical pipelines, with parameters updated via classical optimization from finite-shot measurement estimates \cite{chen2025quantummixed,chen2025quantum}. Here, the quantum circuit serves as a structured mixing or attention-like block on classically computed token embeddings encoded into quantum states.
Across generative and unsupervised objectives, practical cost is typically controlled by the number of circuit evaluations required for readout and gradients rather than by circuit depth alone. Depth-dependent expressivity-trainability trade-offs for implicit deep generative circuits are discussed in Sec.~\ref{sec:qdnn}.
Quantum generative adversarial networks (QGANs) train PQC generators to match target distributions via samples \cite{lloyd2018quantum,dallaire2018quantum,zoufal2019quantum,leyton2021robust,hu2019quantum}.
Expectation-value-sampler models use prescribed observable families for classical data, with tunable observables reducing shot costs \cite{shen2025variational}, while self-supervised hybrids employ variational quantum encoders in contrastive pipelines adapted to finite-shot measurement statistics \cite{jaderberg2022quantum}.
One proposal is to train certain quantum generative models classically when the training objective is classically tractable, while using the QPU primarily for deployment-time sampling, thereby shifting quantum cost to regimes where sampling is the bottleneck \cite{recioarmengol2025trainclassicaldeployquantum,kurkin2025universalitykerneladaptivetrainingclassically}; Fig.~\ref{fig:qdl_parads}(d) summarizes this training and deployment loop.

Alternative formulations include density and mixed-state QNN models that prepare mixtures of trainable unitaries (density QNNs) to improve expressivity–trainability trade-offs, especially on quantum hardware \cite{coyle2025training}, and Bayesian QDL frameworks aimed at providing principled predictive uncertainty estimates in quantum-enhanced models \cite{zhao2019bayesian}.
Diffusion variants use mixed-states, replacing scrambling unitaries with depolarizing channels and learning denoising via parameterized circuits and projective measurements \cite{kwun2025mixedstate}.
Energy-based approaches model Gibbs distributions with hybrid training \cite{amin2018quantum,zoufal2021variational,adachi2015application,melko2019restricted,benedetti2017quantum,shingu2021boltzmann}, including continuous-variable \cite{bangar2025continuous}, and sample-efficient \cite{coopmans2024sample} variants.
Related thermodynamic tasks use variational quantum thermalizers for approximate thermal-state preparation \cite{verdon2019quantumhamiltonian}, while quantum autoencoders compress quantum states \cite{romero2017quantum}.

In hybrid quantum RL, PQCs represent stochastic policies or value functions, trained via gradients including policy-gradient methods \cite{jerbi2021parametrized, chen2020variational, jerbi2021quantum,lockwood2020reinforcement} and deep Q-learning variants \cite{skolik2022quantum}, studied numerically on benchmark tasks.
Hybrid RL can be employed as a design mechanism for hybrid pipelines under hardware constraints by jointly learning a latent observation representation tailored to the QPU architecture \cite{nagy2025hybrid}.
Hybrid tensor network (TN) and variational circuit constructions have also been used to scale to higher-dimensional inputs and larger environments, reporting parameter savings and competitive performance in specific benchmarks \cite{chen2022variational}.

\subsubsection{Differentiation and gradient access}\label{subsubsec:gradient_access_hybrid}

Gradient-based training estimates gradients $\nabla_{\boldsymbol{\theta}} \mathcal{L}(\boldsymbol{\theta})$ from repeated circuit evaluations and measurements.
Unlike classical automatic differentiation \cite{baydin2018automatic} via single backward passes through deterministic primitives, hardware PQCs expose objective value only through finite-shot measurement statistics, with gradient access mediated by circuit evaluations in which cost and statistical accuracy are set jointly by the differentiation rule, observable count, and shot budget \cite{schuld2019evaluating,mitarai2018quantum,sweke2020stochastic}.

The cornerstone of hardware-native gradient estimation is the parameter-shift rule, rooted in quantum optimal control techniques \cite{li2017hybrid}, and subsequently adapted for PQCs \cite{mitarai2018quantum,schuld2019evaluating}.
For gate $U(\mu)=e^{-i\mu G_\mu}$ with generator $G_\mu$ having two distinct eigenvalues $\pm r$ and $G_\mu^2 = r^2 \mathbb{I}$, such as qubit rotations with $G_\mu = \sigma_k /2$ for a Pauli $\sigma_k$, the derivative of an expectation value $\langle {O} \rangle$ with respect to $\mu$ can be obtained from two shifted circuit evaluations:
\begin{equation}
    \partial_\mu \langle {O} \rangle = r \left( \langle {O} \rangle_{\mu + s} - \langle {O} \rangle_{\mu - s} \right), \qquad s=\frac{\pi}{4r}.
\end{equation}
Here, $\langle {O} \rangle_{\mu \pm s}$ denotes the expectation value at shifted $\mu$, while holding all other parameters fixed. On hardware, each $\langle O\rangle_{\mu\pm s}$ is replaced by a finite-shot estimator, rendering the resulting gradient estimator stochastic even when the underlying identity is exact. Additionally, exact parameter-shift rules for linear-optical photonic gates in Fock space have been derived for Fock-state inputs \cite{pappalardo2025photonic}.
For multi-eigenvalue generators, such as Hamiltonian coefficients, generalized parameter-shift rules express derivatives as linear combination of expectations \cite{wierichs2022general,kyriienko2021generalized}. For generators with complex spectra, stochastic parameter-shift rules \cite{Banchi2021measuringanalytic} circumvent the proliferation of shift terms by sampling offsets from a fixed distribution to produce an unbiased gradient estimator.

Beyond access to $\nabla\mathcal{L}$, geometry-aware approaches require access to an information metric on the variational manifold. Quantum natural gradient methods \cite{stokes2020quantum,wierichs2020avoiding,dellanna2025quantum,minervini2025quantum,sohail2025quantum} use a quantum information metric to precondition parameter updates for steepest descent directions on the variational manifold.
For a parameterized state $|\psi(\boldsymbol{\theta})\rangle$, preconditioned gradients are
\begin{equation}
    \tilde{\nabla}\mathcal{L} = g^{-1} \nabla\mathcal{L},
\end{equation}
where $g$ is the regularized quantum Fisher information metric tensor estimated from circuit data. Estimating a dense $g$ requires $\mathcal{O}(m^2)$ metric elements for $m$ parameters, motivating structured approximations and stochastic estimators.
To bypass explicit metric evaluation, quasi-Newton natural-gradient variants, such as the quantum Broyden adaptive natural gradient \cite{Fitzek2024optimizing}, approximate the preconditioner from iterative secant updates.
Alternatively, randomized natural-gradient variants estimate a usable preconditioner from randomized measurements with substantially reduced overhead compared to dense estimation \cite{kolotouros2024random}. Beyond ideal unitary circuits, these natural-gradient methods have also been generalized to noisy and nonunitary regimes, natively integrating metric estimation and regularization into the oracle model \cite{koczor2022quantum}.

\subsubsection{Resource-aware codesign principles}\label{subsubsec:resou_codesign}

Hybrid QDL design is a codesign problem in which architectural capacity needs to be balanced with trainability and measurement cost under hardware constraints. A central limitation is the concentration of measure. In the standard measurement-based setting without the ability to store and reuse quantum states coherently, i.e., without quantum memory, each partial derivative is typically estimated from fresh state preparations via repeated circuit executions and measurements. Achieving backpropagation-like scaling where gradient computation cost scales linearly with depth in fully coherent gradient-propagation schemes proposed to date~\cite{abbas2023backprop} generally requires access to multiple copies of intermediate quantum states, which is not available in standard measurement-based training. Complementarily, structured circuit families can reduce the circuit-count overhead for gradient estimation \cite{Bowles2025backpropagation}.
Moreover, if the number of trainable parameters is very large, parameter-shift-based gradient evaluation can become measurement-dominated, where shot accumulation time exceeds all other costs, even before accounting for shot noise.

Deep unstructured random ans\"atze approximate unitary 2-designs \cite{harrow2009random,brandao2016local}, inducing the barren-plateau phenomenon where gradient variance is exponentially suppressed with system size for broad global cost functions \cite{mcclean2018barren,holmes2022connecting,wang2021noise,cerezo2021cost,pesah2021absence,larocca2025barren}. Hence expressivity \cite{wu2021expressivity} needs to be balanced against trainability: increasing representational capacity can worsen gradient concentration \cite{du2020expressive,friedrich2023quantum}, leading to trade-offs between expressivity and gradient-measurement efficiency in deep PQCs \cite{chinzei2025trade}.
Recent results provide tight upper and lower bounds on loss and gradient concentration for broad classes of parameterized circuits and arbitrary observables, and show that these quantities can be estimated efficiently, furnishing diagnostics and sufficient conditions that can rule out barren plateaus in structured settings \cite{Letcher2024tightefficient}. These considerations clarify that gradient access needs to be specified jointly with circuit structure, observable choice, and shot budgets when assessing the practicality of gradient-based training.

Task-informed initialization, drawing on problem-specific priors \cite{liu2023mitigating}, and staged depth growth, incrementally increasing circuit layers \cite{grant2019initialization,skolik2021layerwise} sustain training within locally optimizable regions.
Matching entangling patterns to device connectivity and native gates minimizes compilation overhead \cite{kandala2017hardware}, while tailored strategies further reduce routing costs \cite{ji2024improving,ji2025algorithm}. Moreover, QEM proves effective only when its sampling overhead yields a net gain in gradient signal-to-noise ratio \cite{kandala2017hardware,endo2021hybrid,cai2023quantum,murali2019noise}.
Codesigning observable selection, grouping, estimation strategies, and shot allocation with the objective and architecture minimizes total circuit executions while satisfying accuracy thresholds \cite{verteletskyi2020measurement,huang2020predicting,huggins2021efficient,hadfield2022measurements,gresch2025guaranteed}. Additionally, shot noise provides regularization for optimization dynamics in specific regimes \cite{liu2025stochastic}. Finally, as capacity variations often produce nonmonotonic generalization and mismatched inductive biases can dominate outcomes, symmetry-aligned encodings and ans\"atze, along with capacity diagnostics, enable precise calibration of data and shot budgets \cite{caro2022generalization,meyer2023exploiting,gil2024understanding}.

\subsection{Quantum deep neural networks}\label{sec:qdnn}

Quantum deep neural networks (QDNNs) are parameterized quantum models that scale representation depth by using design variables such as alternating data-encoding and trainable layers or hierarchical circuits. This section focuses on depth-driven theory, including expressivity, scaling limits, and training dynamics, as well as motifs that enable stable training in deep regimes.

Motivated by the success of classical deep neural networks, QDNNs \cite{beer2020training, abbas2021power, jia2019quantum, perez2020data} have been studied since the early development of variational quantum algorithms (VQAs) \cite{peruzzo2014variational,mcclean2016theory}.
Deep regimes can expand the accessible function classes and frequency spectra via repeated embeddings and layers \cite{shin2023exponential,schuld2021effect}. Encoding-dependent generalization bounds complement this expressivity perspective \cite{caro2021encoding}.
Theoretical analysis has adapted classical neural network theory to the quantum setting.
Recent work shows that certain ensembles of QDNNs converge to Gaussian processes in appropriate large-dimension or large-width limits for specific observable choices \cite{garcia2025quantum,girardi2025trained,melchor2025quantitative}. In these regimes, gradient descent can admit a kernel description analogous to the classical NTK framework, motivating quantum NTKs that capture aspects of training dynamics and representation learning when training remains close to initialization \cite{liu2022representation,nakaji2023quantum,incudini2023quantum,tang2022graphqntk}. Related analyses identify a quantum lazy-training regime in which gradient descent remains close to initialization and learning is effectively kernelized \cite{Abedi2023quantumlazytraining}. These results separate circuit-controlled expressivity from kernel-controlled optimization and provide depth-scaling diagnostics when ensemble and measurement assumptions hold \cite{anschuetz2024unified,melchor2025quantitative}.
Additionally, overparameterization can improve optimization and generalization in studied regimes, analogous to classical DL phenomena and counter to worst-case intuitions \cite{arocca2023theory,Peters2023generalization}.
These overparameterization benefits have an analogue in quantum kernel methods: related double-descent behavior in quantum kernel regression reinforces that generalization is nonmonotone in model capacity \cite{kempkes2025doubledescentquantumkernel}.

QDNNs naturally extend to generative pipelines. Architectural families and benchmarking protocols are surveyed in Sec.~\ref{subsubsec:archi_parad_for_hyb}. Here we focus on depth-dependent expressivity and trainability phenomena in generative settings. Implicit circuit Born machines provide a canonical example in which deeper circuits act as richer implicit generative families learned from measurement samples \cite{liu2018differentiable}. Deep-circuit generative priors for inverse problems provide an application-driven instantiation of this depth-as-capacity viewpoint \cite{xiao2024quantum}. For implicit quantum generative models implemented by deep circuits, depth increases the class of representable distributions, but it can simultaneously exacerbate loss concentration and gradient suppression depending on the chosen divergence and how it is estimated from samples \cite{du2020expressive}.
\textcite{rudolph2024trainability} formalize trainability barriers arising in quantum generative modeling from the interplay between sample-based divergence estimation and circuit depth, showing that commonly used divergences can induce new barren-plateau mechanisms, while trainable low-body losses can fail to distinguish higher-order correlations, creating a depth-dependent tension between fidelity and trainability.

To mitigate deep-regime pathologies, the field has increasingly adopted structural inductive biases inspired by classical success stories.
Residual and skip connections, which are central to classical deep residual learning \cite{he2016deep}, have quantum analogues that improve gradient propagation and/or enrich effective representations in numerical and theoretical studies \cite{kashif2024resqnets,wen2024enhancing}. Reduced-domain parameter initialization scales the initial parameter range inversely with the square root of depth and yields depth-explicit trainability guarantees in the stated settings \cite{wang2024trainability}.
Recent work explicitly targets deep quanvolutional stacks and analyzes architectural mechanisms for stabilizing training as depth grows \cite{kashif2025deep}. Beyond purely unitary ans\"atze, dynamic circuits with intermediate measurements and classical feedforward have been proposed as depth-scalable architectures that can avoid barren-plateau mechanisms in the stated models while retaining expressivity \cite{deshpande2024dynamic}.
Similarly, permutation-equivariant QNN constructions admit setting-specific guarantees linking symmetry constraints to improved trainability proxies \cite{schatzki2024theoretical}.
Beyond feedforward depth, recurrent QNN formulations have been proposed for sequence learning \cite{bausch2020recurrent}, and recent work explores architectures designed for variable-size inputs \cite{vargas2025quantum}.
A central open problem is to identify scalable, reliably trainable QDNN families that remain beyond known efficient classical simulation under the stated access, noise, and observable model, outside restricted or warm-started regimes~\cite{meyer2025trainabilityquantummodelsknown, mhiri2025unifyingaccountwarmstart, puig2025warmstarts}.

\subsection{Quantum algorithms for deep neural networks}\label{sec:q_for_dnn}

Quantum algorithms for deep neural networks (DNNs) aim to accelerate the training and inference of their \textit{classical} counterparts under explicit input and data-access models. Distinct from PQC-based approaches that define new trainable hypothesis classes, these algorithms seek to reproduce the input-output mapping of a specific, predefined classical network, such as a CNN~\cite{krizhevsky2012imagenet}, ResNet~\cite{he2016deep}, or Transformer~\cite{vaswani2017attention}, with a theoretical speedup. Many proposals build on QLA primitives, e.g., HHL \cite{harrow2009quantum} and QSVT~\cite{gilyen2019quantum,martyn2021grand} to accelerate linear-algebraic subroutines under FTQC assumptions \cite{preskill1997faulttolerantquantumcomputation}. End-to-end complexity is conditional on the access-and-readout model $\mathcal{A}$, particularly QRAM availability and its substantial architectural overhead as analyzed in Sec.~\ref{subsubsec:data_acces_qram}. The resource assessment there applies directly to the algorithms surveyed below.

\subsubsection{Nonlinearity and the measure-and-reprepare paradigm}\label{sec:q_for_dnn_foundational}

Applying an element-wise nonlinearity to a classical vector encoded via amplitude encoding, while remaining within the same amplitude-encoding model, cannot in general be achieved exactly by a deterministic unitary alone since unitaries act linearly, so implementing a layerwise classical nonlinearity typically requires non-unitary primitives such as measurement, post-selection, or probabilistic subroutines. A foundational work by ~\textcite{kerenidis2019quantum} introduced a quantum algorithm for implementing and training a CNN via a ``quantum CNN" construction that explicitly incorporates nonlinearities and pooling. Their approach combines quantum routines for the linear parts of the network with a measurement-mediated mechanism to implement nonlinear activation and pooling steps. To handle the output, they introduced a novel tomography protocol with $\ell_\infty$-norm guarantees with query complexity $\mathcal{O}(\log N_{\mathrm{out}} / \epsilon^2)$ calls to the state-preparation unitary and associated measurements, where $N_{\mathrm{out}}$ is the dimension of the output vector and $\epsilon$ is the maximum allowable error per component.

The example above illustrates the \textit{measure-and-reprepare} paradigm. To apply the nonlinearity, a measurement is performed at the end of each layer, the classical result is processed, and a new quantum state is prepared for the next layer. The need for intermediate measurement and repreparation introduces an explicit per-layer overhead, including precision-dependent sampling and tomography costs, which can dominate at large depth $L$ and undermine coherent end-to-end scaling. Related transformer-inference proposals based on QLA similarly assume a pretrained classical model. When a classical output vector is required, they obtain it by measurement or tomography, and for multilayer settings this entails iterating the subroutine over tokens and layers~\cite{guo2024quantum}.

\subsubsection{The coherent approach}\label{sec:q_for_dnn_coherence}

An alternative to intermediate measurements is to implement the nonlinearity coherently, i.e., without layerwise readout. A coherent approach approximates the activation via a polynomial and applies it using QSVT.
\textcite{guo2024nonlinear} formalize nonlinear amplitude transformation in an oracle model, constructing block-encodings from state-preparation unitaries to apply polynomial transforms coherently. Subsequent refinements, such as randomized QSVT~\cite{wang2025randomized}, aim to reduce block-encoding overheads and ancilla complexity, trading them for randomized sampling and different polynomial-degree dependences. \textcite{rattew2025accelerating} propose a fully coherent multilayer inference construction analyzing multiple quantum data-access regimes for ResNet-like architectures. Their analysis highlights that skip connections can help stabilize the propagation of vector norms across depth, improving the stability of the coherent inference procedure.

While coherent approaches can yield polylogarithmic-in-$N$ dependence under strong input oracles for the encoded vector dimension $N$, the overall cost is governed by state-preparation complexity and the polynomial degree required to approximate nonlinearities to target error, and by conditioning parameters in the underlying QLA primitives, which typically scales exponentially with depth to maintain constant end-to-end precision. This highlights a central trade-off: end-to-end coherence is achieved at the cost of algorithm- and access-model-dependent depth overhead, which may become prohibitive at large depth under stringent precision guarantees. Consequently, such algorithms are most compelling when the required depth and approximation degree remain moderate. This suggests that a scalable, fully coherent quantum algorithm for general deep networks with polynomial dependence on depth under realistic access models remains an open problem.

\subsubsection{Component-wise acceleration for modern architectures}\label{sec:q_for_dnn_frontier}

Given the limitations of full-network coherent simulation discussed above, particularly the depth-exponential overhead, a more pragmatic direction has been to focus on accelerating individual, computationally expensive components of modern architectures. This modular approach accepts the current limitations while still offering potential speedups for critical subroutines. For instance, the attention mechanism, a central computational bottleneck in Transformers, has been a major target. As detailed in a recent survey by \textcite{zhang2025survey}, approaches here span both PQC-based paradigms and QLA-based paradigms, with complexity–resource trade-offs and data-access assumptions playing a decisive role. \textcite{cherrat2024quantumvision} propose quantum vision-transformer variants in which patch vectors are loaded via amplitude-encoding style data loaders and processed by quantum orthogonal layers, yielding quantum attention mechanisms that (under their access model) can offer asymptotic runtime and parameter-count advantages relative to classical attention scaling in the number and dimension of patches. \textcite{gao2023fast} present a fast quantum algorithm specifically for computing a sparse attention matrix using Grover-search–based techniques, achieving a polynomial speedup for that sparse-attention setting rather than a generic QLA reduction of dense $QK^\top$ and Softmax.
Similarly, the paradigm has been extended beyond discriminative models. \textcite{Xiao2025quantumdeeponet} adapted the DeepONet architecture, designed for solving partial differential equations, to the quantum domain. They proposed a quantum DeepONet formulation aimed at reducing the evaluation complexity (e.g., in the input dimension). Meanwhile, \textcite{wang2025efficientquantumalgorithmsdiffusion} developed a quantum algorithm for diffusion models~\cite{sohl2015deep}, focusing on accelerating the iterative denoising step by using quantum differential equation solvers and QLA to apply the denoising function.

\subsubsection{Alternative paradigms and challenges}\label{sec:q_for_dnn_critique}

An entirely different path is to bypass the direct simulation of the network's forward pass. \textcite{zlokapa2021quantum} give a quantum algorithm for approximately training overparameterized classical networks in a linearized NTK-type regime by reducing the training update to structured linear-system inversion. Any end-to-end speedup over classical gradient-based training additionally requires efficient quantum state preparation and readout for the relevant data distribution. Related NTK or quantum NTK-based constructions have been developed for structured settings, for example, by~\textcite{tang2022graphqntk} for graph data. In the infinite-width limit where training a DNN is equivalent to a kernel method problem under lazy-training scaling assumptions, this approach elegantly sidesteps the layer-by-layer complexity, though it is restricted to the regime where the NTK theory holds.

The practical relevance of asymptotic speedups hinges on the full access and readout contract. In particular, many QLA and QSVT-based proposals implicitly assume QRAM-style superposition access and/or efficient state preparation, which can dominate end-to-end cost (see Sec.~\ref{subsubsec:data_acces_qram}).
Moreover, QDL faces a moving target problem~\cite{gundlach2025quantum}, i.e., speedup claims that ignore access costs or benchmark only against weak baselines can erode under stronger classical competitors (see Sec.~\ref{subsubsec:class_bound}).
As emphasized in a unifying perspective by~\textcite{liu2024towards}, provable quantum advantage often requires explicit access models and end-to-end accounting of state preparation, measurement, and classical post-processing, and typically hinges on problem structure that keeps key algorithmic parameters controlled. The immense overheads hidden within the polylogarithmic factors of QLA and QSVT, combined with the stringent demands of fault-tolerance and the monumental challenge of realizing QRAM, mean that practical advantage remains distant. Nevertheless, recent work~\cite{rattew2025accelerating} has begun to provide more realistic cost models by analyzing multilayer performance under weakened data-access assumptions, showing that the speedup can degrade from exponential to polynomial when QRAM-like loading is not available, demonstrating a critical step towards assessing viability.

\subsection{Quantum-inspired classical algorithms}\label{subsec:quant_inspi_class}

Quantum-inspired classical algorithms leverage quantum information theoretic concepts to design methods executable entirely on conventional hardware. Representative families include (i) dequantization frameworks, which replicate purported quantum speedups when granted the same oracle access and sampling assumptions as the quantum algorithm \cite{tang2019quantum,chia2022sampling,tang2022dequantizing,kerenidis2016quantum,aaronson2015read}; (ii) TN architectures, which exploit entanglement structures as inductive biases for compression and simulability \cite{schollwock2011density,orus2014practical,stoudenmire2016supervised,cichocki2016tensor}; and (iii) quantum-inspired network architectures, which incorporate motifs such as complex-valued representations and unitary layers into classical pipelines \cite{trabelsi2017deep,scardapane2020complex}.
These approaches constitute an essential baseline family for QDL within the access-model framework of Sec.~\ref{subsec:comp_symbiosis}.

\subsubsection{Dequantization}\label{subsubsec:dequantization}

Early QML proposals suggested that coherent access to data could accelerate linear-algebraic primitives underlying learning, including matrix inversion \cite{harrow2009quantum,rebentrost2014quantum} and eigenvalue estimation \cite{Lloyd2014}.
Dequantization makes the input model explicit.
As detailed in Sec.~\ref{subsubsec:data_acces_qram}, several exponential speedups rely on QRAM-style data oracles.
Classical algorithms with comparable sample and query (SQ) access can achieve similar complexity for several low-rank linear-algebraic primitives and approximate output goals, typically up to polynomial factors in rank, condition number, and inverse precision, shifting the burden of proof to end-to-end, access-model-matched comparisons \cite{tang2022dequantizing,aaronson2015read,chia2022sampling,mande2025lower}.

A canonical example is \textcite{tang2019quantum}, who constructed a quantum-inspired classical analogue of the recommendation system of \textcite{kerenidis2016quantum}.
Under length-squared sampling access and a low-rank promise (rank $k$ for an $m \times n$ matrix), the classical runtime becomes poly$(k)$ and polylog$(mn)$ up to additional polynomial dependence on accuracy- and conditioning-related parameters, eliminating the apparent exponential separation under the same oracle assumptions \cite{tang2019quantum,tang2021quantum,chia2022sampling}.
Subsequent work generalized this perspective beyond recommendation-style tasks to additional SQ-access linear-algebraic primitives and to dequantizations of certain QSVT-related transformations under structural promises, emphasizing that precision and conditioning can dominate the true complexity \cite{chia2022sampling,gharibian2022dequantizing}.
These frameworks can also be made robust to approximate SQ access, with complexity acquiring explicit polynomial dependence on the SQ-approximation error parameters \cite{le2025robust}.

Crucially, the principle that strict structural constraints enable efficient classical emulation extends beyond data access models to the QNN architectures themselves. For instance, recent work demonstrates that for certain unitary QCNN architectures and locally structured datasets, the induced decision rule can be efficiently classically emulated given access to local measurement data \cite{bermejo2024quantum}.
More broadly, increasing evidence shows that highly constrained ans\"atze, such as those restricted to small dynamical Lie algebras \cite{Goh2025} or built with strong symmetry-constrained constructions \cite{Anschuetz2023efficientclassical}, inherently admit efficient classical simulation or have low effective dimension. Ultimately, just as low-rank data promises allow classical algorithms to match quantum recommendation systems, these rigid architectural constraints often undermine the necessity of a quantum processor.

However, dequantization has principled limits. Using communication-complexity methods, \textcite{mande2025lower} derive lower bounds for SQ-access quantum-inspired algorithms on several matrix problems, implying regimes where at least quadratic quantum-classical separations persist under the stated access model and accuracy parameters.
Moreover, \textcite{cotler2021revisiting} show that SQ access can be strictly stronger than receiving copies of the corresponding quantum state, so SQ-based dequantization baselines may be inappropriate for native quantum data settings where the learner's interface is restricted to state preparation and measurement outcomes. In practice, these dequantized analogues can incur large polynomial overheads in conditioning and accuracy parameters, limiting competitiveness outside favorable regimes \cite{Arrazola2020quantuminspired}.

\subsubsection{Tensor network architectures\label{subsubsec:tn_architectures}}

\begin{figure}[tb]
    \centering
    \includegraphics[width=\columnwidth]{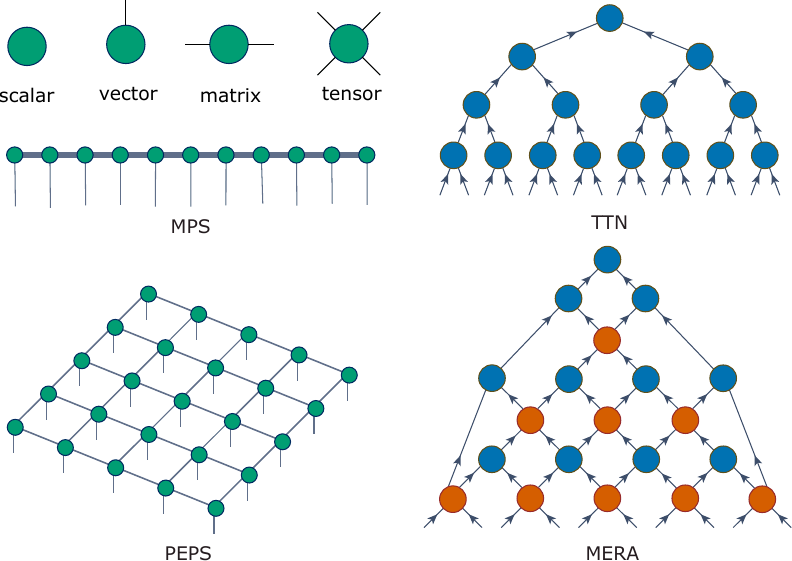}
    \caption{Fundamental tensor network (TN) architectures. Shown are matrix product state (MPS; also known as tensor train), projected entangled pair state (PEPS), tree tensor network (TTN), and the multiscale entanglement renormalization ansatz (MERA). The top row illustrates increasing tensor order (scalar, vector, matrix, tensor) as a visual legend for node types. Adapted from \textcite{berezutskii2025tensor}.}
    \label{fig:tensor_networks_arc}
\end{figure}

Tensor networks provide a concrete bridge between quantum many-body structure and classical DL by representing high-order tensors through networks of low-rank cores \cite{cohen2016expressive,berezutskii2025tensor,orus2014practical,rieser2023tensor}.
In QDL, TNs serve two roles: (i) as standalone models, where the network geometry and bond dimensions impose an entanglement-motivated inductive bias on classical data \cite{stoudenmire2016supervised}, and (ii) as tensorized parameterizations that replace dense weight tensors by factored TN forms to reduce memory and parameter counts while preserving expressive structure \cite{novikov2015tensorizing, wu2023fromtn}.
The mechanism is controlled by two coupled quantities: the bond dimensions, which set the representational capacity, and the contraction complexity, which sets the computational cost \cite{orus2014practical}.
In quantum physics, TN efficiency is motivated by low-entanglement structure, formalized in area-law behavior for broad families of ground states \cite{eisert2010colloquium}.
In ML, an analogous motivation is that many natural signals, such as images and time series, exhibit structured, predominantly local correlations \cite{ruderman1994statistics,simoncelli2001natural}, so low-bond TN priors can act as controlled low-rank constraints rather than generic overparameterization.
When utilizing role (ii), parameter and compute costs are explicitly governed by a rank and bond hyperparameter \cite{cichocki2016tensor}, which typically yields parameter efficiency and an implicit regularization effect by limiting the effective correlation capacity that the layer can represent. However, benefits are regime dependent. TN constraints can be helpful when the task admits low-rank structure, but they do not provide uniform improvements in worst-case learnability and generalization guarantees without additional alignment between data structure and network geometry \cite{wu2025nofreelunch,jahromi2024variational}.
Figure~\ref{fig:tensor_networks_arc} summarizes four fundamental TN architectures, which we detail below.

The matrix product state (MPS) \cite{verstraete2004density,verstraete2008matrix,cirac2021matrix,bengua2015optimal,perez2006matrix,schollwock2011density}, also known in ML as the tensor train decomposition \cite{oseledets2011tensor,grasedyck2013literature}, factorizes an order-$N$ tensor into a 1D chain with bond dimension $\chi$, yielding $O(N d \chi^2)$ parameters and typical contraction costs scaling polynomially in $\chi$.
MPS can serve as tractable generative models, such as Born-machine variants, when the dominant dependencies are effectively one-dimensional \cite{han2018unsupervised,wall2021generative}.
As a learning model, the 1D geometry biases toward ordered structure, but capturing long-range correlations typically requires the bond dimension $\chi$ to grow rapidly with problem size ~\cite{stoudenmire2016supervised,novikov2015tensorizing,harvey2025sequence}. Recent analysis further shows that objectives defined by global observables can exhibit vanishing derivatives, whereas local-observable objectives can remain trainable, and identifies regimes admitting efficient classical evaluation \cite{basheer2025trainability}. Relatedly, variational schemes for preparing short-range MPS on quantum hardware have been developed to reduce circuit depth \cite{jaderberg2026variational}.

The projected entangled pair state (PEPS) \cite{verstraete2004renormalization,verstraete2008matrix,cirac2021matrix,stoudenmire2018learning} generalize MPS to 2D lattices, matching spatial locality and motivating applications to image-like data. PEPS can be viewed as locality-respecting graphical models over 2D neighborhoods, loosely reminiscent of convolutional inductive biases but without the defining CNN mechanisms such as weight sharing and explicit translation equivariance.
They have also been investigated for supervised learning and for modeling spatially structured datasets \cite{cheng2021supervised}.
Their central limitation is computational. In the worst case, exact contraction of 2D PEPS is \#P-hard \cite{schuch2007computational}, so practical use relies on approximate contraction schemes whose cost can dominate training and inference.

Tree tensor networks (TTNs) are a hierarchical generalization of MPS, organizing tensors in a tree-like structure \cite{shi2006classical,stoudenmire2018learning}. This multi-scale architecture has been used successfully as a hierarchical classifier or a feature extractor analogous to pooling layers in a classical CNN \cite{huggins2019towards}. 
TTNs are efficiently contractible and offer a compromise between the limited 1D expressivity of MPS and the costly contraction of PEPS. They can capture non-local correlations along the tree's branches, imposing an inductive bias well-suited for data with a natural compositional or hierarchical structure.
The main limitation of TTNs is their rigidity: the imposed tree structure may not align with the translational or rotational symmetries of many natural datasets. 

The multiscale entanglement renormalization ansatz (MERA) \cite{vidal2007entanglement,evenbly2009algorithms} augments hierarchical structure with explicit disentanglers and isometries to model correlations across scales, originally targeting critical states with logarithmic corrections to area laws.
As a classical architecture it can encode scale structure but incurs higher contraction costs.
As a quantum inspiration, its connectivity patterns motivate QCNN-style ans\"atze and associated trainability analyses ~\citep{cong2019quantum, pesah2021absence}.

Selecting a TN architecture requires balancing the data's correlation structure with the TN's inherent geometry. The bond dimension $\chi$ is a primary hyperparameter governing expressivity for a fixed TN parameterization, with larger values capturing more complex correlations at the cost of elevated training and inference demands for the classical model.
Training and model selection remain practical bottlenecks. TN learning can be performed via sweep-style local updates or via gradient-based optimization through differentiable contractions. Both are sensitive to conditioning, gauge choices (redundant parameterizations of the same TN), and approximation error, so reliable performance typically requires careful regularization and topology and rank selection \cite{stoudenmire2016supervised,liao2019differentiable,evenbly2018gauge}.

Constructive mappings relate several neural-network families to TN representations, providing an explicit classical lens on correlation capacity via entanglement-inspired measures \cite{chen2018equivalence,glasser2018neural,levine2019quantum}.
For RBMs, the induced TN structure is controlled by the hidden-visible interface. Local (short-range) RBMs satisfy area-law entanglement bounds, whereas dense RBMs can realize substantially stronger scaling \cite{chen2018equivalence,glasser2018neural}.
These results have a direct benchmarking implication for QDL on classical data. Arguments based on entanglement-like expressivity need to be tested against classical models whose induced TN capacity matches the same correlation regime.
These correspondences are primarily representational, not algorithmic. A TN representation does not, in general, imply efficient contraction or efficient inference.
Depth-separation results further delineate regimes where low-bond TN surrogates are inadequate, showing that depth and nonlocal connectivity can compactly represent correlation families, including families generated by polynomial-size quantum circuits, that shallow constructions cannot \cite{gao2017efficient,deng2017quantum}.

\subsubsection{Quantum-inspired network architectures}

Beyond TN-based compression, which exploits quantum entanglement structure, a distinct family of fully classical baselines imports mathematical motifs from quantum theory into neural architectures without invoking coherent quantum resources. They test whether purported ``quantum" gains on classical data are better explained by classical inductive biases that can be implemented and optimized on conventional hardware.

Complex-valued and hypercomplex parameterizations introduce phase-like degrees of freedom and structured coupling between channels.
Complex linear layers can be represented as real block-structured maps on doubled channels, but the resulting constraints and nonlinearities can still change optimization and inductive bias relative to generic real-valued networks \cite{trabelsi2017deep,bassey2021survey}.
Quaternion and hypercomplex networks extend this idea by tying groups of channels through algebraic products, often improving parameter efficiency and sometimes empirical performance in settings with correlated feature structure \cite{gaudet2018deep,parcollet2020survey,parcollet2018quaternion}.
Representative instantiations include quantum-measurement motivated complex-valued models for conversational emotion recognition \cite{li2021quantum}, entanglement-inspired embedding constructions for matching \cite{zhang2025quantuminspired}, and superposition-inspired spiking networks \cite{sun2021quantum}.

Gate-inspired qubit neuron models update normalized amplitude-like hidden states via rotation-style transformations and produce probabilities through squared-amplitude readout \cite{mori2006qubit,li2013hybrid,ganjefar2018optimization}.
Subsequent variants extend the state space beyond qubits but remain best interpreted as constrained classical sequence models akin to modern structured parameterizations \cite{bakshi2025quantum}.
Quantum probability and density-matrix formalisms represent uncertainty via positive semidefinite operators and scoring hypotheses using quantum-inspired divergences \cite{tiwari2019towards,leporini2022efficient,yan2021quantum}.
These architectures sharpen QDL evaluation by enlarging the classical comparator set to include quantum-structured priors that require no quantum hardware.

\subsection{Optimization as prerequisite in QDL}\label{subsec:optim_as_prereq}

Training QDL models requires minimizing a loss function $\mathcal{L}(\boldsymbol{\theta})$ derived from quantum observables \cite{beer2020training,liang2021hybrid,romero2021variational,peruzzo2014variational,mcclean2016theory}.
Optimization proceeds via oracle access to shot-limited estimates of $\mathcal{L}$ and, when available, $\nabla \mathcal{L}$.
On quantum hardware, device noise biases the objective, while finite-shot estimation induces stochasticity and drift induces nonstationarity, making optimizer choice a resource-allocation problem over evaluations, shots, classical compute, and wall-clock latency \cite{scriva2024challenges,Kungurtsev2024iterationcomplexity}.
Optimizer comparisons require matched search and tuning budgets with reported evaluation and shot costs \cite{ostaszewski2021structure,benitez2025bayesian,moussa2022hyperparameter,moussa2024hyperparameter,herbst2024optimizing}.

\subsubsection{Local optimizers}

Local optimizers update parameters using information from a local neighborhood of the current iterate, e.g., gradients, local sampling, or a search distribution centered near current parameters, and thus can converge to non-global stationary points in high-dimensional, nonconvex landscapes.
In QDL, the choice of local optimizer is governed less by mathematical differentiability than by the measurement overhead and stochastic variance inherent in the hybrid cost model.
Stochastic loss and gradient estimates from the QPU feed a classical training stack for parameter updates and job submission \cite{benedetti2019parameterized,broughton2020tensorflow}.

When gradient estimators are accessible, first-order methods are the natural choice. Gradient descent \cite{cauchy1847methode,curry1944method} and stochastic variants such as Adam \cite{Kingma2014} serve as standard baselines. However, their practical convergence behavior depends strongly on shot-noise variance and batching strategies \cite{scriva2024challenges}. In low-noise regimes, quasi-Newton methods which approximate second-order curvature information, such as limited-memory BFGS \cite{liu1989limited}, can provide strong local refinement but degrade rapidly under stochasticity. To stabilize gradient-based training, practitioners often employ symmetry constraints \cite{bian2024symmetry}, initialization heuristics \cite{puig2025warmstarts,mhiri2025unifyingaccountwarmstart}, or operational strategies like classical surrogate pretraining followed by quantum fine-tuning \cite{dborin2022matrix}. While these strategies successfully localize the search to trainable subspaces, they do not eliminate the underlying global nonconvexity.

Conversely, when gradients are prohibitively costly or unreliable, derivative-free search dominates. Lightweight local baselines, such as Nelder-Mead~\cite{nelder1965simplex} and pattern search~\cite{hooke1961direct}, remain viable despite their poor scalability with dimension.
For broader exploration, population-based algorithms, such as particle swarm optimization ~\cite{kennedy1995particle,bonyadi2017particle}, differential evolution ~\cite{storn1997differential}, and covariance matrix adaption evolution strategy~\cite{hansen2006cma}, propose candidates from population statistics and best-so-far solutions. Their global behavior requires restarts or hybridization \cite{auger2005restart,muller2009particle}.
In practice, derivative-free and surrogate-assisted routines are also used for outer-loop ansatz selection and tuning under shot-limited evaluations \cite{nguyen2022quantum,sagingalieva2023hybrid}.
Controlled benchmarks indicate that optimizer rankings are instance- and noise-regime dependent and sensitive to hyperparameter tuning \cite{bonet2023performance}. Specialized applications further emphasize this nuance; for example, chemistry-motivated VQE variants benefit significantly from structure-aware routines that accelerate excitation-operator selection and parameter optimization \cite{jager2025fast}.

\subsubsection{Global optimizers}

Global-exploration heuristics aim to reduce sensitivity to initialization by broader search, but for the nonconvex objectives they provide no general, finite-time global-optimality guarantees \cite{horst2013global,stephens1998global}.
Random search~\cite{masri1980global,munson1959ire} provides a minimal-overhead baseline.
A canonical model-based alternative is Bayesian optimization~\cite{kushner1964new,garnett2023bayesian,shahriari2015taking,frazier2018bayesian}, which uses a surrogate, often a Gaussian process, and an acquisition rule such as expected improvement to balance exploration and exploitation~\cite{anton2024review,movckus1974bayesian,jones1998efficient,huang2006sequential,sobester2004parallel,keane2006statistical,knowles2006parego,mockus1991bayesian}. Bayesian optimization is attractive when evaluations are scarce and expensive, as in shot-limited circuit execution \cite{ji2025dataefficient}, but degrades in high dimensions unless strong structure is exploited.
A Bayesian-related method is the data-based online nonlinear extremum-seeker~\cite{verstraete2015optical}, which builds a surrogate model but exhibits more favorable scaling with dimensionality and iteration count than standard Bayesian optimizers. Finally, deterministic partitioning strategies such as multilevel coordinate search \cite{huyer1999global} provide a complementary route to global exploration by iteratively subdividing and sampling promising regions.

\subsubsection{Effects of noise and drift}

Noise in QDL optimization arises from distinct mechanisms that should be separated operationally \cite{herbst2023beyond}. First, finite-shot measurement induces stochastic estimators of the implemented objective and can set the dominant precision cost \cite{scriva2024challenges}. Second, physical device noise biases the implemented objective relative to the ideal target, so guarantees become oracle-assumption dependent \cite{Kungurtsev2024iterationcomplexity}. Finally, hardware calibration drift introduces nonstationarity that complicates stopping and benchmarking under fixed budgets \cite{zhang2023DISQ}.
Accordingly, noise-aware optimization emphasizes variance control and explicit accounting of evaluation cost. Simple replication and batching with robust aggregation provides a simple baseline, while more advanced shot-adaptive dynamically couple the optimizer's step selection directly to the measurement precision \cite{kubler2020inductive,sweke2020stochastic}.

A prominent stochastic approach is simultaneous perturbation stochastic approximation, which approximates gradients via two objective evaluations per iteration, independent of dimensionality in the number of objective calls per iteration, although estimator variance and required iteration counts can still grow with dimension~\cite{spall1992multivariate,spall1998implementation,wiedmann2023anempirical}.
For surrogate-based globalization, noisy or constrained Bayesian optimization formulations can improve sample efficiency when evaluations are scarce \cite{letham2017constrained,oliveira2019bayesian}. Alternatively, population-based heuristics can be adapted to navigate stochastic fitness landscapes, though their ultimate convergence performance remains highly sensitive to both the specific problem instance and the underlying parameter dimensionality \cite{rakshit2017noisy,beyer2007robust,arnold2000efficiency,richter2020model,levitan1995adaptive,beyer2006evolution,arnold2004performance}.

\subsection{Theoretical advantages and fundamental limitations}\label{subsec:theor_advan}

Two fundamental questions determine the theoretical standing of any QDL proposal: what can be learned under a specified access-and-readout model, and what end-to-end resources are required to reach a target accuracy. Because both depend strictly on the framework $(\mathcal{D}, \mathcal{R}, \mathcal{A})$ established in Sec.~\ref{subsec:comp_symbiosis}, an advantage claim is meaningful only when these parameters are fixed and the classical comparator class is explicitly stated. To operationalize this, we introduce a resource taxonomy separating consumable complexity axes from structural hypothesis-class constraints. Using this taxonomy, we analyze statistical learnability and complexity-theoretic separations, delimit provable quantum advantage against classical dequantization boundaries, and characterize trainability as a scaling constraint that directly couples expressivity to classical simulability.

\subsubsection{Taxonomy of operational quantum advantage}\label{subsubsec:taxon_of_quant_advan}

\begin{figure*}[tb]
\centering
\includegraphics[width=\linewidth]{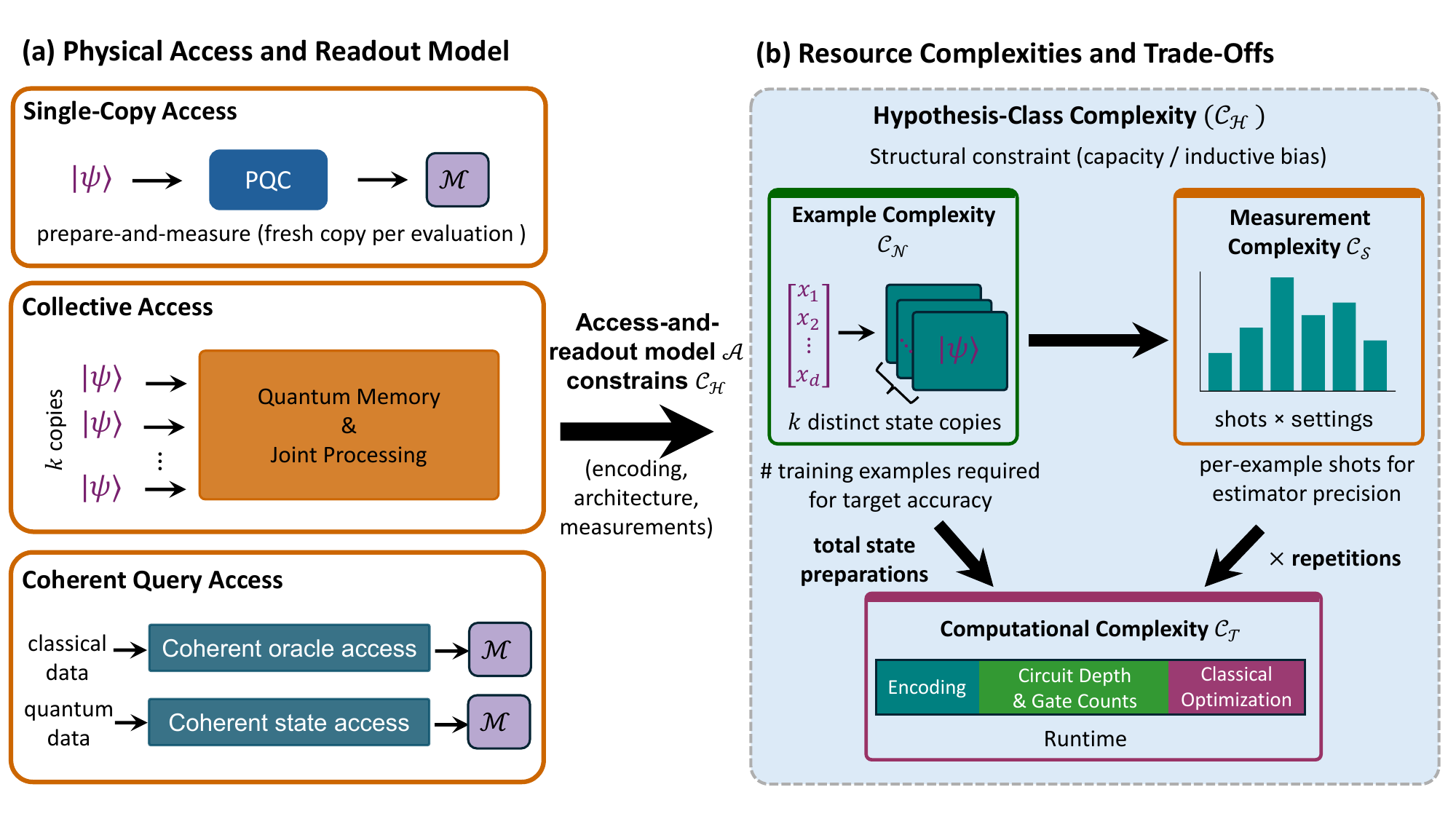}
\caption{Resource complexities under an explicit access model $\mathcal{A}$. (a) three physical access regimes, including single-copy prepare-and-measure; collective access, where $k$~copies of $|\psi\rangle$ per query are processed jointly via quantum memory ($k$ denotes the number of copies jointly processed per example instance); and coherent oracle or QRAM-style state access. (b) $\mathcal{A}$ constrains hypothesis-class complexity $\mathcal{C_H}$, which in turn mediates trade-offs among three consumable resource axes: example complexity $\mathcal{C_N}$, measurement complexity $\mathcal{C_S}$ (total circuit repetitions, i.e., shots $\times$ settings, typically incurred per training example for a fixed estimator precision), and computational complexity $\mathcal{C_T}$. The number of total state preparations per objective evaluation is $\mathcal{C_N} \times \mathcal{C_S}$ and scales as $k~\mathcal{C_N} \times \mathcal{C_S}$ under collective access.}
\label{fig:taxon_quant_advantage}
\end{figure*}

Not all quantum advantages are alike: a reduction in sample complexity is a fundamentally different claim from a runtime speedup. Rigorous advantage statements operationalize Definition~3 by first fixing the physical access-and-readout model $\mathcal{A}$, which is the logical prior of any resource comparison, and then separating consumable resource axes from structural constraints \cite{arunachalam2017guest, chang2025primerquantummachinelearning, huang2021information}.
Figure~\ref{fig:taxon_quant_advantage} makes this explicit: the left panel catalogues three physically distinct access regimes, while the right panel maps the resulting consumable resources against the hypothesis-class constraint $\mathcal{C_H}$ under $\mathcal{A}$.
The access model must be stated before any resource is compared.
We distinguish (i) single-copy access (prepare-and-measure), where each evaluation consumes a fresh state copy; (ii) collective access, which leverages quantum memory and joint processing to perform entangling measurements across $k$ simultaneous copies; and (iii) coherent query access, which encompasses both coherent oracle access for classical data (e.g., via QRAM) and coherent state access for quantum data, methodologies that typically presuppose fault-tolerant capabilities. In addition, the readout model is part of the hypothesis class definition; changing what measurements are permitted can change the effective function class represented by the model.

Given a fixed $\mathcal{A}$, we distinguish three consumable resource complexities and one structural constraint:\\
\noindent (1) \emph{Example complexity} $\mathcal{C_N}$: the number of training examples or, for quantum data, distinct state copies, required to achieve prediction error at most $\epsilon$ with high probability under the stated access model.\\
\noindent (2) \emph{Measurement complexity} $\mathcal{C_S}$: the number of circuit repetitions required by the chosen estimator to evaluate losses, gradients, or observables to the precision needed for learning; operationally, it scales with ``shots $\times$ settings" and is incurred per example.\\
\noindent (3) \emph{Computational complexity} $\mathcal{C_T}$: the end-to-end runtime resources for training and inference, including state preparation, circuit depth and gate counts, classical postprocessing, and optimizer steps.\\
\noindent (4) \emph{Hypothesis-class complexity} $\mathcal{C}_H$: a structural property of the induced hypothesis class determined by the encoding, architecture, and readout, which constrains the attainable trade-offs among $\mathcal{C}_N$, $\mathcal{C}_S$, and $\mathcal{C}_T$.

As indicated by the central arrow in Fig.~\ref{fig:taxon_quant_advantage}, the access model $\mathcal{A}$ constrains $\mathcal{C}_H$ through what encodings, circuit families, and measurement procedures are physically permitted.
Operationally, $\mathcal{C}_H$ captures capacity and inductive-bias constraints, and it can be instantiated via standard capacity proxies such as covering numbers \cite{caro2022generalization}, which bound the achievable trade-offs among $\mathcal{C}_N$, $\mathcal{C}_S$, and $\mathcal{C}_T$ under the stated access model \cite{vapnik1998statistical,abbas2021power}. Throughout, $\mathcal{C}_H$ is not a consumable cost but a constraint on what error levels and trade-offs are attainable under the chosen model family; local geometric quantities used in trainability analyses are discussed below in Sec.~\ref{subsubsec:train_limit_and_symbi}.
These resource complexities realize the abstract resource contract $\mathcal{R}$ introduced in the computational symbiosis framework (Sec.~\ref{subsec:comp_symbiosis}), with $\mathcal{C}_T$ and $\mathcal{C}_S$ as primary consumable resources and $\mathcal{C}_H$ constraining achievable trade-offs under the chosen access-and-readout model $\mathcal{A}$.

These quantities are not independent. Figure~\ref{fig:taxon_quant_advantage}(b) makes explicit two common couplings: $\mathcal{C}_N$ contributes to the total number of state preparations required across training, while $\mathcal{C}_S$ determines the number of repetitions (shots, and possibly multiple settings) needed per evaluated quantity to control estimator precision. Consequently, in variational workflows, the end-to-end runtime $\mathcal{C}_T$ typically contains a term proportional to the number of state preparations multiplied by per-evaluation repetitions, plus classical overhead and the number of optimization steps. Conversely, reductions in $\mathcal{C}_S$ can require stronger measurements or additional quantum processing in the readout, thereby altering the admissible measurement map and the induced hypothesis class $\mathcal{C}_H$. Accordingly, improvements in $\mathcal{C}_T$ do not imply improvements in $\mathcal{C}_N$ or $\mathcal{C}_S$, and apparent gains along one axis can be offset by increases along another.

\subsubsection{Statistical learnability}\label{subsubsec:stati_learn}

Statistical learnability formally quantifies the sample complexity $\mathcal{C}_N$ and measurement complexity $\mathcal{C}_S$ required to achieve generalization under $\mathcal{A}$, assuming the learning algorithm can successfully optimize the stated objective. Generalization is the foundational objective of any learning protocol: it provides the mathematical guarantee that a model achieving low empirical error on finite training data will also achieve a predictably small expected error on unseen data drawn from the same underlying distribution $\mathcal{D}$.
This cleanly isolates a model's information-theoretic capacity from its \emph{trainability}, which represents a fundamentally distinct bottleneck analyzed separately in Secs.~\ref{subsec:optim_as_prereq} and~\ref{subsubsec:train_limit_and_symbi}.

Generalization guarantees can be formalized in the probably approximately correct (PAC) framework through bounds on the sample complexity needed to achieve low error with high probability.
Building on classical foundations \cite{kearns1994on}, \textcite{arunachalam2018sample} show that in broad distribution-independent settings the quantum and classical sample-complexity scalings match up to constant factors (see also \cite{arunachalam2017guest}), with details depending on the SQ-access model.
Tight characterizations in PAC and agnostic models reinforce that coherent access to training examples in the stated learning model does not generically yield asymptotic improvements in the distribution-independent regime where the learner must succeed for all data distributions \cite{arunachalam2018sample}.
A related caution is that simulation hardness is not equivalent to learning hardness: depending on the data-generating and label model, a classical learner may achieve low prediction error via surrogate hypothesis classes or concentration structure even when exact simulation of the underlying quantum model is intractable \cite{huang2021power,schreiber2206classical}. Moreover, in the stated learning model and measurement interface, polynomial-time learning procedures given sample access to quantum outputs and an appropriate measurement interface exist for certain shallow circuit families \cite{huang2024learning}. The central implication is that being quantum does not generically reduce data requirements without additional structure or stronger access assumptions.
Nevertheless, separations can arise in distribution-specific regimes and under computational assumptions relative to restricted classical baselines. \textcite{Sweke2021quantumversus} construct distribution-learning tasks exhibiting a quantum-classical separation under cryptographic assumptions within the specified learning model, and \textcite{lewis2025quantum} provide separations under explicit structural assumptions and restricted baselines. Beyond cryptographic constructions, \textcite{molteni2026exponential} prove exponential quantum-classical separations for learning partially unknown observables from classical (measured-out) data in the PAC framework, and they delineate regimes in which the same observable-learning setting becomes efficiently classically learnable.

Generalization theory for QNNs remains incomplete in highly overparameterized regimes. \textcite{gil2024understanding} show that QNNs can fit random quantum states and random labels, and they provide constructions in which QNNs can fit arbitrary labels to quantum states, behavior that is often poorly predicted under standard uniform convergence analyses used in current bounds. In parallel, recent theory has developed approximation and generalization-error bounds for PQCs and QNNs in structured settings \cite{Manzano2025approximation,ohno2025generalization}, bounds showing improved sample efficiency when the target lies in a low-complexity subspace of the induced feature space \cite{caro2022generalization}, optimizer-dependent generalization bounds \cite{zhu2025optimizer}, and tight analyses clarifying which parameter dependencies can and cannot appear in standard bounds \cite{wang2025tight}.
Moreover, geometric and information-theoretic perspectives relate generalization to the quantum Fisher information metric \cite{haug2024generalization}, margin-like notions \cite{hur2024understanding}, and Rényi-information-based quantities \cite{banchi2021generalization}.

For learning tasks targeting properties of quantum states or processes, the access model becomes decisive.
\textcite{huang2021information,huang2022quantum} show that exponential separations are possible between protocols that permit quantum memory and entangling joint measurements across multiple copies and protocols restricted to single-copy measurements for tasks such as predicting large families of observables and related property-learning problems.
\textcite{oh2024entanglement} extend entanglement-enabled learning separations to bosonic continuous-variable channel-learning settings. These results establish an information-theoretic hierarchy within their stated models, but they rely on multi-copy coherent processing, which is substantially more demanding than single-copy prepare-and-measure access.
Finally, for QDL workflows requiring classical outputs, measurement overhead is a central bottleneck, as detailed in Sec.~\ref{subsubsec:quant_measu}.
Allowing additional quantum processing in the readout can enlarge the effective hypothesis class relative to strictly local readouts \cite{panadero2024regressions}, while single-shot limitations can make measurement overhead information-theoretically unavoidable under the stated embedding and measurement model \cite{recio2025single}. These considerations motivate the time-to-solution analysis for variational workflows under finite shots and noise.

\subsubsection{Computational complexity}

Computational complexity $\mathcal{C}_T$ concerns the runtime axis: the end-to-end resources required to train and deploy a specified model to target accuracy under an explicit access and readout model. Unlike statistical learnability, which is governed by information constraints on examples and measurements, computational separations ask whether quantum resources permit asymptotically faster learning or inference than any classical procedure restricted to the same access and readout interface.
For generative workloads, this runtime axis naturally splits into learning, which estimates parameters under a specified loss function, and inference, which samples from the trained model. Recent work constructs families of generative quantum models that are argued to be efficiently trainable yet for which the inference task is classically hard to simulate under standard sampling-hardness assumptions, and reports beyond-classical demonstrations in this regime \cite{huang2025generative}. Complementary limitations identify trade-offs between classical sampling hardness, often linked to anticoncentration, and average-case trainability for common estimator choices \cite{herbst2025limits}.

Unconditional separations are known in restricted circuit classes. \textcite{bravyi2018quantum} establish separations between constant-depth quantum and constant-depth classical circuits for a relational task, demonstrating that shallow quantum computation can exceed classical shallow computation without unproven complexity assumptions. In a learning-theoretic direction, \textcite{pirnay2024unconditional} prove an unconditional separation in a PAC distribution-learning setting between constant-depth quantum and classical hypothesis classes, while \textcite{pirnay2023superpolynomial} obtain a superpolynomial quantum-classical separation for density modeling under standard cryptographic assumptions in a fault-tolerant regime.

For supervised learning, stronger separations typically rely on cryptographic or complexity assumptions. \textcite{liu2021rigorous} construct classification problems tied to discrete-log-type hardness in which efficient quantum learning is possible while efficient classical learning is ruled out for specified baseline classes under the stated assumptions.

A complementary line isolates computational barriers and capabilities via oracle models that abstract broad families of learning algorithms. In the quantum statistical query (QSQ) model \cite{arunachalam2020quantum}, the learner accesses expectation values of specified observables up to a tolerance, capturing ``learning from statistics" in a quantum access setting.
\textcite{du2021learnability} relate noisy variational QNN training to the QSQ framework, motivating QSQ as a resource-aware lens for near-term learnability claims.
Subsequent work sharpened the role of entanglement and statistics in learning hierarchies \cite{arunachalam2023on} and extended QSQ beyond predictors to learning quantum processes \cite{Wadhwa2025learningquantum}.
However, SQ and QSQ lower bounds yield barriers that are robust to algorithmic details: even learning output distributions can be information-theoretically feasible yet computationally inaccessible to broad SQ-type learners. In particular, \textcite{hinsche2023one} show hardness of distribution learning for circuit output distributions under minimal non-Clifford resources, and \textcite{nietner2025averagecase} prove average-case SQ hardness for learning output distributions of random local circuits across depth regimes.

A distinct route to runtime improvements targets the fully fault-tolerant regime, where QLA and differential-equation primitives can yield speedups for structured training dynamics under stringent input access, conditioning, and precision assumptions \cite{liu2024towards}. Closely related proposals for generative workloads potentially hinge on strong structure together with explicit assumptions about data loading and readout \cite{wang2025efficientquantumalgorithmsdiffusion}.
Between NISQ heuristics and fully fault-tolerant algorithms lies an early fault-tolerant regime in which limited QEC extends feasible circuit depth by trading reduced effective noise against overhead \cite{dangwal2025variational}.

\subsubsection{The classical boundary}\label{subsubsec:class_bound}

Any claimed quantum advantage should be evaluated against the strongest classical algorithms available under the same access and readout interface.
Tensor-network, dequantization, and quantum-inspired baselines are reviewed in Sec.~\ref{subsec:quant_inspi_class}; here we summarize their role as comparator families.
For QDL on classical data, the boundary is often defined by classical surrogates that approximate the induced hypothesis class, classical simulation methods that approximate the quantum subroutine, and classical learning procedures that recover predictors without simulating the underlying quantum dynamics.
Representative surrogate families include TN reductions, which can efficiently approximate variational circuits when the relevant entanglement complexity remains controlled \cite{shin2024dequantizing}, and random-feature reductions, which can approximate families of quantum kernels or implicit feature maps when the induced feature structure admits efficient classical approximation \cite{sahebi2025dequantization,Sweke2025potential}.
The boundary is also set dynamically by advances in hardware; exascale-class state-vector simulations now reach 50 qubits \cite{de2025universal}, reinforcing the need to evaluate the classical comparator against contemporary capabilities rather than historical baselines.

Entanglement alone is not a reliable proxy for computational hardness, and many state-of-the-art simulators are controlled by nonstabilizerness (the resource enabling universal quantum computation beyond Clifford gates), often called ``magic". In this context, \textcite{leone2022stabilizer} analyze stabilizer R\'enyi entropy (SRE) as a computable measure of distance from stabilizer structure to quantify non-Clifford resources, formalizing regimes in which near-stabilizer simulation can be effective even when entanglement is not small.
Complementary approaches exploit operator structure, and in some settings noise structure, to accelerate simulation in restricted regimes. Pauli-propagation frameworks provide a representative example \cite{rudolph2025pauli,rall2019simulation,GonzalezGarcia2025paulipath}.

Classical surrogates can also arise at the level of hypothesis classes. \textcite{Sweke2025potential} characterize conditions under which random Fourier features yield an efficient classical surrogate for PQC-based regression models, in terms of the task, circuit-induced feature structure, and the approximation parameters. In such regimes, the PQC model admits an implicit feature-map description that can be approximated efficiently by classical random-feature methods. Accordingly, surpassing strong classical kernel baselines on classical data requires circuit-induced features that evade efficient random-feature approximations under the same access and readout interface.
Quantum kernel advantages may persist in structured settings where the feature map encodes task symmetries that are difficult to capture for the chosen classical kernel class under the same interface \cite{Wang2021towards}, but recent work on neural quantum kernels \cite{rodriguez2025neural} and on the expressivity of embedding quantum kernels \cite{gil2024expressivity} emphasizes that any such advantage requires balancing expressivity against inductive bias for generalization under finite data. Shadow-based protocols \cite{jerbi2024shadows} can further blur the boundary by enabling classical postprocessing and, in some settings, classical evaluation of learned predictors for many observables from a fixed set of quantum measurements after the quantum data-acquisition stage.
As an illustrative boundary case study, topological data analysis admits explicit classical benchmarks and dequantizations that materially narrow the parameter window in which a superpolynomial advantage could plausibly persist under stated assumptions \cite{Gyurik2022towardsquantum,berry2024analyzing}.
Collectively, these results underscore that advantage claims depend on explicit access models and have to survive both simulation baselines and learning-based classical surrogates.

\subsubsection{Trainability limits}\label{subsubsec:train_limit_and_symbi}

This section focuses on whether a variational or hybrid workflow can achieve a target accuracy within a resource contract $\mathcal{R}$, accounting for finite-shot estimation, device noise, drift, compilation, and classical control overhead.
Accordingly, advantage claims for trainable QDL modules should be stated in end-to-end terms that account explicitly for the shot budget $\mathcal{C}_{S}$, per-circuit latency, classical post-processing, and the number of optimization steps, rather than circuit depth or gate counts alone \cite{liu2023can}.

A central obstruction is unfavorable optimization geometry coupled with measurement-limited updates. Building upon the gradient variance suppression mechanisms defined in Sec.~\ref{subsubsec:resou_codesign}, refined analyses underscore that blanket statements about barren plateaus are misleading without explicit qualifiers regarding cost locality, initialization, and architecture \cite{thanasilp2023subtleties,holmes2022connecting,Zhao2021analyzingbarren}.
Vanishing gradients are not the sole failure mode: trap-dominated landscapes can obstruct optimization even when gradients do not concentrate in a barren-plateau sense \cite{anschuetz2022quantum,nemkov2025barren}.
Recent theory shows that wide-QNN losses and their derivatives admit Wishart-process limits (a random matrix theory framework for analyzing correlated Gaussian observables) under broad conditions, yielding architecture-dependent characterizations of gradient statistics and local minima beyond the original concentration picture \cite{anschuetz2024unified}.
Moreover, as formulated in Sec.~\ref{subsubsec:resou_codesign}, gradient estimation is fundamentally measurement-limited, imposing an explicit statistical trade-off between ansatz expressibility and finite-shot measurement overhead.

Noise further limits variational performance. Under local noise models, limitations constrain the regimes in which variational algorithms can outperform efficient classical approximations as depth increases and noise accumulates \cite{depalma2023limitations}. Any mitigation or robustness strategy is therefore assessed under $\mathcal{R}$, since reduced effective noise can shift costs to additional circuit repetitions and classical post-processing.
A distinct and fundamental constraint is the inability to reuse unknown quantum information: limits on quantum backpropagation clarify why classical state-reuse assumptions do not carry over, reinforcing the need to include repeated state preparation and measurement in the end-to-end training cost~\cite{abbas2023backprop}.

Mitigating these limitations requires navigating the trainability-simulability tension discussed in Sec.~\ref{subsec:comp_symbiosis}, aiming to reduce measurement demand and stabilize optimization while preserving a hypothesis class that is not trivially classically emulable under the stated $(\mathcal{R},\mathcal{A})$.
Structural routes achieve this by imposing inductive bias through symmetry or locality-respecting parameterizations \cite{schatzki2024theoretical} and by using controllability- and overparameterization-based diagnostics to identify regimes with more favorable loss geometry \cite{arocca2023theory}.
Algorithmic approaches instead redistribute computational load: train-classical-deploy-quantum paradigms shift optimization to classically tractable surrogates while reserving quantum hardware for inference under explicit hardness assumptions \cite{recioarmengol2025trainclassicaldeployquantum}. Alternatively, density-learning constructions \cite{coyle2025training} and complexity curricula \cite{recio2024learning} modify the model family itself to improve convergence.

\section{Implementations and Challenges}\label{sec:imple_and_chall}

This section surveys enabling hardware platforms and software frameworks, reviews representative demonstrations, and analyzes dominant near-term constraints and failure modes.

\subsection{The QDL implementation stack}\label{subsec:qdl_imple_stack}

\begin{figure}[tb]
    \centering
    \includegraphics[width=\linewidth]{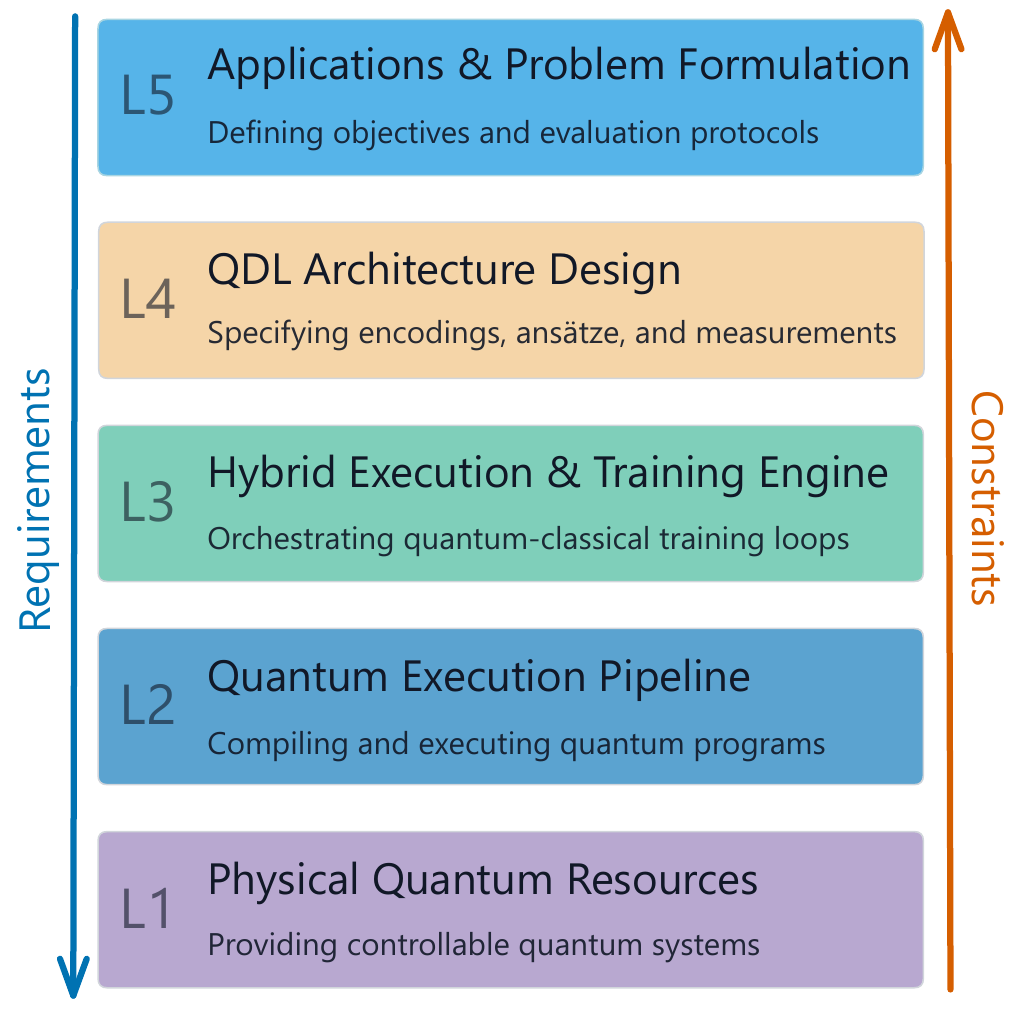}
    \caption{Five-layer implementation stack for QDL, illustrating top-down requirements flow and bottom-up constraint propagation.}
    \label{fig:qdl_stack}
\end{figure}

The practical realization of QDL relies on continuous feedback between algorithm design and hardware constraints.
Figure~\ref{fig:qdl_stack} organizes this implementation hierarchy into five distinct layers. A top-down perspective (L5 $\rightarrow$ L1) illustrates how high-level application demands drive lower-level specifications, while a bottom-up perspective (L1 $\rightarrow$ L5) traces the upward propagation of physical hardware constraints.

L1 (physical quantum resources) encompasses the qubit modality and the associated control electronics. Device properties, including qubit connectivity, coherence time, and native gate sets, bound the feasible circuit depth, while gate durations, measurements, active resets, and control latencies govern sampling rates. Calibration drift imposes temporal limits on iterative training workloads by restricting the window over which calibrations and noise models remain approximately stationary. 
L2 (quantum execution pipeline) translates abstract logical circuits into physical instruction schedules, enabling execution on hardware. This layer handles gate synthesis, routing, and qubit mapping \cite{ji2025algorithm}. Compilation strategies that account for device calibration and noise properties can improve the fidelity of the realized circuit at a fixed logical depth.
Some workflows bypass gate abstractions to optimize analog control pulses directly, treating waveform degrees of freedom as trainable parameters in a pulse-level control loop \cite{tao2025unleashing,tao2025design,Melo2023pulseefficient}.
L3 (hybrid execution and training engine) orchestrates the iterative training loop, managing the latency-critical interplay between quantum circuit execution, gradient estimation, and classical parameter updates. These workloads benefit from low-latency integration of QPUs with classical high-performance computing for error management and statistical analysis, a workflow often termed quantum-centric supercomputing \cite{lanes2025framework}, in which QPUs act as tightly coupled coprocessors and performance can be dominated by orchestration overheads, such as latency, batching, and scheduling, in addition to circuit depth.
Building upon the execution layers, L4 (QDL architecture design) specifies the concrete architecture, such as encoding schemes, ans\"atze, and measurement operators, to be trained.
Finally, L5 (applications and problem formulation) defines the loss landscapes and evaluation protocols that drive requirements through the stack.

Treating layers independently can obscure systems-level bottlenecks in QDL. The stack is characterized by complex interdependencies: top-down, application metrics (L5) dictate training budgets and compilation precision (L3/L2); bottom-up, the physical noise floor and calibration drift (L1) bound the feasible circuit depth and realizable model capacity at L4. For instance, an architecturally sound ansatz at L4 may become untrainable after mapping onto the restrictive topology or noise profile inherited from L1.
A potential longer-term vision is a fully differentiable and data-driven stack where control, compilation, and architecture are optimized jointly. Recent work on pulse control \cite{li2025robust,sarma2025designing}, differentiable formulations of analog quantum computing and control problems \cite{leng2022differentiable,clayton2024differentiable}, and automated architecture search \cite{rapp2024reinforcement} indicates early progress toward such holistic optimization.

\subsection{Quantum hardware platforms}\label{subsec:hardware_platforms}

The DiVincenzo criteria \cite{divincenzo2000physical} enumerate the minimum physical prerequisites for scalable quantum computing: well-characterized qubits, state initialization, long coherence relative to control times, a universal gate set, and qubit-specific measurement. While several leading platforms satisfy these requirements at the component level \cite{google2025quantum, BSIQuantum2025}, such as high-fidelity single- and two-qubit gates and readout, NISQ QDL implementations are limited by reliably executable depth due to system reliability and error rates insufficient for large-scale FTQC. Motivated by performance-centric roadmapping that ties hardware progress to end-to-end algorithmic success rates rather than qubit count alone~\cite{barends2025performance}, we define four critical execution-centric metrics for QDL:\\
\noindent (1) Error-depth budget: gate and measurement infidelities, leakage, and coherence time ($T_1$ and $T_2$) relative to control times jointly bound the feasible effective depth of a QDL ansatz, while the width is strictly bounded by the physical qubit count and topology.\\
\noindent (2) Repetition rate and control latency: The repetition rate denotes the frequency at which quantum circuits are sampled (shots per second), while control latency represents the delay in the classical-quantum communication interface. In the context of QDL, these parameters dictate the total wall-clock training time, as the iterative nature of gradient estimation requires large numbers of circuit evaluations per epoch.\\
\noindent (3) Connectivity and routing overhead: Connectivity describes the topology of direct entangling interactions available between physical qubits. In QDL architecture design, sparse connectivity necessitates the insertion of auxiliary SWAP gates to realize non-local correlations (e.g., in strongly entangling layers); this routing overhead consumes the limited coherence budget and compounds the stochastic error per algorithmic-level operation~\cite{ji2025algorithm,murali2019noise}.\\
\noindent (4) Calibration stability: This represents the temporal invariance of hardware parameters, such as gate fidelities and resonant frequencies, over extended periods. For QDL workloads that span multiple hours or days, high stability is crucial to ensure a consistent loss landscape, preventing the optimization trajectory from being derailed by hardware drift.

Table \ref{tab:hardware-metrics-qdl} summarizes the characteristics most relevant to QDL feasibility and codesign opportunities.
This diversity creates a rich algorithm-hardware codesign space in which QDL architectures, compilation strategies, and training protocols can be tailored to platform-specific execution constraints.

\begin{table*}[t]\small
\caption{Modality-level hardware characteristics most relevant to QDL feasibility and codesign. Connectivity refers to native two-qubit (or native entangling) interaction structure; QDL bottleneck denotes a typical dominant limitation for end-to-end hybrid training at current scales. CV: continuous-variable. For annealers, minor embedding refers to mapping a logical problem graph onto the fixed hardware graph.}
\label{tab:hardware-metrics-qdl}
\centering
\newcommand{\colA}{0.17\textwidth}
\newcommand{\colB}{0.20\textwidth}
\newcommand{\colC}{0.32\textwidth}
\newcommand{\colE}{0.26\textwidth}
\newcommand{\cell}[2]{\parbox[t]{#1}{\raggedright #2}}
\begin{ruledtabular}
\begin{tabular}{llll}
\cell{\colA}{\textbf{Modality}} &
\cell{\colB}{\textbf{Primitive}} &
\cell{\colC}{\textbf{Connectivity}} &
\cell{\colE}{\textbf{QDL bottleneck}} \\
\hline
\cell{\colA}{Superconducting} & \cell{\colB}{digital gates} & \cell{\colC}{sparse 2D (routing)} & \cell{\colE}{routing overhead, drift/crosstalk} \\
\cell{\colA}{Trapped ions} & \cell{\colB}{digital gates} & \cell{\colC}{all-to-all (in-modular); modular links} & \cell{\colE}{wall-clock throughput} \\
\cell{\colA}{Neutral atoms} & \cell{\colB}{digital gates + analog} & \cell{\colC}{reconfigurable between shots} & \cell{\colE}{entangling error, atom loss} \\
\cell{\colA}{Photonics} & \cell{\colB}{optical sampling; CV} & \cell{\colC}{interferometric meshes; delay networks} & \cell{\colE}{optical loss, calibration} \\
\cell{\colA}{Annealers} & \cell{\colB}{analog sampling} & \cell{\colC}{fixed graph (minor embedding)} & \cell{\colE}{embedding, non-equilibrium bias} \\
\cell{\colA}{Semiconductor spin} & \cell{\colB}{digital gates} & \cell{\colC}{local coupling} & \cell{\colE}{variability, coupling range} \\
\end{tabular}
\end{ruledtabular}
\end{table*}

\subsubsection{Superconducting circuits}\label{subsubsec:superc_circuits}

Superconducting processors use lithographically fabricated superconducting circuits with Josephson junctions \cite{Josephson1962,Josephson1974} as nonlinear inductors, producing weakly anharmonic oscillators (e.g., transmons \cite{koch2007charge}) addressable by microwave control \cite{clarke2008superconducting,krantz2019quantum,kjaergaard2020superconducting,gu2017microwave}. Circuit quantum electrodynamics \cite{blais2021circuit} enables microwave control and dispersive readout, with effective measurement fidelity depending on the chosen measurement basis \cite{ji2025exploring,pommerening2020what}.
The observation of macroscopic quantum tunneling \cite{devoret1985measurements} and energy quantization \cite{martinis1985energy}, recognized by the 2025 Nobel Prize in Physics \cite{nobel2025physics}, established experimentally that engineered superconducting circuits can exhibit coherent quantum dynamics at macroscopic scales, an enabling prerequisite for modern qubit engineering and control \cite{nakamura1999coherent}.
Recent roadmaps emphasize performance-centric scaling, prioritizing end-to-end algorithmic success rates and operational stability over qubit count alone \cite{barends2025performance}.

This platform features fast gate operations \cite{abughanem_ibm_2025} supporting high-throughput circuit evaluation, but sparse planar connectivity, which requires routing operations and increases circuit depth \cite{ji2022calibration}.
Hardware-supported gate sets combine single-qubit rotations with an entangling interaction such as CNOT or CZ \cite{abughanem_ibm_2025}. Pulse-level access supports hardware-tailored optimization and error mitigation \cite{ji2023optimizing,ji2024synergistic} and direct pulse-parameter optimization framed through quantum optimal control \cite{wittler2021integrated,koch2022quantum,Magann2021ContPQCQOC}.
Key limitations encompass two-qubit gate error, measurement error, crosstalk, and calibration drift, which can degrade long-running training stability. Data-efficient noise characterization can provide the context-specific error information targeting the interplay between algorithm and hardware to inform compiler decisions and routing heuristics \cite{ji2025dataefficient}.

The superconducting platform has been a primary experimental testbed for end-to-end variational QDL, including gradient-based training \cite{pan2023experimental}, quantum adversarial learning \cite{ren2022experimental}, transfer learning \cite{mari2020transfer}, and generative-model benchmarks \cite{zoufal2019quantum,hamilton2019generative,riste2017demonstration}.
More recent advancements include the experimental demonstration of quantum continual learning \cite{zhang2026experimental}, addressing catastrophic forgetting in sequential tasks and parameter-efficient quantum anomaly detection for general image datasets \cite{wang2025parameter}.

\subsubsection{Trapped ions}

Trapped-ion processors confine ions in electromagnetic traps, using long-lived internal hyperfine or optical transitions as qubits \cite{bruzewicz2019trapped,haffner2008quantum}. 
Entangling operations are generated by laser-driven coupling to shared motional modes, so that collective phonons act as an interaction bus and M{\o}lmer-S{\o}rensen-type gates provide a standard native entangling primitive \cite{sorensen1999quantum,molmer1999multiparticle}.
Following the proposal by \textcite{cirac1995quantum}, early experiments demonstrated a fundamental two-qubit quantum logic gate in this platform \cite{monroe1995demonstration}.
Demonstrations report single-ion coherence times exceeding one hour \cite{wang2021single} and two-qubit gate fidelities at or above the $99.9\%$ under optimized conditions \cite{strohm2024ion}. Current roadmaps emphasize fast mid-circuit measurement, measuring qubits during circuit execution rather than only at the end, and real-time feed-forward using measurement outcomes to control subsequent gates as central primitives for many QEC and lattice-surgery style protocols \cite{paetznick2024demonstration,benito2025scaling}, while alternative measurement-free logical toolboxes are also being explored to mitigate slow measurement \cite{butt2026demonstration}.

All-to-all connectivity within a module eliminates SWAPs, enabling dense, highly expressive ans\"atze, but microsecond two-qubit gates yield lower throughput \cite{haffner2008quantum,bruzewicz2019trapped}, making the wall-clock time for large-batch training significantly higher.
Dominant performance bottlenecks include motional heating, laser phase noise, and scaling-induced spectral crowding of the motional modes \cite{bruzewicz2019trapped,strohm2024ion}.
This platform has served as a critical testbed for high-fidelity, effectively all-to-all connectivity within modules and for QEC architecture studies. The platform's dense connectivity enables the implementation of complex entangling layers for generative modeling and classification tasks that exploit the native M{\o}lmer-S{\o}rensen structure \cite{Chen2024benchmarkingtrapped}. The ability to perform mid-circuit measurements and feed-forward can enable sample-efficient learning strategies and logical-layer training \cite{jones2025architecting}, potentially mitigating throughput constraints by enabling superior per-shot information density and error suppression.
Benchmarks compare QNNs across trapped-ion and superconducting devices, showing strong noise sensitivity \cite{lakhdar2025benchmarking}.

\subsubsection{Quantum annealers}

Quantum annealers, introduced as an open-system computation model in Sec.~\ref{subsubsec:models_of_comput}, implement programmable Ising-type Hamiltonians in superconducting flux-qubit hardware, enabling high-throughput sampling of near-ground-state configurations \cite{kadowaki1998quantum, albash2018adiabatic,hauke2020perspectives}. 
Contemporary commercial implementations, such as D-Wave's Advantage2 featuring Zephyr topology (a specific qubit connectivity graph) with over 4,400 flux qubits, utilize thousands of superconducting flux qubits to physically realize these dynamics, providing a natural testbed for large-scale benchmarking of Ising optimization \cite{quinton2025quantum,willsch2022benchmarkingannealers,heidari2024quantum}, simulation \cite{vodeb2025falsevacuumdecay,kingbeyond2025}, and sampling tasks \cite{willsch2020qsvm,cavallaro2020qsvm,schulz2025lda,salloum2024quantum,wilson2021quantum,benedetti2018quantum,perdomo2018opportunities}.

This platform offers high-throughput repeated sampling but operates on a fixed sparse hardware graph, so nonnative couplings require minor embedding \cite{yarkoni2022quantum,pelofske2025comparing,willsch2022benchmarkingannealers}.
Open-system dynamics with thermal relaxation and schedule-dependent freeze-out (premature cessation of quantum dynamics before reaching ground state) yield an output distribution that can be modeled by an effective-temperature Gibbs form but exhibits systematic, problem- and schedule-dependent deviations \cite{amin2015searching,marshall2019power,nelson2022high}.
Utilization in learning workloads follows two distinct paradigms: treating annealers as approximate physical Gibbs samplers for training energy-based generative models \cite{higham2023quantum,adachi2015application,benedetti2017quantum} or as global optimizers for discriminative networks, where discrete network parameters are encoded into the Hamiltonian to minimize the training objective \cite{abel2022completely,zhang2025quantumsequel}.
Benchmarking studies emphasize sensitivity to embedding overhead and device-specific systematics \cite{pelofske2025comparing}. Credible performance claims therefore require controlling calibration drift and bias \cite{chancellor2022error}, and validating that the realized sampling distribution is consistent with the intended target model for sampling-based uses, under the stated access model and matched classical baselines \cite{nelson2022high}.

\subsubsection{Photonics}

Photonic quantum processors \cite{romero2024photonic,lee2025photonic,wang2020integrated,kok2007linear} encode information in optical modes using discrete variables (DV) where information is encoded in countable photon numbers or continuous variables (CV) paradigms where information is encoded in field amplitude and phase quadratures. In DV approaches, qubits are realized via dual-rail \cite{psiquantum2025manufacturable} or time-bin encodings \cite{vagniluca2020efficient}, while CV paradigms encode information in the canonical field quadratures such as amplitude and phase of optical bosonic modes. Linear-optical interferometers implement unitary mode transformations, but scalable universality requires an additional effective nonlinearity. DV schemes achieve entangling operations via measurement-induced nonlinearities with ancillas and feed-forward, whereas CV architectures require at least one non-Gaussian resource beyond Gaussian optics. Recent roadmaps emphasize integrated photonics and system-level codesign of sources, routing, and detection, with measurement-based and fusion-based architectures \cite{bartolucci2023fusion} providing prominent scaling roadmaps \cite{aghaee2025scaling,perez2025large}.

The characteristics are high-rate sampling throughput and loss-limited effective depth.
Time-multiplexed photonic samplers demonstrate fast end-to-end sampling in large interferometric networks, making photonics attractive for sampling-centric learning primitives \cite{madsen2022quantum}. Connectivity is implemented through programmable interferometers. Spatial meshes provide flexible mode mixing over a fixed number of modes \cite{clements2016optimal}, while time-multiplexed designs synthesize large structured interaction graphs via delay networks and component reuse \cite{madsen2022quantum}; frequency-multiplexed architectures analogously realize large mode graphs through reconfigurable operations in the spectral domain \cite{lu2023frequency}. The dominant constraint is photon loss, including coupling, propagation, and detection inefficiencies. In postselected DV settings, the probability that all carriers survive decays rapidly with circuit depth, directly limiting the feasible layer count for variational constructions \cite{arrazola2021quantum}.

Photonics is particularly well-matched to generative \cite{sedrakyan2024photonic,salavrakos2025error} and reservoir computing paradigms \cite{garcia2023scalable,zia2025quantum,aadhi2025scalable,sakurai2025quantum,rambach2025photonic} that exploit fixed optical dynamics as high-dimensional feature generators. In these approaches, the optical quantum module is non-trainable or only weakly parameterized, and optimization is concentrated in a classical readout or lightweight outer loop, as exemplified by experimental photonic extreme learning machines \cite{suprano2024experimental,pierangeli2021photonic} and boson sampling-enhanced predictors \cite{rambach2025photonic}. Beyond sampling demonstrations, integrated photonic processors have enabled experimental quantum feature maps for kernel methods \cite{yin2025experimental}. Recent progress also includes hybrid quantum-classical photonic neural networks with cavity-assisted interactions for universal logical processing \cite{basani2025universal}. Additional advancements encompass photonic QCNNs with adaptive state injection \cite{monbroussou2025photonic}. Accordingly, credible performance claims should be formulated and evaluated end-to-end, accounting for loss, calibration drifts, and classical baselines under matched access models, an approach operationalized by recent community benchmarking efforts \cite{notton2025establishing}.

\subsubsection{Neutral atoms}

Neutral atom processors \cite{saffman2025quantum,bluvstein2025fault} use optical tweezers to trap hundreds to thousands of individual atoms in dynamically reconfigurable 2D or 3D geometries, encoding qubits in long-lived hyperfine ground states \cite{saffman2010quantum,scholl2021quantum,pichard2024rearrangement,manetsch2025tweezer}. Entangling interactions are commonly mediated by Rydberg excitation, where strong van der Waals interactions yield a Rydberg blockade mechanism that enables fast multi-qubit correlations and two-qubit gates \cite{evered2023high,jaksch2000fast}.
The platform has demonstrated rapid scaling, including commercial implementations such as QuEra \cite{wurtz_aquila_2023}, Pasqal \cite{scholl2021quantum}, and Atom Computing \cite{wintersperger2023neutral}.
Notably, the continuous operation of coherent arrays at the $\sim 3000$-qubit scale \cite{chiu2025continuous} and the trapping of over $6000$ highly coherent tweezer-trapped qubits \cite{manetsch2025tweezer} indicate rapid progress in system engineering alongside improvements in logical-level operation \cite{bluvstein2024logical,bluvstein2025fault}.

The characteristics are programmable geometry with distance-mediated interactions and cycle-time-limited end-to-end throughput. Rydberg-mediated entangling operations occur on microsecond timescales, but iteration rate is often dominated by state preparation, readout, and array assembly and rearrangement overheads \cite{wintersperger2023neutral,pichard2024rearrangement}.
Reconfigurability is a key advantage. Atom positions can be adapted between circuit evaluations to better match a target interaction graph, reducing SWAP-driven routing overhead for structured architectures \cite{pichard2024rearrangement}. Within a single execution, however, the effective interaction graph remains constrained by blockade radius, parallelization limits, and control crosstalk, so connectivity is highly programmable but not arbitrary at fixed time \cite{saffman2025quantum}.
Uniquely, the platform supports both digital gate-model operation and analog Hamiltonian simulation, enabling hybrid digital-analog workflows that trade compiled depth for hardware-native correlation generation \cite{lu2024digital}.
Dominant limitations for long training runs include finite Rydberg lifetime, laser amplitude and phase noise, atom loss (loss of trapped atoms from tweezers during execution) and recapture policies, and measurement-induced errors and drift across repeated cycles \cite{scholl2021quantum}.

Neutral atoms are particularly well matched to QDL paradigms that exploit structured interaction graphs and analog dynamics as learning primitives. Graph-based learning via quantum evolution kernels uses programmable Rydberg-array dynamics to define kernels on graphs \cite{henry2021quantum}, with recent extensions incorporating attributed graphs by encoding node and edge information through local detunings in the Rydberg Hamiltonian \cite{djellabi2025attributed}.
Reservoir computing finds a natural platform match here: the Rydberg array's complex analog dynamics act as a fixed, high-dimensional feature generator, with only the classical readout trained. Platform-specific realizations include foundational Rydberg-reservoir theory  \cite{bravo2022quantum} and increasingly large-scale neutral-atom implementations \cite{kornjavca2024large}.

\subsubsection{Semiconductor spin qubits}

Semiconductor spin-qubit processors encode quantum information in spin degrees of freedom, spanning single-electron (or hole) spins in gate-defined quantum dots and dopant-based donors \cite{loss1998quantum,kane1998silicon,burkard2023semiconductor} as well as optically addressable defect-center spins such as nitrogen-vacancy (NV) centers in diamond \cite{atature2018material,awschalom2021development,Fischer2025NV}. Semiconductor spin qubits leverage CMOS-aligned fabrication for high integration densities and cryogenic cointegration \cite{li2018crossbar,bartee2025spin}. Milestones include industrial multi-qubit modules \cite{george202412,steinacker2025industry}, universal multi-qubit control \cite{philips2022universal,edlbauer202511}, and early QEC \cite{takeda2022quantum}. Elevated-temperature operation has been demonstrated above 1K for electron spins \cite{huang2024high,petit2022design} and above 4K for hole spins \cite{camenzind2022hole}.
In parallel, defect spins emphasize optical networking for modular scaling \cite{awschalom2021development}.

For QDL, semiconductor spin qubits offer fast physical gate times with predominantly local connectivity and calibration-intensive control \cite{burkard2023semiconductor,stano2022review,bartee2025spin}.
Single-qubit gates use electron or electric-dipole spin resonance, while two-qubit gates employ nearest-neighbor exchange, achieving high fidelities \cite{burkard2023semiconductor,noiri2022fast,stano2022review}.
Nonlocal operations rely on shuttling \cite{de2025high,nagai2025digital} or resonator and photon links \cite{burkard2023semiconductor,awschalom2021development}.
Hole spin qubits provide strong spin-orbit coupling, enabling fast all-electrical control. Recent work emphasizes operating points that mitigate charge-noise sensitivity while retaining strong driveability \cite{hendrickx2024sweet,secchi2025hole,fanucchi2025giant,john2025robust}. Defect spins offer remote entanglement and networking but face throughput limits from photon collection and rates \cite{atature2018material,awschalom2021development}.
QDL demonstrations on spin-based platforms are presently concentrated in small-scale learning primitives and algorithmic subroutines \cite{wang2022experimental,wen2019experimental,pan2014experimental}, rather than end-to-end trained deep variational models on scalable semiconductor arrays. Representative examples include quantum kernel estimation in nuclear magnetic resonance \cite{sabarad2024experimental} and subroutines such as quantum principal component analysis \cite{xin2021experimental,li2021resonant}. Reservoir learning has also been explored, but much of the current evidence remains simulation- or theory-driven, including studies where dissipation shapes performance \cite{mifune2025effects}.

\subsection{Software tools and frameworks}\label{subsec:software_frameworks}

This section reviews the software ecosystems supporting the construction, training, and deployment of QDL models, focusing on actively maintained open-source frameworks that provide practical interfaces for hybrid workflows. The landscape can be broadly categorized into general-purpose QML frameworks, hardware-optimized platforms, and photonic or continuous-variable specialized toolchains. Ten widely used QDL software frameworks are summarized in Table~\ref{tab:qml-software-comparison}.

General-purpose QML frameworks prioritize interoperability with classical ML libraries and support hybrid training loops. PennyLane~\cite{bergholm2018pennylane} offers a unified differentiable interface for quantum circuits, enabling seamless integration with PyTorch~\cite{Paszke2019}, TensorFlow/Keras~\cite{Abadi2016}, and JAX~\cite{bradbury2018jax}. It combines classical autodifferentiation with quantum-specific gradient rules and provides analysis tools such as the quantum metric tensor for studying optimization landscapes. TensorFlow Quantum~\cite{broughton2020tensorflow} embeds Google’s Cirq~\cite{CirqDevelopers_2025} circuits as TensorFlow ops, allowing quantum subroutines to be trained within Keras pipelines using classical backpropagation and quantum gradient estimators. Qiskit Machine Learning~\cite{qiskit_ml} extends the Qiskit \cite{qiskit} ecosystem with variational model builders and scikit-learn style interfaces~\cite{scikitlearn_api}, while its TorchConnector enables gradient-based training via PyTorch. Similarly, sQUlearn~\cite{kreplin2025squlearn} provides scikit-learn compatible quantum classifiers and regressors, interfacing with Qiskit, PennyLane, and the Qulacs simulator~\cite{qulacs} for accelerated circuit evaluation.

Hardware-optimized and performance-oriented platforms emphasize efficient simulation and execution on heterogeneous hardware. NVIDIA’s CUDA-Q~\cite{slysz2025hybrid} adopts a single-source programming model designed for quantum-classical co-execution in high-performance computing (HPC) environments, with native GPU acceleration. TensorCircuit-NG~\cite{Zhang2023tensorcircuit} leverages just-in-time compilation and TN contraction, supporting automatic differentiation through JAX, TensorFlow, and PyTorch backends. Within the QPanda ecosystem~\cite{dou2022qpanda,zou2025qpanda3}, VQNet~\cite{chen2019vqnet} embeds trainable quantum operators into classical learning pipelines, and VQNet-2.0~\cite{bian2023vqnet} extends this with hybrid automatic differentiation and cross-platform deployment capabilities.

Photonic and continuous-variable toolchains cater to non-qubit encodings. Strawberry Fields~\cite{Killoran2019strawberryfields} supports the design and simulation of Gaussian and non-Gaussian continuous-variable photonic circuits. Piquasso~\cite{Kolarovszki2025piquassophotonic} is a full-stack photonic programming and simulation platform supporting both discrete- and continuous-variable circuits, with optional high-performance backends and differentiable simulation, e.g., TensorFlow and JAX integration, to enable photonic QML prototyping and benchmarking. For discrete-variable photonics, Perceval~\cite{Heurtel2023percevalsoftware} provides a dedicated environment for linear-optical circuit construction and simulation, with hardware-oriented interfaces. More recently, DeepQuantum~\cite{he2025deepquantum} has emerged as a PyTorch-based platform that aims to unify hybrid training workflows across both qubit-based and photonic backends.

Across these frameworks, key practical considerations include simulation capabilities (GPU acceleration, HPC support, TN methods), hybrid workflow integration (support for Torch, JAX, or Keras), and quantum hardware support (circuit-based or photonic backends). Table~\ref{tab:qml-software-comparison} provides a detailed comparison along these dimensions, highlighting the trade-offs and specialization of each ecosystem.

There is also a supporting software layer for QDL, including compilation, optimization, and noise mitigation, which is increasingly central to hardware-facing QDL studies. Representative examples include compilers such as TKet \cite{sivarajah2020t}, circuit optimization toolchains such as ZX-calculus \cite{kissinger2019pyzx}, and backend-agnostic QEM libraries such as Mitiq \cite{LaRose2022mitiqsoftware}.
Benchmarking suites include Benchpress, which compares software stacks across large circuit families and device-motivated constraints \cite{nation2025benchmarking}.
The Munich Quantum Toolkit \cite{burgholzer2025munich,burgholzer2025mqt} provides a modular suite of design-automation tools, spanning intermediate representations, compilation, verification, and benchmarking.

\newcommand{\yes}{{\checkmark}}
\newcommand{\no}{{$\times$}}
\newcommand{\partiall}{{$\circ$}}

\begin{table*}[htbp]
\caption{Comparison of ten mainstream QDL software frameworks. The evaluation spans classical simulation capabilities, integration with classical libraries for hybrid workflows, and support for diverse quantum hardware paradigms. GPU-enabled: officially supported GPU-accelerated simulation backend; HPC-enabled: documented multi-node and distributed execution; TN: documented tensor network (TN) approximate simulators integrated in the stack.}
\label{tab:qml-software-comparison}
\renewcommand{\arraystretch}{1.15}
\setlength{\tabcolsep}{2pt}
\small
\begin{tabular}{l ccc ccc cc}
\toprule
\multirow{2}{*}{\textbf{Software}} & 
\multicolumn{3}{c }{\textbf{Classical Simulation}} & 
\multicolumn{3}{c }{\textbf{Hybrid Workflow}} & 
\multicolumn{2}{c}{\textbf{QPU Integration}} \\
\cmidrule(lr){2-4} \cmidrule(lr){5-7} \cmidrule(lr){8-9}
& GPU-enabled & HPC-enabled & TN
& \makecell{PyTorch} & \makecell{JAX} & \makecell{Keras} 
& \makecell{Circuit-based} & \makecell{Photonics} \\
\midrule
PennyLane & \yes & \yes & \yes & \yes & \yes & \yes & \yes & \yes \\
TensorFlow Quantum & \yes & \yes & \no & \no & \no & \yes & \yes & \no \\
Qiskit Machine Learning & \yes & \yes & \yes & \yes & \no & \no & \yes & \no \\
CUDA-Q & \yes & \yes & \yes & \yes & \no & \no & \yes & \no \\
sQULearn & \yes & \partiall & \no & \yes & \no & \no & \yes & \no \\
TensorCircuit-NG & \yes & \yes & \yes & \yes & \yes & \yes & \yes & \no \\
VQNet 2.0 & \yes & \yes & \no & \partiall & \no & \no & \yes & \no \\
Strawberry Fields & \yes & \partiall & \yes & \yes & \no & \yes & \no & \yes \\
Perceval & \no & \no & \no & \no & \no & \no & \no & \yes \\
DeepQuantum & \yes & \partiall & \yes & \yes & \no & \no & \yes & \yes \\
\bottomrule
\end{tabular}
\vspace{8pt}
\begin{minipage}{0.9\textwidth}
\footnotesize
\yes~Supported \quad 
\partiall~Limited \quad 
\no~Unsupported
\end{minipage}
\end{table*}

\subsection{Experimental demonstrations}\label{subsec:exp_demos}

This section surveys QDL demonstrations on physical hardware, tracing a progression from early primitive validations to hierarchical NISQ training, scaling efforts, and the current frontier of practical advantage.

\subsubsection{Early proofs-of-concept}

Early experiments largely served to validate learning-relevant primitives and separations under controlled assumptions. 
Before variational hybrid training became standard, few-qubit photonic and nuclear magnetic resonance platforms demonstrated entanglement-assisted classification and support vector machine primitives \cite{cai2015entanglement,li2015experimental}.
In parallel, early experiments began probing learning-relevant speedups \cite{lee2019experimental,saggio2021experimental}, typically as restricted-model query or communication improvements or constant-factor effects rather than end-to-end DL speedups.

A representative example is the oracle-based learning-parity-with-noise demonstration on a five-qubit superconducting processor by \textcite{riste2017demonstration}, which illustrates a query-complexity separation on hardware within the corresponding oracle model. Although far from data-driven DL, such results helped clarify that the \emph{meaning} of advantage depends on the access model and the precisely fixed classical baseline required for falsifiability.
In parallel, small-scale experiments explored perceptron-like and classifier primitives on gate-based devices. For example, \textcite{tacchino2019artificial} experimentally tested a quantum-perceptron concept on a small processor; these demonstrations primarily test end-to-end training interfaces and readout-induced nonlinearities, rather than providing performance evidence on realistic datasets.
Concurrently, early hardware demonstrations of variational generative modeling \cite{hu2019quantum,hamilton2019generative} and kernel-style classification \cite{havlivcek2019supervised,peters2021machine} established the practical hybrid training stack and highlighted sensitivity to shot noise and device imperfections.

\subsubsection{Training hierarchical models on NISQ processors}

Once hybrid training became the default interface, the central experimental challenge shifted to building \emph{multilayer} models without losing trainability under finite-shot gradient estimation and device noise. \textcite{skolik2021layerwise} proposed and tested in simulation a layerwise training strategy, in which only subsets of parameters are updated while earlier layers are held fixed; such schemes can improve optimization behavior in practice.

A notable experimental milestone is the demonstration that layered QNNs can be trained with gradient-based updates on superconducting hardware using hardware-in-the-loop differentiation, with the forward process executed on hardware and the backward process implemented on a classical computer \cite{pan2023deep}. However, this approach does not constitute a scalable, fully quantum backpropagation primitive. Extending it to wider or deeper models would require either large-scale classical simulation of the backward pass or measurement-intensive reconstruction of intermediate quantum information, both of which can scale exponentially with subsystem size in the worst case \cite{abbas2023backprop}. These works sharpen the practical questions for experimental QDL: how depth, shot budgets, noise, and drift trade off against achievable performance and reproducibility.

\subsubsection{Pushing the frontiers: scale, depth, and robustness}

Recent experiments have pushed QDL along three axes: scale (qubit count or modes), executable depth, and robustness. On the scale axis, reservoir-style approaches are particularly attractive because the quantum substrate is not trained by circuit-parameter gradients; instead, fixed quantum dynamics provide a high-dimensional feature map, and only a classical readout is optimized. This avoids quantum-gradient evaluation characteristic of variational models under finite-shot noise, though repeated evaluations are still required to estimate reservoir features and fit the readout. A prominent example is the large-scale experimental demonstration of quantum reservoir computing on a neutral-atom analog quantum computer, scaling to systems of up to 108 qubits \cite{kornjavca2024large}. Note that this is an analog-reservoir regime rather than deep variational circuit training. While such reservoir pipelines are not end-to-end trained variational deep networks, they provide a concrete and practically relevant route by which large qubit counts can contribute functionally to temporal learning tasks.

Executable depth records are best interpreted primarily as hardware capability rather than as QDL performance evidence; for example, deep coherent circuit demonstrations such as quantum signal processing sequences inform depth headroom and control limits \cite{bu2025exploring}. Within learning-centric experiments, depth remains constrained by a combined budget of decoherence, coherent control errors, and sampling noise, which motivates architectures and training protocols that are explicitly depth-aware.

Robustness has also emerged as a critical figure of merit. Empirical studies have reported training instability and sensitivity to hardware noise and finite-shot sampling overhead in variational deep RL workflows on superconducting platforms \cite{franz2023uncovering}. Separately, early experimental work has demonstrated that quantum models can exhibit adversarial vulnerability in close analogy to classical networks \cite{ren2022experimental}, motivating systematic robustness benchmarking \cite{zhang2025experimental}. Mitigation strategies include depth reduction via approximate data encoding \cite{west2024drastic} and finite-shot noise reduction via variance-regularized training objectives \cite{kreplin2024reduction}.

\subsubsection{Benchmarking and reproducibility}\label{subsubsec:bench_repro}

As experimental demonstrations mature, benchmarking has become a dominant methodological bottleneck. The central difficulty is that nascent quantum models are often compared against classical baselines that are not equivalently tuned, or under mismatched computational budgets. Without a disclosed hyperparameter-search protocol and training-time budget, performance comparisons are rarely informative \cite{bowles2024better}. Benchmark-quality studies should therefore report the resource contract alongside the full hyperparameter and wall-clock budget used to tune both quantum and classical models.
At a minimum, a credible pipeline-level advantage claim requires: (i) a functional-contribution test, e.g., ablation showing the quantum subroutine changes the outcome beyond what is explained by classical preprocessing and noise; (ii) budget-matched comparison to tuned classical baselines, including transparent hyperparameter and training-time budgets; (iii) repeated runs across calibrations to quantify variability; and (iv) explicit costing of verification, especially as experiments approach beyond-simulability regimes where full classical replication is unavailable and only restricted, property-based certificates may be feasible under stated assumptions.

While most current QDL experiments remain in or near classically reproducible regimes, hidden code sampling provides a concrete example of such property-based verification logic by checking statistical properties of output distributions without full simulation \cite{deshpande2025peaked}.
Recent benchmarking efforts further bridge this gap, including time-series prediction studies that compare against strong classical sequence baselines \cite{fellner2025quantum}, reinforcement-learning comparisons conducted in controlled simulation settings \cite{zare2025performance}, and large-scale hardware image-classification experiments that aim to provide more systematic, reproducible benchmarks \cite{gharibyan2025quantum}. However, reproducibility remains hindered by device drift and calibration variability, underscoring the need for explicit characterization of time-dependent error modes and drift-aware protocols to substantiate \cite{buonaiuto2024effects,proctor2020detecting}. Platform diversity expands accessible paradigms and benchmarking targets, yet it also amplifies noncomparable claims absent standardized resource models and reporting conventions \cite{lall2025review}.

\subsubsection{Toward practical quantum advantage}

The experimental arc reveals a consistent pattern: successive demonstrations have widened the operational envelope of QDL, yet to our knowledge, no published experiment simultaneously satisfies all six conditions of Definition 3 under an explicit, budget-matched classical comparator at a scale beyond the best available classical simulation. The limiting factor is not a lack of isolated positive results, but the scarcity of claims that are simultaneously falsifiable, budget-matched, and fully disclosed, and reproducible across calibrations within a single, explicitly stated resource contract. Closing this gap requires treating verification as a first-class experimental deliverable: as experiments move beyond regimes where full classical replication is feasible, the evidentiary standard must shift from end-to-end simulation to protocol-level certification under clearly specified access-and-readout model assumptions.

Within this constrained landscape, near-term pathways to practical advantage diverge strictly based on the data modality. For \emph{quantum-native data} and \emph{structured generative} tasks, state-preparation overheads are either absent or offset by sampling hardness, allowing theoretical separations to be grounded directly in information-theoretic constraints. Conversely, for \emph{classical-data QDL}, severe encoding bottlenecks dictate that the most plausible routes to advantage are systems-level and targeted rather than monolithic. Representative loci include: (i) amortized inference-time benefits, where a fixed quantum subroutine is executed at high repetition rate and the relevant metric is throughput at fixed accuracy; (ii) weakly trained or gradient-free quantum feature dynamics, such as fixed-circuit or reservoir-style primitives, that trade expressivity for reduced gradient-estimation overhead, while still requiring explicit sampling budgets; and (iii) hardware codesign, where device-native primitives, including midcircuit measurement and reset, analog blocks, and parallelism, are integrated with compilation, mitigation, and classical postprocessing to reduce constant-factor overheads in an end-to-end contract.
These divergent application domains are surveyed in Secs.~\ref{subsec:appli_domai} and \ref{subsec:comparative_analysis}, while the corresponding trajectory toward scalable, fault-tolerant implementations is developed in Sec.~\ref{subsec:future_directions}.

\subsection{Challenges and opportunities}\label{subsec:chall_and_oppor}

Scalable advantage claims for deep quantum learners are constrained by both hardware resource budgets and the scaling of optimization and statistical estimation costs.
Hardware constraints bound compiled depth and executable two-qubit layers, per-iteration measurement budgets, and throughput within calibration-drift windows. Moreover, many QDL architectures face scaling barriers, most prominently barren plateaus and input-output bottlenecks, that can render training or inference resource-prohibitive.

\subsubsection{Hardware constraints}\label{subsubsec:hardware_constraints}

While general-purpose quantum metrics are well-documented~\cite{lall2025review}, QDL workloads are constrained by four coupled budgets: effective compiled depth $D_{\mathrm{eff}}$, shot time $t_{\mathrm{shot}}$, iteration latency $T_{\mathrm{iter}}$, and stability window $\tau_{\mathrm{stable}}$.
The depth budget $D_{\mathrm{eff}}$ is defined by the compiled circuit on the target topology, comprising both the trainable ansatz and the data-encoding subcircuit. For common feature-map and data re-uploading strategies, encoding depth often scales with input dimension and can consume a substantial fraction of the coherence budget before processing begins. Consequently, the maximum depth is bounded by accumulated gate errors and decoherence (Sec.~\ref{subsec:hardware_platforms}).

Wall-clock feasibility is governed by finite throughput and system latencies. The iteration time $T_{\mathrm{iter}}$ decomposes as:
\begin{equation}
T_{\mathrm{iter}} \approx S t_{\mathrm{shot}} + t_{\mathrm{control}} + t_{\mathrm{compile}} + t_{\mathrm{load}} + t_{\mathrm{classical}},
\end{equation}
where $S$ is the total number of circuit executions (shots) per optimization iteration. $t_{\mathrm{control}}$ and $t_{\mathrm{compile}}$ denote overhead from the control stack and circuit compilation, respectively. $t_{\mathrm{load}}$ represents the bandwidth-bound latency required to transfer classical features to quantum control parameters, and $t_{\mathrm{classical}}$ encapsulates the compute-bound classical workload, including bitstring post-processing, gradient estimation, and optimizer updates.

Calibration drift renders the implemented channel time-dependent over a stability window ($\tau_{\mathrm{stable}}$).
Rigorous QDL protocols therefore require time-resolved diagnostics, such as interleaved reference circuits, to detect and quantify drift during training \cite{proctor2025benchmarking,proctor2020detecting}.
Fault tolerance mitigates depth constraints via logical encoding but shifts bottlenecks to syndrome-extraction bandwidth and decoding latency without guaranteeing stable learning-loop performance.

These budgets imply that robust QDL progress is often constrained by end-to-end systems feasibility: the full training loop should remain executable within $D_{\mathrm{eff}}$, $t_{\mathrm{shot}}$, $T_{\mathrm{iter}}$, and $\tau_{\mathrm{stable}}$. At a minimum, studies are recommended to report:
(i) compiled circuit depth, including two-qubit layer count and compilation overhead;
(ii) the number of trainable parameters, objective terms, and the estimator used per term;
(iii) total QPU calls and shots per optimization step with wall-clock cost, including any gradient-estimation overhead;
(iv) drift and calibration management during training;
and (v) the classical resources and baselines used for training and evaluation.

\subsubsection{Scalability challenges}\label{subsubsec:scalability_challenges}

Scalability means increasing qubit number, circuit depth, and model capacity while keeping training and inference resources competitive with tuned classical baselines under comparable access assumptions. Three coupled barriers dominate feasibility in practice: (i) optimization signal degradation, including shot-noise-limited gradients and barren-plateau regimes, (ii) limited scaling of error-mitigation strategies along iterative learning trajectories, and (iii) input-output overheads from data loading and finite-shot readout.

As established in Sec.~\ref{subsubsec:resou_codesign}, highly expressive ans\"atze can exhibit barren plateaus. Operationally, this can imply rapidly increasing shot requirements to resolve gradients at fixed precision. Crucially for hardware implementations, this scaling barrier is severely compounded by realistic local noise models \cite{wang2021noise,singkanipa2025beyond} and necessitates refined Lie-algebraic diagnostics \cite{ragone2024lie,larocca2025barren} to identify scalable regimes. Moreover, improved trainability guarantees need not imply computational advantage, and can coincide with regimes admitting efficient classical descriptions \cite{cerezo2025does}.
Mitigations based on circuit and objective structure \cite{pesah2021absence}, initialization \cite{grant2019initialization}, and non-local parametrization \cite{broers2024mitigated} can improve trainability, and can also be aided by measurement design \cite{sack2022avoiding}, but they need to be evaluated jointly with the induced resource budgets and with matched baselines (see Secs.~\ref{subsubsec:resou_codesign} and~\ref{subsubsec:class_bound}). Landscape statistics can also change qualitatively with parameter count: overparameterization phase transitions have been analyzed and can improve trainability in specific model families, but they come with increased optimization and measurement burdens \cite{arocca2023theory}, and noise can materially modify these regimes \cite{garcia2024effects,mele2024noise}.

As introduced in Sec.~\ref{subsubsec:nisq_fault}, QEM can incur sampling overheads that grow rapidly with circuit depth for broad classes of noise models \cite{cai2023quantum,takagi2022fundamental}. In QDL, these overheads severely compound with the repeated measurement demands of iterative training \cite{schumann2024emergence,wang2021noise,wang2024can}. Recent lower bounds sharpen this limitation by showing that, for broad classes of noise models and mitigation protocols, the required sampling overhead can scale very unfavorably, even when mitigation is otherwise well-defined \cite{takagi2023universal,quek2024exponentially}

Input-output constraints can dominate scaling. The encoding and access-model overheads are developed in Sec.~\ref{subsec:data_encoding}.
Data-encoding choices can also materially affect trainability by controlling induced ansatz expressibility and entanglement, tightening the coupling between input design and optimization scaling \cite{kashif2023unified}.
For many classical tasks, data-loading costs can dominate total runtime, reducing or eliminating any net advantage from quantum processing \cite{aaronson2015read}. On the output side, classical information extraction is limited by shot noise. Estimating expectation values and gradients to precision $\epsilon$ requires $\Omega(\epsilon^{-2})$ samples for bounded-variance estimators, with additional factors depending on the number of measured terms and the chosen gradient estimation strategy \cite{schuld2019quantum}.
These constraints delimit regimes where scalable advantage is most plausible, including quantum-native data and objectives that avoid classical-to-quantum loading \cite{huang2022quantum}, and structured generative settings in which learning can be efficient while sampling and inference are argued to remain classically hard, as reported experimentally on a 68-qubit superconducting processor \cite{huang2025generative}.

\section{Applications and Comparative Analysis}\label{sec:appli_and_compa}

This section surveys QDL application domains and then synthesizes a cross-domain protocol for quantum-classical comparison using those domains as empirical evidence.

\subsection{Application domains}\label{subsec:appli_domai}

\begin{figure*}[tb]\centering
\includegraphics[width=\linewidth]{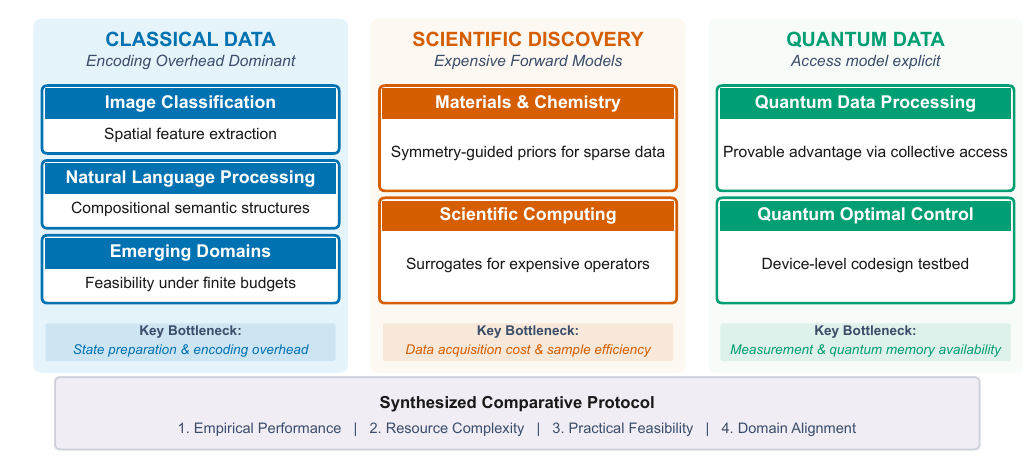}
\caption{Taxonomy of QDL application regimes by data interface. Classical data pipelines embed classical features into quantum circuits and are typically bottlenecked by state-preparation overheads. Scientific discovery integrates quantum modules into label-expensive workflows with costly forward models for both classical and quantum data, where sample efficiency is paramount. Quantum data pipelines operate directly on quantum-native states or processes under an explicit access model, with bottlenecks set by the readout interface and the availability of coherent quantum memory. The lower panel summarizes the four-pillar evaluation protocol used in Sec.~\ref{subsec:comparative_analysis}.}
\label{fig:applications_qdl}
\end{figure*}

QDL applications are categorized by data regime, distinguishing classical inputs processed by hybrid subroutines from inherently quantum data accessed coherently, as shown in Fig.~\ref{fig:applications_qdl}. In classical regimes, evaluation emphasizes empirical feasibility and parameter efficiency constrained by encoding overheads and shot noise. Conversely, in quantum-data settings, provable separations emerge for specific task families under explicit coherent or collective access models, with classical comparators restricted to information obtainable from measurement outcomes. Scientific discovery, typically involving both regimes, bridges these extremes by imposing rigorous resource accounting on expensive forward models, motivating the four-pillar comparison protocol in Sec.~\ref{subsec:comparative_analysis}.

\subsubsection{Image classification}\label{subsubsec:app_image}

Image classification is widely used as a diagnostic and comparative benchmark for QDL \cite{kharsa2023advances,senokosov2024quantum}.
Classical inputs and mature classical vision architectures provide well-understood baselines. Quantum pipeline costs are often dominated by data encoding and sampling, making this domain a controlled testbed for quantum feature maps, inductive biases, and finite-shot trainability under device noise.
Rigorous evaluations require carefully tuned classical baselines and, where relevant, quantum-inspired classical surrogates (e.g., TN or kernel methods) under matched resource contracts. Reported empirical improvements are, at present, more often explained by baseline choice, inductive-bias effects, and hybridization rather than by a demonstrated end-to-end quantum advantage under budget-matched tuning.

Early pipelines downsample images into patches and encode features into PQC parameters \cite{farhi2018classification,li2020quantum,grant2018hierarchical}, which weaken or only partially preserve spatial inductive biases and can make performance strongly dependent on the chosen feature map \cite{schuld2021effect,perez2020data,havlivcek2019supervised}.
Many QDL models draw architectural inspiration from classical CNNs that are well established for vision tasks. QCNNs formally defined in Sec.~\ref{subsubsec:archi_parad_for_hyb} impose multiscale convolution-pooling structure \cite{cong2019quantum,herrmann2022realizing,hur2022quantum} and can avoid barren-plateau behavior in certain locality-structured settings \cite{pesah2021absence}, but, as discussed in Sec.~\ref{subsubsec:dequantization}, under additional locality or limited-entanglement assumptions, they may also admit efficient classical simulation in broad regimes \cite{bermejo2024quantum}.

Hybrid CNN pipelines insert small PQCs into classical vision stacks \cite{liu2021hybrid,fan2023hybrid,long2025hybrid}. Closely related are quanvolutional pipelines \cite{henderson2020quanvolutional} defined in Sec.~\ref{subsubsec:archi_parad_for_hyb} in which shallow patchwise circuits produce spatial feature maps consumed by classical backbones, including recent extensions with trainable kernels \cite{kashif2025deep,bhatia2023federated}, and transfer-learning schemes that compress images via classical embeddings before quantum refinement \cite{mari2020transfer, kim2023classical, azevedo2022quantum,sugunapriya2025variational}. However, for classical-image workloads at currently accessible sizes, these architectures typically employ small numbers of qubits and shallow circuits, and are therefore often compatible with efficient classical simulation in the studied parameter regimes.
Alternative models for image classification aiming to explore less conventional quantum learning paradigms have been proposed, such as quantum extreme learning machines \cite{de2025harnessing} or transformer-like proposals \cite{xue2024end,cherrat2024quantumvision,guo2024quantum}.

Most studies use traditional vision datasets such as MNIST \cite{LeCun1998}, Fashion-MNIST \cite{xiao2017fashion}, and CIFAR \cite{krizhevsky2009learning}. A smaller study considers domain-specific scientific images such as high-energy physics experiments \cite{chen2022quantumqcnn} and medical imaging \cite{landman2022quantum,chow2025quantum,mathur2021medical}.
While select studies report parameter efficiency compared to small classical networks \citep{liu2025lean}, the stability, reproducibility, and magnitude of reported gains are often sensitive to baseline choice and hyperparameter tuning \cite{tasnim2025quantum,basilewitsch2025quantum}. To date, no study has established a scalable and reproducible quantum advantage over state-of-the-art classical vision baselines under an explicit, budget-matched resource contract for practically relevant image-classification workloads, and performance is frequently limited by data-loading costs, sampling noise, and device errors.

\subsubsection{Natural language processing}\label{subsubsec:app_nlp}

NLP \cite{chowdhary2020natural} studies discriminative and generative modeling of classical text data.
Quantum NLP refers to NLP pipelines where the quantum component changes the learning or inference interface in a way that is explicit in the resource contract \cite{widdows2024quantum}.
A technically specific motivation is provided by compositional semantic formalisms in which grammatical structure induces multilinear maps. For such models, sentence meaning reduces to a TN contraction compilable into circuits.
This transparency enables falsifiable questions about inductive bias and parameter efficiency \cite{Widdows2024QNLP,nausheen2025,guarasci2022quantum,varmantchaonala2024quantum,pallavi2025quantum}.
Given Transformer dominance \cite{vaswani2017attention,devlin2019} and classical NLP's strong scaling laws, near-term quantum NLP is best viewed as a platform for methodology development and controlled benchmarking rather than as a competitor to large-scale classical models.

In compositional distributional semantics, the DisCoCat framework \cite{coecke2010,Grefenstette2011} maps grammatical derivations to tensor-diagram contractions that define sentence meanings.
In quantum implementations, these diagrams can be compiled into parameterized circuits (e.g., via ZX-calculus-based rewrites) \cite{coecke2020,meichanetzidis2020quantum} and trained end-to-end using dedicated toolchains such as lambeq \cite{kartsaklis2021lambeq}.
Demonstrations include small-scale sentence classification experiments \cite{Lorenz_2023,Meichanetzidis_2023,silver2024lexiql} and natural language generation \cite{karamlou2022} on superconducting quantum processors, and scalable proposals on trapped ions \cite{duneau2024scalable}. Related graph-based variants incorporate PQC-based attention within message-passing NLP architectures \cite{aktar2025}.
An alternative route is composition-agnostic, mapping text to quantum feature states for kernel estimation \cite{Alexander_2022}. Related proposals study attention-inspired PQCs primitives \cite{li2024quantum,tomal2025quantum,chen2025} and hybrid modules for parameter-efficient large language model (LLM) adaptation \cite{kong2025}.
Circuit-level analyses of transformer subroutines and attention computation \cite{gao2023fast,liao2024} are conditional on explicit access models.

Quantum-inspired NLP uses quantum probability formalisms as classical modeling primitives, representing queries or documents as density matrices and fitting matching models via information-theoretic criteria such as entropy-based objectives \cite{sordoni2013,sordoni2014,sordoni2014learning}. Representative modern variants develop density-matrix-based sentence representations for classification and matching \cite{zhang2018endtoend,Zhang2019,shi2023two,shi2024pretrained}, introduce explicit state-evolution updates for language modeling and sentiment dynamics \cite{fan2024quantum,yan2024quantum}, and explore entanglement-inspired embeddings that encode nonclassical correlations across tokens \cite{chen2023qlm}.

\subsubsection{Materials design and quantum chemistry}\label{subsubsec:app_mater_chemi}

Materials design and quantum chemistry \cite{sajjan2022quantum,mcardle2020quantum} are natural target domains for QDL and are often highlighted as potentially favorable because high-fidelity classical data acquisition is costly and frequently yields small or sparse labeled datasets, making inductive bias and parameter efficiency especially important. Although inputs are encoded in classical representations such as atomic structures, graphs, or descriptors, the supervised targets are inherently quantum-mechanical observables, including energies, forces, spectra, and response properties governed by an electronic Hamiltonian. The dominant bottleneck is the data-generation budget, so methodological claims are most meaningful under matched supervision budgets and accuracy targets.

The quantum contribution is typically introduced through a quantum-parameterized hypothesis class such as a PQC layer \cite{benedetti2019parameterized} or a quantum kernel estimator \cite{schuld2019quantum,havlivcek2019supervised,schuld2021supervised}, used as a module within a hybrid model and evaluated against compute-matched baselines including molecular GNNs \cite{gilmer2017neural}, SchNet-class models \cite{schutt2017schnet}, and equivariant interatomic potentials \cite{batzner20223}. Key confounders are data-loading assumptions, the strength of symmetry-aware baselines, and the gap between chemically relevant accuracy thresholds and finite-shot noise.

A primary scientific difficulty is maintaining chemical accuracy across high-dimensional configurational and chemical spaces, particularly in strongly correlated regimes where bias and generalization are fragile \cite{mcardle2020quantum}. Classical neural wavefunction ans\"atze such as FermiNet \cite{pfau2020abinitio} and PauliNet \cite{hermann2020deep} provide strong references, while on quantum hardware wavefunction-centric hybrids target electronic-structure observables via variational objectives, most prominently VQE \cite{peruzzo2014variational,mcclean2016theory}. Beyond VQE, \textcite{xia2018quantum} explores an RBM wavefunction trained with quantum-assisted optimization in a small-scale setting. Inverse-design workflows optimize a target functional over many-body states prepared on a quantum simulator and then use Hamiltonian learning to extract a geometrically local parent Hamiltonian \cite{kokail2026inverse}.

In property-prediction pipelines, data loading remains a recurring obstacle: mapping high-dimensional descriptors to few-qubit devices can dominate cost unless access assumptions are explicit \cite{tang2021quantum,giovannetti2008quantum}. \textcite{reddy2021hybrid} combine latent-space compression with a quantum regression module, which functions as a specialized inductive-bias component and should be compared under matched budgets. Graph-based inductive biases remain central in molecular ML, and quantum-embedded GNN pipelines have been explored in resource-limited regimes \cite{verdon2019quantum,chen2021hybrid,lu2025quantum}. Reported gains often emphasize parameter efficiency or stability and remain baseline-dependent. Recent work explores data-efficient models \cite{hagelueken2025data} and early drug-discovery prototypes \cite{ghazi2025quantum}, but large-scale chemically accurate quantum advantage remains unestablished.

\subsubsection{Scientific computing}\label{subsubsec:sci_discovery}

Scientific computing focuses on data-driven surrogates for accurate but expensive first-principles simulations. High-fidelity supervision is the limiting resource, favoring hypothesis classes that exploit structural priors such as locality, conservation laws, and symmetries. QDL approaches include surrogate modeling for quantum mechanical problems, quantum model discovery from sparse data, and operator learning with expensive forward models \cite{chen2024enabling,ye2024hybrid,Xiao2025quantumdeeponet}.
Quantum contributions typically take the form of quantum-parametrized hypothesis classes or subroutines within classical training loops.
Quantum surrogates are most informatively assessed when benchmarked against structure-aware classical baselines, particularly physics-informed neural networks \cite{raissi2019physics} and neural operators \cite{lu2021learning,li2020fourier,kovachki2023neural}.

Scientific workloads impose stringent constraints because high-fidelity supervision often requires expensive forward solves and calibration, which limits data-generation budgets even when synthetic sampling is possible. Thus, data efficiency and robustness often matter as much as predictive accuracy. Existing hybrid demonstrations emphasize feasibility studies for surrogates of partial differential equation dynamics \cite{chen2024enabling,ye2024hybrid}. More structured proposals target operator learning \cite{Xiao2025quantumdeeponet}, where the objective is to approximate maps between function spaces.
Related exploratory directions include hybrid quantum physics-informed solvers and quantum neural operators \cite{leong2025hybrid,wang2025quanonet}.
Additionally, many-body Hamiltonians serve as natural testbeds for evaluating these quantum surrogates, given their intrinsic alignment with quantum-mechanical state spaces. Classical neural network quantum states provide strong baselines \cite{carleo2017solving,gao2017efficient}. QDL proposals introduce PQC-based parametrizations and/or quantum-assisted sampling strategies within this variational-learning setting \cite{zhang2025quantum,gardas2018quantum}.

High-energy physics offers realistic benchmarking with large structured classical datasets and stringent calibration requirements \cite{di2024quantum}, making it a demanding benchmarking environment for end-to-end evaluation. Quantum pipelines apply collider-style analyses with circuit feature maps and kernel estimators \cite{wu2021application}, while quantum-inspired classical methods, including TN classifiers, are highly competitive \cite{felser2021quantum}. These comparisons highlight the importance of domain-informed inductive bias and careful benchmarking.
This domain is methodologically demanding for QDL. Defensible conclusions require rigorous baseline tuning, uncertainty-aware evaluation, and explicit end-to-end resource accounting.

\subsubsection{Quantum data processing}\label{subsubsec:app_quantumdata}

As formalized in the access-model taxonomy of Sec.~\ref{subsubsec:taxon_of_quant_advan}, quantum-data workflows are distinguished most sharply by the access-and-readout interface: (i) \emph{measure-first} protocols that map each copy to a classical record and learn by classical postprocessing, and (ii) \emph{coherent or collective} protocols that retain quantum information (e.g., via quantum memory) to enable joint measurements or coherent processing across copies. General learnability bounds and access-model separations are treated in Sec.~\ref{subsubsec:stati_learn}; here we focus on application-level instantiations of these interfaces.

Operationally, the quantum contribution lies in exploiting quantum resources at the interface, including measurement-efficient acquisition protocols with explicit sample-complexity guarantees \cite{huang2020predicting,Grier2024sampleoptimal,chen2024optimal}, quantum-native compression \cite{romero2017quantum,Pepper2019,ma2024quantum}, or fixed quantum dynamical reservoirs \cite{senanian2024microwave,palacios2024role,zhu2025practical}.
Kernel methods belong here only when the inputs are quantum-native and the kernel is estimated from quantum access; kernels for classical inputs are treated under classical-data applications.

Full reconstruction of generic $n$-qubit states requires resources exponential in $n$ in the worst case, so efficient protocols require either additional structure promises (e.g., low rank) or weaker learning goals \cite{gross2010quantum}.
PAC-style learning can instead target prediction of measurement outcomes without full reconstruction \cite{aaronson2007learnability}.
Representative task families that exploit coherent or collective interfaces include collective-measurement protocols for learning tasks \cite{huang2022quantum,chen2022exponential}, unsupervised classification of quantum states \cite{sentis2019unsupervised}, and separations highlighting limitations of fixed-measurement measure-first pipelines in supervised learning \cite{gyurik2025limitations}. A scalable photonic implementation has experimentally reported a task-specific quantum learning advantage in sample complexity under an explicit photonic access model, illustrating this hierarchy in practice \cite{liu2025quantum}.

The classical shadows framework \cite{aaronson2018shadow,huang2020predicting,jerbi2024shadows,Grier2024sampleoptimal} introduced in Sec.~\ref{subsubsec:quant_measu} is a canonical \emph{measure-first} strategy for predicting many observables from randomized single-copy measurements via classical postprocessing.
By constructing a compact classical record from randomized measurements, it enables amortized evaluation across downstream queries with explicit sample-complexity control \cite{huang2020predicting,jerbi2024shadows,Grier2024sampleoptimal}.
In the worst case, shadow tomography without quantum memory provably requires $\Omega(\min(M,2^n))$ copies to estimate $M$ target observables \cite{chen2022exponential} to a fixed accuracy, where $n$ is the number of qubits. This lower bound clarifies when a fixed classical record cannot simultaneously support all downstream queries.
Related constructions extend these ideas to resource-constrained characterization and learning of quantum processes, with costs set by both measurement budgets and circuit resources \cite{Levy2024}.
Complementary analyses clarify when classical postprocessing alone suffices to recover task-relevant structure and when additional quantum access is information-theoretically necessary \cite{de2024classical}.

When data admit a low-complexity model class, generative learning replaces worst-case reconstruction with trainable representations from incomplete measurements \cite{Torlai2018,Carrasquilla2019}. Practical performance depends on optimization stability and evaluation under realistic shot budgets, as examined in recent trainability and benchmarking studies \cite{hibat2024framework,rudolph2024trainability}.
Related directions include learning dynamics and system identification \cite{gentile2021learning,hangleiter2024robustly}, quantum-dynamics modeling \cite{liu2022solving}, representation learning \cite{jayakumar2024learning}, compression \cite{romero2017quantum,Pepper2019,ma2024quantum}, and measurement-efficient online learning with fixed quantum dynamical reservoirs \cite{senanian2024microwave,palacios2024role,zhu2025practical}.
Quantum data processing is the application regime where the access-and-readout model can be stated most explicitly. Practical impact therefore hinges on realistic interface assumptions, including whether quantum memory or only measured outcomes are available, measurement budgets, and rigorous resource-matched benchmarking.

\subsubsection{Quantum optimal control}\label{subsubsec:app_qcontrol}

Quantum optimal control designs time-dependent controls that steer quantum dynamics toward a target objective, and is traditionally addressed with classical gradient-based methods and, more recently, RL approaches \cite{khaneja2005optimal,koch2022quantum}.
As noted in the superconducting platform discussion of Sec.~\ref{subsubsec:superc_circuits}, pulse-level access enables systematic translations between control parameterizations and circuit ans\"atze \cite{Magann2021ContPQCQOC}, a correspondence that connects hardware-level control with variational QDL training.
Accordingly, classical optimal-control and RL pipelines define the comparator stack: empirical claims are most interpretable under matched experimental-call budgets, hardware constraints, and evaluation metrics, such as time-to-target fidelity or robustness. Standard classical benchmarks include GRAPE \cite{khaneja2005optimal}, GOAT \cite{machnes2018tunable}, PEPR \cite{heimann2025}, and Krotov-type updates \cite{konnov1999global,tannor1992control,reich2012monotonically}, while RL has been used for robust pulse design in hardware-facing settings \cite{bukov2018reinforcement,sivak2022model,baum2021experimental,li2025robust,niu2019universal,sarma2025designing}. Pulse-level control can also act as an alternative to gate-level compilation in variational workflows, trading discrete gate synthesis for directly optimized control waveforms and therefore requiring pulse-level resource accounting \cite{leng2022differentiable,meirom2023pansatz,Melo2023pulseefficient,deKeijzer2023pulsebased}.

In this review, we focus on QOC as a QDL application when the policy or control ansatz is quantum-parametrized and trained in a measurement-limited closed loop where updates are driven by measured cost signals. Variational-circuit policies have been proposed for continuous-action control via quantum deterministic policy-gradient variants \cite{wu2025quantum}, and related work explores quantum enhancements to deep RL in large action spaces \cite{jerbi2021quantum}. At the hardware level, pulse-parametrized learning frameworks enable direct optimization of control waveforms within variational learning stacks \cite{liang2022vqpl}. Global pulse parametrizations, such as Fourier-basis schedules, can reduce gradient-variance concentration relative to local piecewise-constant controls \cite{broers2024mitigated}, resulting in a mitigation of barren plateaus. This is related to the control-theoretic diagnostics that link trainability to controllability through the dynamical Lie algebra generated by the ansatz \cite{larocca2022diagnosing}. Hybrid VQAs have also been proposed explicitly for quantum control objectives, providing a direct bridge between VQA methodology and QOC benchmarks \cite{huang2025optimal}. Under this scope criterion, QOC provides a device-level testbed for quantum-classical codesign under explicit end-to-end resource accounting. Outside of it, QOC primarily supplies comparator families and evaluation protocols for learning-based claims.

\subsubsection{Emerging domains}\label{subsubsec:app_emerging}

QDL has been explored in a growing set of emerging sociotechnical and engineering domains, including finance \cite{herman2023quantum,Cherrat2023quantumdeephedging}, cyberphysical systems \cite{ajagekar2021quantum,kaseb2024quantum,schetakis2025quantum,yi2025hybrid,ansere2024quantum}, healthcare \cite{gupta2025systematic,hossain2024adeep,sagingalieva2023hybridqnn}, security- and privacy-constrained inference \cite{yu2023quantum}, logistics \cite{correll2023quantum}, robotics \cite{yan2024quantumrobotics,hohenfeld2022quantum}, and distributed learning \cite{nguyen2025quantum,kwak2023quantum}. Representative industrial vision tasks provide an additional testbed beyond toy datasets \cite{yang2022semiconductor}. Across these areas, the dominant role of such studies is best understood as using the application as a testbed, employing realistic domain problems to stress-test QDL methodology rather than claiming domain-specific advantage. In line with the patterns highlighted for image classification and natural language processing, they primarily probe how hybrid QDL pipelines behave under realistic deployment constraints, rather than establishing domain-specific, scalable quantum advantage.

Despite diverse problem settings, a common architectural pattern recurs. Most works embed a shallow PQC as a learnable nonlinear module, often a feature map or compact classifier head, inside an otherwise classical workflow that provides the main representational capacity. This convergence largely reflects near-term feasibility, including depth and shot budgets, as well as the practicality of integrating PQCs into existing DL toolchains. Consequently, reported benefits are typically finite budget effects, such as parameter reductions, shifts in constant factors, or comparable performance under tight resource ceilings, rather than asymptotic improvements \cite{bowles2024better}.

A recurring limitation in QDL studies is the difficulty of generalizing empirical conclusions across works, as evaluation practices and baselines are not yet standardized and are domain-specific. Many studies use domain-aligned objectives, for example, robustness, latency, privacy, or verification cost, but the resulting comparisons can be sensitive to comparator choice, tuning budgets, and resource accounting. This underscores the need for stronger benchmark discipline in emerging domains: clearly specified evaluation contracts and matched baselines are essential to distinguish genuine quantum utility from artifacts of experimental design. These issues motivate the comparative perspective of Sec.~\ref{subsec:comparative_analysis}, where empirical QDL results are examined under explicit benchmarking and resource matching protocols.

\subsection{Comparative analysis with classical DL}\label{subsec:comparative_analysis}

Cross-domain evidence in Sec.~\ref{subsec:appli_domai} suggests four recurrent evaluation bottlenecks: (i) comparator sensitivity under matched tuning, dominant in vision and language; (ii) end-to-end resource accounting including encoding and sampling, prominent in chemistry and scientific pipelines; (iii) hardware-facing feasibility and verification cost, prominent in control and security; and (iv) structural alignment between the quantum component and the data, often decisive for quantum-native settings.

Motivated by \textcite{schuld2022quantum}’s critique of single-goal quantum advantage framings, we translate the explicit-contract specification of Definition~3 (Sec.~\ref{subsec:comp_symbiosis}; see also Sec.~\ref{subsubsec:taxon_of_quant_advan}) into four practical reporting pillars (Table~\ref{tab:framework}).
While theoretical separations and fundamental limitations are treated in Sec.~\ref{subsec:theor_advan}, here we focus on experimental controls that most strongly influence empirical conclusions:\\
\noindent (1) \textit{Empirical performance} operationalizes the task and data model $\mathcal{D}$ by specifying how quality is measured under the stated distribution and protocol.\\
\noindent (2) \textit{Resource complexity} operationalizes the end-to-end resource contract $\mathcal{R}$ by specifying cost accounting and budget ceilings.\\
\noindent (3) \textit{Practical feasibility} operationalizes the implementability and validation clauses of Definition~3 by documenting hardware- and verification-facing constraints at the target scale.\\
\noindent (4) \textit{Domain alignment} operationalizes the access-and-readout model $\mathcal{A}$ by checking that the quantum component's structural and interface assumptions match the task's information pathway and the stated comparator class.

We treat these four pillars as jointly necessary conditions for interpreting finite-size quantum-classical comparisons. Passing Pillars~1--3 without Pillar~4 is not, by itself, evidence that an observed gain is attributable to quantum-mechanical resources under the declared $(\mathcal{D},\mathcal{R},\mathcal{A})$. Absent controlled scaling under matched budgets, empirical results primarily serve as proof-of-principle demonstrations informing feasibility, robustness, and inductive-bias hypotheses.
To maximize interpretability, empirical studies are most interpretable when they report: baseline families and inclusion rationale; preprocessing and feature pipeline on both sides; hyperparameter search space and allocated tuning budgets; a common stopping rule (e.g., time-to-target under $\mathcal{R}$) with learning curves; and negative controls that isolate the quantum block (e.g., parameter-matched classical modules, randomized features,  or classical surrogates.)
Claims of improvement are conditional on the declared comparator stack and contract $(\mathcal{D},\mathcal{R},\mathcal{A})$, interpreted alongside domain-specific bottlenecks identified in Sec.~\ref{subsec:appli_domai}.

\begin{table*}[tb]
\caption{A four-pillar protocol for rigorous quantum-classical comparison in finite-size empirical studies. Claims are evaluated under an explicit access and readout interface and an end-to-end resource contract \(\mathcal{R}\) (Definition~3). Mechanisms underlying limitations are treated in Sec.~\ref{subsec:theor_advan}.}
\label{tab:framework}
\begin{ruledtabular}
\renewcommand{\arraystretch}{1.15}
\setlist[itemize]{label=--, leftmargin=*, nosep, topsep=0pt, partopsep=0pt, itemsep=0pt, parsep=0pt}
\begin{tabular}{lll}
\textbf{Pillar} & \textbf{Reporting axes} & \textbf{Primary failure modes} \\
\midrule
\parbox[t]{0.30\textwidth}{\raggedright
\textbf{1. Empirical performance}\\
\emph{Does it improve quality under matched evaluation?}} 
&
\parbox[t]{0.34\textwidth}{\raggedright\leavevmode
\vspace{-\baselineskip}
\begin{itemize}
\item Task-specific metric and calibration 
\item Learning curves versus data and compute 
\item Robustness across seeds and splits 
\item Sample-efficiency diagnostics 
\end{itemize}
} &
\parbox[t]{0.34\textwidth}{\raggedright\leavevmode
\vspace{-\baselineskip}
\begin{itemize}
\item Under-tuned or outdated baselines 
\item Unmatched hyperparameter optimization 
\item Hidden pre or postprocessing
\item Selective reporting of best runs 
\end{itemize}} \\~\vspace{-3mm}\\
\midrule
\parbox[t]{0.30\textwidth}{\raggedright
\textbf{2. Resource complexity}\\
\emph{Is it more efficient under end-to-end accounting?}} &
\parbox[t]{0.34\textwidth}{\raggedright\leavevmode
\vspace{-\baselineskip}
\begin{itemize}
\item Total shots and QPU calls 
\item Wall-clock breakdown and time-to-target
\item Energy and monetary cost boundary
\item Compiled depth and two-qubit counts 
\item Classical coprocessing and communication 
\end{itemize}} &
\parbox[t]{0.34\textwidth}{\raggedright
\vspace{-0.5\baselineskip}
\begin{itemize}
\item Reporting circuit metrics without wall-clock cost 
\item Incommensurate access or readout interfaces or training interfaces (e.g., classical backpropagation)
\item State-preparation and encoding omitted 
\item Missing classical surrogate checks 
\end{itemize}} \\~\vspace{-3mm}\\
\midrule
\parbox[t]{0.30\textwidth}{\raggedright
\textbf{3. Practical feasibility}\\
\emph{Can it be trained and executed reliably on quantum hardware?}}&
\parbox[t]{0.34\textwidth}{\raggedright\leavevmode
\vspace{-\baselineskip}
\begin{itemize}
\item Training stability and failure rate 
\item Verification method and its cost 
\item Inference throughput and latency 
\item Compilation overheads 
\end{itemize}} &
\parbox[t]{0.34\textwidth}{\raggedright\leavevmode
\vspace{-\baselineskip}
\begin{itemize}
\item Optimizer stagnation and seed sensitivity 
\item Shot-noise dominated updates 
\item Unsupported intractability 
\end{itemize}} \\~\vspace{-3mm}\\
\midrule
\parbox[t]{0.30\textwidth}{\raggedright
\textbf{4. Domain alignment}\\
\emph{Is the quantum component structurally matched to the task?}} &
\parbox[t]{0.34\textwidth}{\raggedright\leavevmode
\vspace{-\baselineskip}
\begin{itemize}
\item Architecture class and symmetry constraints
\item Modality and interface realism 
\item Structure-sensitive ablations and controls 
\item Inductive-bias diagnostics 
\end{itemize}} &
\parbox[t]{0.34\textwidth}{\raggedright\leavevmode
\vspace{-\baselineskip}
\begin{itemize}
\item Access-model mismatch to the deployment setting 
\item Baselines that ignore the same structure 
\item Claims of inductive bias without controls 
\end{itemize}} \\
\end{tabular}
\end{ruledtabular}
\end{table*}

\subsubsection{Empirical performance}

Building on the reporting criteria of Sec.~\ref{subsubsec:bench_repro}, empirical performance asks whether a QDL model improves task-relevant quality under matched evaluation and comparable tuning effort relative to strong classical baselines. On current benchmark suites, outcomes can be highly sensitive to preprocessing choices and hyperparameter optimization, and reported performance often reflects these design decisions as much as architectural differences \cite{bowles2024better,moussa2024hyperparameter}. Moreover, the relevant classical comparator set is not static as the DL software-hardware stack evolves \cite{jouppi2017datacenter}.
In addition to the elements required by Definition~3, a reproducible comparison benefits from disclosing the full feature pipeline, reporting the hyperparameter-optimization budget used for each model class, and summarizing variability across random seeds and data splits with uncertainty intervals.
When efficiency is part of the claim, results are most interpretable when summarized under a fixed stopping rule, for example time-to-target quality under the stated contract $\mathcal{R}$.
Beyond a single operating point, learning curves versus dataset size and training budget, together with ablations that replace the quantum block by parameter-matched classical components or randomized feature maps, help isolate which aspects of the pipeline contribute materially to observed gains.
Recent large-scale benchmarking indicates that, on standard classical datasets, reported advantages are often regime dependent and can diminish under broader tuning, preprocessing control, or dataset coverage \cite{bowles2024better}; see also time-series case studies in \cite{fellner2025quantum}.
Quantum-kernel benchmarking similarly shows sensitivity to classical optimization choices and dataset selection \cite{schnabel2025quantum,alvarez2025benchmarking}.
In this light, finite-size improvements on classical benchmarks are often most informatively interpreted as probes of inductive bias, feasibility, and robustness, rather than as evidence for sustained performance separation.

\subsubsection{Resource complexity}

Resource complexity asks whether an observed finite-size improvement persists once end-to-end costs are accounted for under the same access and readout interface. For classical baselines, relevant costs include wall-clock time, energy consumption, monetary cost, memory footprint, and accelerator utilization, all of which can shift rapidly as hardware and software coevolve \cite{jouppi2017datacenter,schwartz2020greenai,henderson2020towards,strubell2020energy,patterson2021carbon}.
For quantum pipelines, the operational object of comparison is the end-to-end cost to reach a target quality under a fixed contract $\mathcal{R}$, including state preparation and encoding, circuit execution and compilation overheads, measurement, and outer-loop orchestration.
For energy accounting, $\mathcal{R}$ typically specifies the system boundary, for instance at device or facility level, and whether cooling and control are included, and the measurement or estimation protocol \cite{cordier2025scaling,lorenz2025systematic}.
Laid out at the theory level, energy can itself be treated as a complexity resource; in a query-complexity framework, \textcite{meier2025energy} prove an exponential energy-consumption advantage for Simon's problem and propose criteria for experimental demonstration.
Empirical resource claims are most naturally interpreted against contemporary classical capabilities rather than static historical baselines, with explicit documentation of the chosen comparator stack. Small-scale empirical comparisons further illustrate the sensitivity of energy-to-solution to the chosen boundary and workload \cite{desdentado2024exploring}.

A practical accounting decomposes the pipeline into four coupled contributions:\\
\noindent (i) \emph{Input cost} comprises classical preprocessing and the end-to-end latency of embedding data into the quantum model; this is particularly relevant whenever classical feature compression is used to make encoding feasible.\\
\noindent (ii) \emph{Compute cost} captures the per-circuit footprint after compilation, e.g., depth, two-qubit gate counts, and compilation time, ideally anchored to measured (or explicitly justified) execution latency on the target backend.\\
\noindent (iii) \emph{Output cost} accounts for the total number of circuit repetitions used in training and inference, together with estimator details and the achieved statistical uncertainty at the reported operating point; this is often a dominant term in kernel-style estimators and expectation-value training loops \cite{bowles2024better,schnabel2025quantum}.\\
\noindent (iv) \emph{Loop overhead} aggregates the outer optimization costs, including total QPU calls, optimizer steps, and classical postprocessing and communication, which can dominate wall-clock time in hybrid workflows even when per-circuit execution is fast \cite{guerreschi2017practical}.

Claims of resource advantage are most informative when supported by an end-to-end breakdown of time and energy, and by time-to-target and energy-to-target curves under a fixed stopping rule and declared measurement protocol. Any surrogate baseline used is ideally stated explicitly, and access-model assumptions are typically declared as part of $\mathcal{A}$ and reported alongside the contract $\mathcal{R}$.

\subsubsection{Practical feasibility}

Practical feasibility, applying the verification and hardware constraints of Secs.~\ref{subsubsec:bench_repro} 
and~\ref{subsubsec:hardware_constraints} to the evaluation context, asks whether a QDL architecture can be trained and executed reliably on available hardware, and whether inference can be performed at acceptable cost under the intended deployment contract. This pillar is empirical by design, focusing on what is commonly measured and reported for hardware-facing claims to be interpretable.
First, studies that invoke limited classical verifiability or conditional intractability generally benefit from specifying the verification strategy and its cost. This includes the simulator or emulator class used for validation, the observable set being checked, the statistical confidence reported, and the verification budget relative to the claimed advantage. Without such specification, statements about hardness or unverifiability can be difficult to contextualize methodologically.
Second, training feasibility is often characterized by stability rather than only best-achieved performance. Empirical papers often report failure rates across seeds, sensitivity to initialization and stopping rules, and a gradient signal proxy at the stated shot budgets, together with time-to-target distributions rather than single-run outcomes.
In addition, the training procedure itself is shaped by hardware realism: computational tools commonly used in simulated training of QDL models, most notably backpropagation and related gradient evaluation techniques, are fundamentally incompatible with quantum hardware due to the no-cloning theorem and measurement collapse, meaning gradients are instead estimated through resource-intensive repeated circuit executions or hybrid classical-quantum loops.
When training is performed on hardware, drift sensitivity and device-side throughput constraints are often reported in comparable terms, since these effects directly enter the time-to-solution contract.
Third, deployment-facing claims generally include inference-time costs. Quantum models typically produce stochastic measurement outcomes; inference therefore incurs a shots-per-prediction cost that is reported alongside throughput and latency. Single-shot or few-shot prediction regimes, when claimed, are most interpretable when accompanied by explicit accuracy–confidence tradeoffs under the stated measurement model \cite{recio2025single}. Feasibility claims that omit inference overhead are correspondingly harder to compare to deterministic classical inference baselines.

\subsubsection{Domain alignment}

Domain alignment asks whether the proposed quantum component is structurally matched to the task and to the data interface, and therefore whether the benchmark is capable of testing a plausible quantum contribution. For classical-data domains, the dominant barriers are often encoding overhead and the strength of pretrained classical baselines; for quantum-native domains, the dominant barriers are access-model realism and measurement budgets. Accordingly, inductive-bias arguments can be viewed as design hypotheses that are validated by structure-sensitive controls and baselines.

Data modality and interface realism form a primary consideration. For high-dimensional classical inputs, hybrid designs and classical feature compression are often required to make encoding feasible, and the cost of this compression enters the end-to-end comparison \cite{mari2020transfer}. For quantum-native data, coherent interfaces can be decisive; empirical claims in this regime typically state the access and readout model and benchmark against classical comparators restricted to the same measurement interface \cite{huang2022quantum,cho2024machine,liu2025quantum}.

Structure and symmetry provide another lens for alignment. QDL is particularly well motivated when architectures enforce known structure such as locality or group symmetries, restricting the effective hypothesis class to task-relevant functions and often improving stability and data efficiency in structured regimes \cite{meyer2023exploiting,larocca2022group,nguyen2024theory}. In such settings, alignment is commonly established empirically by baselines that incorporate the same structure and by causal ablations that break the purported prior, rather than by comparing to generic unstructured baselines.

Label scarcity and acquisition cost offer a further dimension. In data-rich regimes such as large-scale vision and language with billions of training examples, large pretrained classical models impose a high bar; conversely, in label-expensive scientific regimes such as materials and chemistry, the premium often shifts toward sample efficiency, uncertainty quantification, and robust generalization, making targeted structure and access-model choices more consequential \cite{mcardle2020quantum}.

\section{Conclusions and Future Directions}\label{sec:concl_and_futur}

\subsection{Summary of key findings}

The evolution of QDL is shaped by three coupled tensions that separate finite-size, proof-of-principle improvements from scalable advantage: (i) expressivity-trainability trade-off, (ii) trainability-simulability constraint, and (iii) quantum-classical interface and certification bottleneck.
The expressivity-trainability trade-off \cite{holmes2022connecting} becomes more diagnostic when recast in physically grounded controllability language: the dynamical Lie algebra generated by an ansatz's implementable generators provides a principled handle on which directions in Hilbert space are actually reachable and on how ``effective expressivity" correlates with loss concentration and training costs \cite{ragone2024lie,kazi2025analyzing}.
This refines the barren-plateau reasoning: exponentially suppressed gradients are well established for broad classes of global objectives and highly expressive circuit ensembles, yet the onset and severity of gradient suppression depend sensitively on symmetry structure, cost locality, noise, and the measurement procedure used to obtain gradients \cite{mcclean2018barren,larocca2025barren}. Accordingly, trainability functions as a scaling constraint rather than a purely algorithmic inconvenience, and it naturally motivates structured inductive biases, such as symmetry-preserving parameterizations, locality-aware architectures, and localized cost constructions, that reduce concentration without silently changing the learning task \cite{cerezo2022challenges}.

Mitigating gradient pathologies, however, tightens the second tension: the same structural restrictions that improve trainability can also move a quantum model toward regimes that admit efficient classical simulation. Locality, shallow depth, restricted connectivity and constrained generator sets can suppress concentration effects \cite{cerezo2021cost,pesah2021absence}, but these constraints often align with known simulation or dequantization mechanisms, shrinking the space of candidate quantum advantages \cite{bermejo2024quantum,shin2024dequantizing}. In this sense, quantum-inspired classical algorithms, including tensor-network models and dequantized linear-algebraic algorithms \cite{tang2019quantum,tang2021quantum}, are not merely ``baselines" but essential foils: they delimit which gains plausibly require device-native quantum resources rather than classical structure exploitation. Credible empirical comparisons therefore hinge on contracts that fix the task specification, access assumptions, and comparator families tightly enough to prevent implicit asymmetries in preprocessing, tuning budgets, or oracle access \cite{schuld2022quantum}.

The third tension unifies what is often treated as a grab bag of practical issues: data access, readout, and verification are coupled by the quantum-classical interface. For classical data, coherent state preparation and task-aligned preprocessing can dominate end-to-end cost and can change which resource axis is actually limiting~\cite{bowles2024better,zhang2024circuit}. This observation helps explain why near-term claims are typically most defensible in quantum-native data regimes, where the dominant overheads shift from loading classical information to measurement, control, and stability constraints \cite{huang2022quantum}. It also clarifies why ``verification" cannot be treated as an afterthought: as experiments move beyond classically reproducible regimes, the strongest classical cross-checks degrade, and evaluation increasingly relies on protocol-level validation and statistically meaningful tests under the stated access model, rather than full classical reproduction of outputs.
A complementary strategy is to pivot from deterministic data-access pipelines to sampling-centric objectives, treating quantum devices as native samplers for families of distributions that are conjectured to be classically hard to sample under matched access assumptions \cite{huang2025generative}.

The empirical record emphasizes that methodology and systems integration are not peripheral details but part of the scientific content of any advantage claim. Reported performance becomes interpretable only when shot budgets, mitigation choices, optimizer settings, hyperparameter search spaces, reproducibility under drift, and tuned baselines are specified together ,so that conclusions are attributable to the declared contract rather than to hidden degrees of freedom. Error mitigation can enable meaningful finite-size demonstrations, but its bias-variance and sampling-cost tradeoffs typically steepen with circuit depth and target accuracy, suggesting that sustained accuracy and reliability for large-scale, deeply iterative learning loops will increasingly align with fault-tolerant error correction rather than mitigation alone. Consequently, the most robust pathway to practical advantage is to navigate these coupled tensions through explicit, verifiable contracts; to target problem classes in which data interface and readout demands match realistic access models; and to articulate fault-tolerant roadmaps in which scaling claims are tied to auditable cost accounting and validation procedures.

\subsection{Future directions and roadmap}\label{subsec:future_directions}

\begin{figure*}[tb]\centering
\includegraphics[width=\linewidth]{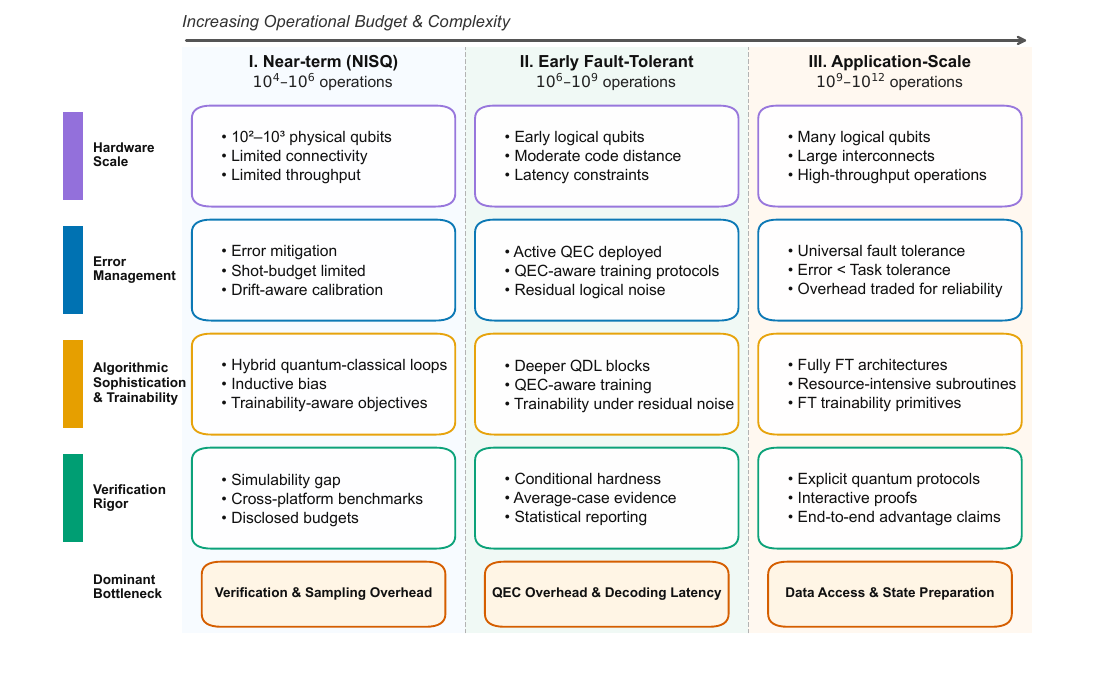}
\caption{Strategic roadmap for QDL, illustrating the coupled evolution of four dimensions: hardware scale, error management, algorithmic sophistication, and verification rigor. We distinguish three capability regimes by heuristic operation budgets and error-handling assumptions \cite{eisert2025mind,preskill2025beyond}. Here, an ``elementary operation" means one primitive physical action in a single compiled circuit execution (one QPU call), counting each native single-qubit gate, two-qubit gate, and single-qubit measurement once (summed over all qubits). Under this convention: (i) the near-term NISQ regime spans $\sim 10^4$--$10^6$ elementary operations per circuit execution, dominated by hybrid training and error mitigation; (ii) an early fault-tolerant regime spans $\sim 10^6$--$10^9$, enabling deeper circuits and explicitly QEC-aware training protocols; and (iii) an application-scale fault-tolerant regime spans $\sim 10^9$--$10^{12}$, where fully fault-tolerant quantum deep architectures and resource-intensive algorithmic subroutines become feasible.}
\label{fig:roadmap_qdl}
\end{figure*}

Progress in QDL requires simultaneous, rather than isolated, advances across four dimensions: (i) hardware scale, encompassing the number, quality, and connectivity of qubits; (ii) error management, spanning strategies from mitigation to full correction; (iii) algorithmic sophistication and trainability, reflecting the depth and inductive bias of deployable QDL architectures; and (iv) verification rigor, addressing the certification of results in regimes beyond classical simulation.
Figure~\ref{fig:roadmap_qdl} maps these dimensions across three operational regimes, explicitly tracking the dominant bottlenecks that constrain end-to-end utility.
The core thesis is that expanding qubit counts without commensurate advances in error management and verification limits effective circuit depth to classically reproducible or too shallow regimes. Verification needs to be elevated to a first-class design constraint, parallel to hardware and algorithm design.

\textit{Near-term: NISQ hardware and verification-aware hybrid workflows.}---The near-term regime is defined by NISQ hardware \cite{preskill2018quantum} with limited depth and throughput, where hybrid quantum-classical models compete with strong quantum-inspired classical baselines.
Error management relies on QEM~\cite{cai2023quantum}; however, because QEM techniques can incur large, and in some settings exponential, sampling overheads as a function of circuit depth and/or target precision, their viability is strictly bounded by the total shot budget and device drift. A central methodological challenge is the verification-simulation gap: as experiments enter regimes that are difficult to classically simulate, verification need to be treated as part of the resource contract rather than as an afterthought.
Accordingly, near-term priorities include: (i) designing trainability-aware architectures with explicit inductive bias (e.g., locality or symmetry) \cite{meyer2023exploiting,ragone2024lie}; (ii) tightening algorithm-hardware codesign to match ans\"atze, measurement strategies, and gradient estimators to device topology and calibrated noise \cite{ji2025algorithm}; and (iii) institutionalizing rigorous, multi-platform benchmarking against tuned classical baselines with disclosed budgets \cite{bowles2024better}.
Applications prioritize quantum-native data, such as learning from physical experiments or quantum sensors, bypassing the prohibitive cost of state preparation \cite{mcardle2020quantum,sajjan2022quantum,guo2024harnessing,de2021materials,merchant2023scaling,gentile2021learning}.
The open question in QDL is whether a practical quantum advantage can be demonstrated and certified under realistic end-to-end constraints, without collapsing into either unverified beyond-simulability behavior or a classically reproducible surrogate.

\textit{Mid-term: early fault-tolerant systems and QEC-aware training.}---The mid-term horizon targets early fault-tolerant quantum computers, with operational budgets expanding to approximately $10^6$ to $10^9$ elementary operations. This regime enables deeper circuits and longer training horizons but introduces the QEC overhead and decoding latencies as the primary bottlenecks, including the logical cost ratio (physical-to-logical qubits) and the real-time syndrome-decoding latency \cite{eisert2025mind,ryan2021realization}.
Research imperatives therefore center on (i) realizing low-overhead codes, such as quantum low-density parity-check codes \cite{breuckmann2021quantum}, on hardware with sufficient connectivity \cite{tillich2014quantum,gottesman2013fault,komoto2025quantum}; (ii) integrating decoding and control into high-throughput hybrid training loops; and (iii) developing QEC-aware training protocols that maintain trainability under residual logical noise by balancing gradient variance against syndrome costs.
Theoretically, because worst-case complexity guarantees are often inaccessible, certification may increasingly rely on conditional average-case hardness evidence under carefully stated computational assumptions \cite{huang2025vast}.
The central open question is whether a net QDL advantage survives the full fault-tolerant resource contract, including overheads and latency, while remaining verifiable beyond classical reach.

\textit{Long-term: FASQ and the frontier of quantum intelligence.}---The long-term horizon encompasses fault-tolerant application-scale quantum (FASQ) systems \cite{preskill2025beyond}, with operation budgets exceeding $10^{9}$--$10^{12}$ operations. In this regime, error management aims for scalable fault tolerance, suppressing logical error rates below task-dependent tolerances at the cost of substantial overhead rather than eliminating them.
This capability could enable the realization of autonomous quantum agents that integrate quantum sensors, quantum memory, and deep networks with minimal classical intermediation.
With gate error rates no longer the primary constraint, the bottleneck transitions to data access and state preparation, where QRAM or encoding costs may dominate for classical data.
Consequently, long-horizon priorities accordingly emphasize (i) data-frugal QDL protocols operating directly on quantum data or compressed classical encodings; (ii) fault-tolerant trainability primitives, such as parameter-shift-compatible logical gate decompositions and gradient estimators that account for QEC overhead, to enable end-to-end optimization at application scale; (iii) quantitative frameworks for rigorously bounding the complexity of quantum advantage and quantum generalization in hybrid systems, clarifying how quantum models interact with next-generation classical systems. A key open question is whether certain QDL advantages exist that defy classical certification, so that verification would require explicitly quantum protocols under an explicit contract rather than purely classical postprocessing.

Across all three regimes, this roadmap is intended as a constraint-driven guide rather than a prescriptive promise. The defining scientific question is no longer merely whether larger devices will enable more complex models; rather, it is whether we can establish advantage claims that are end-to-end, resource-accounted, and strictly verifiable under realistic access, noise, and measurement assumptions.
Achieving this demands coordinated progress across quantum information, learning theory, systems engineering, and those specific application domains where quantum-native data and objectives arise naturally.
Over the past decade, the field has transitioned from theoretical speculation into a rigorous discipline characterized by reproducible experiments, sharpened theory, and emerging design principles. The coming era will determine whether these principles cohere into a reliable, operational advantage. This endeavor, at the confluence of quantum computing and artificial intelligence, constitutes a highly ambitious scientific pursuit, one that fundamentally probes the physical limits of computation and the mechanics of intelligence.

\begin{acknowledgments}

The authors thank Lourens van Niekerk and Peter Röseler for insightful suggestions. Y.J. acknowledges support from the programme ``Profilbildung 2022", an initiative of the Ministry of Culture and Science of the State of North Rhine-Westphalia, within the framework of the project ``Quantum-based Energy Grids (QuGrids)".
Z.-Y. Chen acknowledges support from the National Key Research and Development Program of China (Grant No. 2023YFB4502500).
M.R. was supported by the Dieter Schwarz foundation through the Fraunhofer Heilbronn Research and Innovation Centers HNFIZ.
O.A. and M.K. acknowledge the support from DLR through funds provided by BMFTR (50WM2347, 50MW2247) and by the State of Berlin within the Pro FIT program under Grant No. 10206829.
D.W. acknowledges support from the project J\"ulich UNified Infrastructure for Quantum computing (JUNIQ) that has received funding from the Federal Ministry of Research, Technology and Space (BMFTR) and the Ministry of Culture and Science of the State of North Rhine-Westphalia.
C.S. was supported by the European Union’s Horizon 2020 Research and Innovation Programme, grant agreement 101147319 (EBRAINS 2.0 Project), the Helmholtz Association port-folio theme “Supercomputing and Modeling for the Human Brain”, and the Helmholtz Association’s Initiative and Networking Fund through the Helmholtz International BigBrain Analytics and Learning Laboratory (HIBALL) under the Helmholtz International Lab grant agreement InterLabs-0015, and by HELMHOLTZ IMAGING, a platform of the Helmholtz Information \& Data Science Incubator [X-BRAIN, grant number:  ZT-I-PF-4-061].
L.M. acknowledges support by the ERDF of the European
Union and by ``Fonds of the Hamburg Ministry of Science, Research, Equalities and Districts (BWFGB)", as well as funding by the Cluster of Excellence “Advanced Imaging of Matter” (EXC 2056) Project
No.390715994.

\end{acknowledgments}




\bibliography{refs}

\end{document}